\documentclass[twocolumn]{aastex701}
\newcommand\kms{\ensuremath{{\rm km}~{\rm s}^{-1}}}

\usepackage[version=4]{mhchem}
\usepackage{soul}

\usepackage{CJK}

\begin{document}

\begin{CJK}{UTF8}{}
\CJKfamily{mj}

\title{Revealing Hidden Excitation in the Core of NGC~4696: New Infrared Diagnostic Diagrams with JWST}

\author[0009-0000-4104-9909]{Olivia R. Pereira}
\affiliation{D\'{e}partement de Physique, Universit\'{e} de Montr\'{e}al, Succ. Centre-Ville,
Montr\'{e}al, Qu\'{e}bec, H3C 3J7, Canada}
\email{...}  

\author[0000-0001-7271-7340]{Julie Hlavacek-Larrondo}
\affiliation{D\'{e}partement de Physique, Universit\'{e} de Montr\'{e}al, Succ. Centre-Ville, Montr\'{e}al, Qu\'{e}bec, H3C 3J7, Canada}
\email{...}  

\author[0000-0003-4701-8497]{Michael Reefe}
\affiliation{Kavli Institute for Astrophysics and Space Research,
Massachusetts Institute of Technology,
77 Massachusetts Avenue, Cambridge, MA 02139, USA}
\email{}

\author[0000-0003-4503-6333]{Gary J. Ferland}
\affiliation{Department of Physics and Astronomy, University of Kentucky,
505 Rose Street, Lexington, KY 40506, USA}
\email{}

\author[0000-0002-3173-1098]{Hyunseop Choi (최현섭)}
\affiliation{Department of Astronomy, University of Michigan, 1085 S. University, Ann Arbour, MI 48109, USA}
\email{...}  

\author[0009-0008-7156-4678]{Mathieu Marquis}
\affiliation{D\'{e}partement de Physique, Universit\'{e} de Montr\'{e}al, Succ. Centre-Ville,
Montr\'{e}al, Qu\'{e}bec, H3C 3J7, Canada}
\email{...}  

\author[0000-0003-2405-7258]{Jorge Barrera-Ballesteros}
\affiliation{Instituto de Astronom\'{i}a, Universidad Nacional Aut\'{o}noma de M\'{e}xico,
A.P. 70-264, 04510 CDMX, M\'{e}xico}
\email{}

\author[0000-0002-2478-5119]{Benjamin Vigneron}
\affiliation{D\'{e}partement de Physique, Universit\'{e} de Montr\'{e}al,
Succ. Centre-Ville, Montr\'{e}al, Qu\'{e}bec, H3C 3J7, Canada}
\email{}

\author[0000-0001-5880-0703]{Ming Sun}
\affiliation{Department of Physics and Astronomy, University of Alabama in Huntsville,
301 Sparkman Drive, Huntsville, AL 35899, USA}
\email{}

\author[0000-0003-1398-5542]{Rebecca E. A. Canning}
\affiliation{Institute of Cosmology and Gravitation, University of Portsmouth,
Dennis Sciama Building, Portsmouth, PO1 3FX, UK}
\email{}

\author[0000-0001-6495-7731]{Gregory Taylor}
\affiliation{Department of Physics and Astronomy, University of New Mexico,
Albuquerque, NM 87131, USA}
\email{}

\author[0000-0003-0475-9375]{Lo\"ic Albert}
\affiliation{Institut Trottier de recherche sur les exoplan\`etes and D\'epartement de Physique, Universit\'e de Montr\'eal, 1375 Avenue Th\'er\`ese-Lavoie-Roux, Montr\'eal, QC H2V 0B3, Canada}
\email{}

\author[0000-0003-2388-8172]{Francesco D'Eugenio}
\affiliation{Kavli Institute for Cosmology, University of Cambridge,
Madingley Road, Cambridge, CB3 0HA, UK}
\affiliation{Cavendish Laboratory, University of Cambridge,
19 JJ Thomson Avenue, Cambridge, CB3 0HE, UK}
\email{}

\author[0000-0002-2808-0853]{Megan Donahue}
\affiliation{Department of Physics and Astronomy, Michigan State University,
East Lansing, MI 48824, USA}
\email{}

\author[0000-0002-9378-4072]{Andrew C. Fabian}
\affiliation{Institute of Astronomy, University of Cambridge,
Madingley Road, Cambridge, CB3 0HA, UK}
\email{}

\author[0000-0001-8608-0408]{John S. Gallagher}
\affiliation{Department of Physics and Astronomy, Macalester College,
1600 Grand Avenue, Saint Paul, MN 55105, USA}
\email{}

\author[0000-0002-7326-5793]{Marie-Lou Gendron-Marsolais}
\affiliation{D\'{e}partement de Physique, de G\'{e}nie Physique et d'Optique,
Universit\'{e} Laval, Qu\'{e}bec, QC G1V 0A6, Canada}
\email{}

\author[0000-0002-2421-1350]{Pierre Guillard}
\affiliation{Sorbonne Universit\'{e}, CNRS, UMR 7095, Institut d'Astrophysique de Paris,
98bis bd Arago, 75014 Paris, France}
\email{}

\author[0000-0002-3680-5420]{Minghao Guo (郭明浩)}
\affiliation{Department of Astrophysical Sciences, Princeton University, Princeton, NJ 08540, USA}
\email{...}  

\author[0000-0001-5600-0534]{Nina Hatch}
\affiliation{School of Physics and Astronomy, University of Nottingham,
University Park, Nottingham, NG7 2RD, UK}
\email{}

\author[0000-0002-4460-9892]{Ralf Kotulla}
\affiliation{Department of Astronomy, University of Wisconsin--Madison,
475 N. Charter Street, Madison, WI 53706, USA}
\email{}

\author[0000-0001-5262-6150]{Yuan Li (黎原)}
\affiliation{Department of Astronomy, University of Massachusetts,
Amherst, MA 01003, USA}
\email{}

\author[0000-0002-4985-3819]{Roberto Maiolino}
\affiliation{Kavli Institute for Cosmology, University of Cambridge,
Madingley Road, Cambridge, CB3 0HA, UK}
\affiliation{Cavendish Laboratory, University of Cambridge,
19 JJ Thomson Avenue, Cambridge, CB3 0HE, UK}
\affiliation{Department of Physics and Astronomy, University College London,
Gower Street, London WC1E 6BT, UK}
\email{}

\author[0000-0003-2475-124X]{Allison Man}
\affiliation{Department of Physics \& Astronomy, University of British Columbia,
6224 Agricultural Road, Vancouver, BC V6T 1Z1, Canada}
\email{}

\author[0000-0001-5226-8349]{Michael A. McDonald}
\affiliation{Kavli Institute for Astrophysics and Space Research,
Massachusetts Institute of Technology,
77 Massachusetts Avenue, Cambridge, MA 02139, USA}
\email{}

\author[0000-0002-2622-2627]{Brian R. McNamara}
\affiliation{Department of Physics and Astronomy, University of Waterloo,
200 University Avenue West, Waterloo, ON N2L 3G1, Canada}
\email{}

\author[0000-0001-6638-4324]{Valeria Olivares}
\affiliation{Departamento de F\'{i}sica, Universidad de Santiago de Chile,
Av. Victor Jara 3659, Santiago 9170124, Chile}
\affiliation{Center for Interdisciplinary Research in Astrophysics and Space Exploration (CIRAS),
Universidad de Santiago de Chile, Santiago 9170124, Chile}
\email{}

\author[0009-0003-0932-2487]{Marine Prunier}
\affiliation{D\'{e}partement de Physique, Universit\'{e} de Montr\'{e}al,
Succ. Centre-Ville, Montr\'{e}al, Qu\'{e}bec, H3C 3J7, Canada}
\affiliation{Max-Planck-Institut f{\"u}r Astronomie, K{\"o}nigstuhl 17, D-69117 Heidelberg, Germany}
\email{}

\author[0000-0002-1510-4860]{Christopher S. Reynolds}
\affiliation{Department of Astronomy, University of Maryland,
College Park, MD 20742-2421, USA}
\affiliation{Joint Space Science Institute (JSI), University of Maryland,
College Park, MD 20742-2421, USA}
\email{}

\author[0000-0003-2001-1076]{Carter Rhea}
\affiliation{Dragonfly Focused Research Organization, 150 Washington Avenue, Santa Fe, 87501, NM, USA}
\affiliation{Centre de Recherche en Astrophysique du Québec (CRAQ), Québec, QC G1V 0A6, Canada}
\email{}

\author[0000-0001-7597-270X]{Annabelle Richard-Laferri\`{e}re}
\affiliation{Institute of Astronomy, University of Cambridge,
Madingley Road, Cambridge, CB3 0HA, UK}
\email{}

\author[0000-0001-5208-649X]{Helen R. Russell}
\affiliation{School of Physics and Astronomy, University of Nottingham,
University Park, Nottingham, NG7 2RD, UK}
\email{}

\author[0000-0001-9633-5750]{Philippe Salom\'{e}}
\affiliation{LERMA, Observatoire de Paris, PSL Research University, CNRS,
Sorbonne Universit\'{e}, 75014 Paris, France}
\email{}

\author[0000-0001-8176-7665]{Prathamesh Tamhane}
\affiliation{Department of Physics and Astronomy, University of Alabama in Huntsville,
301 Sparkman Drive, Huntsville, AL 35899, USA}
\email{}

\author[0000-0001-5223-1888]{Auriane Thilloy}
\affiliation{D\'{e}partement de Physique, Universit\'{e} de Montr\'{e}al,
Succ. Centre-Ville, Montr\'{e}al, Qu\'{e}bec, H3C 3J7, Canada}
\email{}

\author[0000-0002-5445-5401]{Grant R.~Tremblay}
\affiliation{Harvard-Smithsonian Center for Astrophysics,
60 Garden Street, Cambridge, MA, USA}
\email{}

\author[0000-0002-3514-0383]{G. Mark Voit}
\affiliation{Department of Physics and Astronomy, Michigan State University,
East Lansing, MI 48824, USA}
\email{}

\author[0000-0002-6413-4142]{Stephen A. Walker}
\affiliation{Department of Physics and Astronomy, University of Alabama in Huntsville,
301 Sparkman Drive, Huntsville, AL 35899, USA}
\email{}

\begin{abstract}
The warm molecular and ionized gas in filamentary nebulae of cool-core brightest cluster galaxies (BCGs) provide a sensitive probe of the poorly understood exchange of energy between the hot intracluster medium (ICM), cold molecular gas, and active galactic nucleus (AGN) feedback. Optical line ratios commonly yield similar low-ionization nuclear emission-line region (LINER)-like classifications across these systems, but do not resolve how the excitation of warm molecular gas varies between compact nuclear structures and surrounding filaments. We present JWST/NIRSpec integral-field spectroscopy of the central $3\arcsec\times3\arcsec$ ($620\times620~{\rm pc^2}$) of the Centaurus cluster BCG, NGC~4696. The near-infrared (NIR) data resolve H$I$ and ro-vibrational H$_2$ emission across compact nuclear apertures, a circumnuclear swirl, and filamentary structures. H$_2$ excitation diagrams yield characteristic low-excitation temperatures of $T_{\rm ex}\simeq1.8-2.2\times10^3$~K, while curvature at higher energies requires hotter or non-thermal components. The H$_2$/Pa$\alpha$ balance varies strongly: filamentary regions have H$_2$ 1--0 S(1)/Pa$\alpha$ ratios 4--8 times higher than the compact nuclear apertures. In contrast, matched-aperture optical diagnostic diagrams show substantially less regional separation, with all regions occupying a low-ionization locus. We thus introduce new spatially resolved NIR diagnostic diagrams that reveal distinct molecular excitation conditions across the nucleus, swirl, and filaments. Ratios are broadly consistent with contributions from AGN-related photoionization, ICM-related cooling radiation, and non-radiative heating, although the present data do not uniquely determine their relative importance. These results demonstrate how spatially resolved NIR spectroscopy reveals excitation variations hidden in conventional optical classifications, providing a new route to constrain the multiphase cooling--feedback cycle in cluster cores.
\end{abstract}


\section{Introduction}

Understanding how the most massive galaxies nurture, disrupt, or recycle their gas is crucial to understanding galaxy formation and evolution. Brightest cluster galaxies (BCGs) are the most luminous galaxies in their host clusters and reside near the minima of the cluster gravitational potentials, where they interact strongly with the surrounding intracluster medium (ICM) through processes dominated by active galactic nucleus (AGN) feedback \citep{fabian_observational_2012, mcnamara_mechanical_2012}. The hot, X-ray-emitting ICM loses energy via radiation and should, in theory, cool rapidly, leading to massive star formation in cluster cores. However, mechanical energy released by the central supermassive black hole in the form of jets heats the ICM, inflating cavities and driving processes such as turbulence, shock waves and sound waves that inhibit runaway cooling \citep[e.g.][]{mcnamara_mechanical_2012, gaspari_constraining_2013}. This feedback loop is thought to be self-regulating, with cold gas condensing from the ICM via thermal instabilities and raining back onto the AGN to fuel its continued activity, maintaining balance on kiloparsec to sub-parsec scales \citep{voit_precipitation-regulated_2015, tremblay_galaxy-scale_2018}. Metals produced by star formation and stellar evolution enrich the ICM, and their distribution is then reshaped by large-scale motions such as bulk sloshing and uplift, which move gas away from the centre, form eddies and filamentary structures, and promote dust grain formation and energy redistribution in cluster cores \citep[e.g.][]{churazov_evidence_1999, panagoulia_searching_2013, tremblay_galaxy-scale_2018, ghodsi_molecular_2024}.

Within the cluster's core, gas exists in multiple phases and is subject to competing heating and cooling flows. The hot ICM is enriched by metals produced in star formation and supernovae, and sharp metallicity gradients observed with Chandra and XMM-Newton reveal ongoing enrichment and redistribution processes \citep{panagoulia_searching_2013}. Furthermore, the observed central depression of iron and other reactive metals may reflect a combination of resonant scattering effects and a genuine abundance drop, with the latter likely associated with metals locked into dust grains that further alter the cooling properties of the gas \citep{sanders_resonance_2006, Olivarex_metalicity_2026}.
 
NGC~4696 is the brightest cluster galaxy of the nearby Centaurus system, whose estimated halo mass is M$_{200} \sim 2 \times 10^{13}$~M$_\odot$ \citep{matilsky_x-ray_1985, allen_rosat_1994}, provides one of the best nearby laboratories in which to study these interconnected processes. Located at a stellar redshift of $z=0.01003$ \citep[][from MUSE spectroscopy; we adopt a nominal luminosity distance of $43.5$~Mpc and a physical scale of $206$~pc~arcsec$^{-1}$, approximately consistent with $H_0=70$~km~s$^{-1}$~Mpc$^{-1}$, $\Omega_{\rm m}=0.3$, and $\Omega_\Lambda=0.7$]{collaboration_bulk_2025}, it dominates the gravitational and baryonic environment of its host cluster. The adopted distance and physical scale could be approximately $25\%$ smaller owing to departures from pure Hubble expansion.  The galaxy's assembly history, reflected in the bimodal metallicity distribution of its globular cluster system (Fe/H $\approx -1.0$ and $-0.25$), points to a history of mergers and dynamical interactions within a turbulent environment \citep{churazov_evidence_1999, federle_turbulent_2024}. Recent XRISM/Resolve observations further show that the intracluster gas in the central $\lesssim 30$ kpc is undergoing coherent bulk motions and velocity gradients, with the ICM significantly blueshifted relative to the stellar body of NGC~4696, consistent with sloshing of the core gas relative to the BCG and large-scale gas flows driven by past mergers \citep{collaboration_bulk_2025}.

NGC~4696 hosts a powerful central AGN and an extended multiphase filamentary nebula, making it a nearby example of the broader cool-core BCG population discussed above. Its radio jets inflate X-ray cavities with radii of order a few kiloparsecs, while its $\sim40$~kpc nebula contains thread-like filaments spanning nearly six decades in temperature, from $\sim5\times10^7$~K X-ray plasma to cold molecular gas \citep{sanders_constraints_2010,fabian_optical_1982,crawford_extended_2005,johnstone_discovery_2007,sparks_imaging_1989,sanders_very_2016,fabian_hst_2016,Olivares_2019,hamer_discovery_2019}.
Such nebulae occur preferentially in systems with central entropy $K_0\lesssim30~{\rm keV\,cm^2}$ and short ICM cooling times \citep[e.g.,][]{Cavagnolo_Donahue_Voit_Sun_2009,Kirkpatrick_2009,panagoulia_searching_2013,Olivares_2019,olivares_hx-ray_2025}. They are highly multiphase, containing gas from the hot X-ray phase at $T\sim10^7$~K to cold molecular gas at $T\sim30$--$50$~K, with close spatial correspondence among X-ray emission, optical-line emission, and CO- and HCN-traced molecular components \citep[e.g.][]{Salome_2008,edge_detection_2001,russell_driving_2019,olivares_hx-ray_2025}.

In many systems, these filaments extend over several kiloparsecs while remaining remarkably narrow, with characteristic widths of $\sim60$--$70$~pc. Their coherent morphologies often closely trace radio bubbles and X-ray cavities, and individual filaments can exhibit ordered kinematics along their lengths. Magnetic tension and magnetic draping may therefore play an important role in supporting filaments and enabling their survival within the hot ICM \citep{Fabian_Johnstone_Sanders_Conselice_Crawford_Iii_Zweibel_2008}. Although cool-core BCG filaments can contain substantial cold-gas masses, star formation is often spatially offset from both the densest molecular gas and the brightest line-emitting structures. Feedback and environmental processes, including ram pressure and turbulence, can strongly influence where and when star formation occurs; correspondingly, while depletion times in cluster centres can resemble those in normal disk galaxies, the geometry and timing of star formation relative to the filamentary gas differ from those in standard disks \citep[e.g.][]{Donahue_2011, McDonald_2015,Olivares_2019}.

The excitation of these filaments remains uncertain. In many cool-core BCGs, low-ionization, LINER-like optical spectra and strong molecular emission cannot be explained by pure stellar H~{\sc ii}-region photoionization alone \citep[e.g.,][]{Heckman_1989, Edwards_2009,Donahue_2011,hamer_discovery_2019, Olivares_2019}. In the central few kiloparsecs and in the extended multiphase filaments that drape around radio bubbles and X-ray cavities, the line emission may reflect a combination of cooling-ICM radiation, AGN-driven turbulence, particle heating, and AGN photoionization \citep[e.g.][]{Crawford_1999,ferland_collisional_2009,russell_driving_2019}. Optical IFU and long-slit studies show that BCG nuclear spectra are diverse: some nuclei have clear AGN-like or star-forming signatures, whereas many exhibit low-ionization spectra likely produced by a mixture of photoionization and additional heating \citep[e.g.][]{voit_donahue_1997, Crawford_1999,Edwards_2009}. On larger scales, the tight morphological and kinematic correspondence among X-ray, optical, and molecular tracers demonstrates the multiphase nature of the filaments, but their low-ionization optical spectra and strong H$_2$ emission still require non-stellar heating contributions such as shocks, turbulence, or particle heating \citep[e.g.][]{mcdonald_origins_2011,mcdonald_optical_2012,mcdonald_massive_2012,mcdonald_hst_2014,tremblay_multiphase_2012,ferland_collisional_2009,Donahue_2011,russell_driving_2019}.

For NGC~4696 specifically, plausible contributors include photoionization by the AGN or by cooling intracluster gas, excitation and ionization by high-energy particles such as cosmic rays \citep{Mateusz_2018_Halpha}, and mechanical heating associated with AGN-inflated bubbles \citep{canning_deep_2011,Mateusz_2018_Halpha,polles_excitation_2021}. Distinguishing among these processes is essential because they imply different couplings among the hot ICM, the central AGN, and the cold molecular phase. Deep HST H$\alpha$ imaging revealed a striking swirl of ionized gas within the black hole sphere of influence \citep{fabian_hst_2016}; this structure offers an especially valuable setting in which to test how excitation changes from the circumnuclear environment into the surrounding filament network.

Previous infrared studies provide particularly direct evidence that the line-emitting gas in cool-core BCGs requires an energy source beyond ordinary star formation. Using \textit{Spitzer}/IRS, \citet{Donahue_2011} measured polycyclic aromatic hydrocarbons (PAHs), strong pure-rotational H$_2$ S(1), S(2), and S(3) lines, [Ne~{\sc ii}], [Ne~{\sc iii}], and warm dust continuum emission over $24$--$70~\mu{\rm m}$ in nine cool-core BCGs: 2A~0335+09, Abell~478, Abell~1068, Abell~1795, Abell~1835, Abell~2597, Hydra~A, MS~0735.6+7421, and PKS~0745$-$19. They found that the H$_2$ and neon lines are over-luminous relative to their total infrared emission, raising the question of luminosity relative to ordinary star-forming galaxies and implying that an additional non-stellar heating channel is required for the molecular gas and at least part of the ionized component.

Earlier, \citet{Salome_2008} combined CO and near-infrared observations of Perseus to show that its filaments contain cold molecular gas traced by CO, warm H$_2$ seen in the near-infrared, and hotter $\sim10^4$~K ionized gas traced by H$\alpha$ and other optical lines; their warm-H$_2$ emission could not be reproduced by simple fast shocks alone. In the Centaurus cluster, \citet{mittal_herschel_2011} used \textit{Herschel} to detect [C~{\sc ii}], [O~{\sc i}], and [N~{\sc ii}], finding that [C~{\sc ii}] emission extends over several kiloparsecs and closely follows the morphology and kinematics of H$\alpha$. This result links the far-infrared, optical, and soft-X-ray phases to a common power source within the filaments. Building on these data, \citet{polles_excitation_2021} modelled intracluster filaments around BCGs and showed that attenuated EUV and soft-X-ray radiation from cooling ICM gas, combined with modest additional heating such as turbulence or particle heating, can reproduce much of the optical-to-infrared filament spectrum, including strong H$_2$ emission, while tensions remain in the CO emission and detailed cloud structure.

Together, these studies establish that warm molecular and ionized gas are central to the cooling--feedback cycle in BCGs, but they leave open how the relevant excitation channels vary within individual structures on tens-to-hundreds-of-parsecs scales. This limitation is particularly important for NGC~4696: far-ultraviolet observations \citep[FUV;][]{danforth_far-uv_2016} and high-spectral-resolution X-ray observations \citep{anderson_searching_2016} have constrained ionization mechanisms, microturbulence, and AGN-driven bulk motions in the galaxy, but they do not resolve the temperature, density, or excitation structure of the warm molecular phase. Resolving that phase is essential for linking the cold filamentary molecular gas to the hot ICM. The near-infrared surface-brightness profile of NGC~4696 \citep[e.g.,][]{mur_near-infrared_2006} and the correspondence between its optical filaments and X-ray emission \citep{crawford_extended_2005} reinforce the need for a unified description of its interwoven physical processes.

JWST/NIRSpec IFU spectroscopy now provides the spatially resolved near-infrared coverage required to probe the warm ionized and molecular gas in the core of NGC~4696. The observations used here cover $1.7$--$3.1~\mu{\rm m}$, including hydrogen recombination lines and warm ro-vibrational H$_2$ transitions that diagnose the physical, chemical, and excitation state of the gas. By resolving these tracers across the nucleus and inner filament network, we test whether the warm-gas emission follows a common excitation sequence or instead reflects spatially varying combinations of AGN-related photoionization, ICM-related radiation, and non-radiative heating.

These observations also complement the broader multiphase-feedback picture involving molecular-gas uplift, turbulent heating, and AGN-driven fountain flows \citep{tremblay_galaxy-scale_2018,ghodsi_molecular_2024,leftley_jwstnirspec_2024}. \citet{hlavacek-centaurus-2026} show that the H$\alpha$ swirl is a rotating circumnuclear disk connected to the surrounding filament network, consistent with a pathway that channels gas from kiloparsec-scale cooling filaments to the central $\sim100$~pc and ultimately to the supermassive black hole. \citet{marquis_mapping_2026} provides a complementary analysis of gas and stellar kinematics across the NIRSpec field of view and in comparison with previous observations.

Near-infrared spectroscopy provides a particularly powerful view of multiphase gas because it simultaneously accesses \ion{H}{1} recombination lines and ro-vibrational transitions of H$_2$. Hydrogen recombination lines, including Pa$\alpha$, trace ionized gas through radiative recombination and provide a reference for the strength of the ionized component, while the near-infrared H$_2$ 1--0 ro-vibrational lines arise from highly excited ro-vibrational levels of the ground electronic state and trace warm molecular gas. The brightest transitions, such as H$_2$ 1--0 S(1), S(3), and S(5), sample upper-level energies of several thousand kelvin and are therefore sensitive to the heating and excitation processes operating at molecular-gas interfaces \citep{turner_quadrupole_1977, black_fluorescent_1987, hollenbach_molecule_1989}. Unlike cold molecular-gas tracers such as CO, these lines do not measure the bulk molecular reservoir; rather, they probe the comparatively small gas fraction that is warm or non-thermally excited.

The relative strengths and excitation structure of the H$_2$ lines provide insight into the dominant energy source. In photodissociation regions, far-ultraviolet pumping can populate ro-vibrational H$_2$ levels and produce fluorescent line ratios, with collisions modifying the emergent spectrum at sufficiently high density \citep{black_fluorescent_1987, sternberg_infrared_1989}. Conversely, shocks, turbulent dissipation, and suprathermal-particle heating can collisionally excite H$_2$ and yield strong thermal or multi-temperature molecular emission \citep{hollenbach_molecule_1989, guillard_h2_2009, ferland_collisional_2009}. Ratios between H$_2$ and \ion{H}{1} recombination lines, together with ratios among H$_2$ transitions of different excitation energy, therefore distinguish variations in the molecular-to-ionized-gas balance from changes in the internal excitation of the warm molecular phase. Near-infrared H$_2$/recombination-line diagnostics have consequently been applied to star-forming nuclei, AGN, and low-ionization systems, where elevated H$_2$ relative to recombination emission frequently signals an important non-stellar heating contribution \citep[e.g.][]{ogle_jet-powered_2010,  rodriguez-ardila_2011, davies_fueling_2014}. 

In this paper, we present a JWST/NIRSpec IFU analysis of the central $3\arcsec\times3\arcsec$ ($\sim620\times620$~pc$^2$) of NGC~4696, focused on spatially resolved emission-line ratios and H$_2$ excitation. We compare these near-infrared measurements with matched-aperture MUSE optical diagnostics to test whether the compact nuclear, Swirl, and filamentary structures can be distinguished in their excitation properties. We then interpret the NIR diagnostic diagrams using \textsc{CLOUDY} reference photoionization sequences \citep{cloudy25}, building on companion analyses of the circumnuclear disk and gas/stellar kinematics \citep{hlavacek-centaurus-2026,marquis_mapping_2026}.

The paper is organized as follows. Section~\ref{sec:observations} describes the observations and data reduction. Section~\ref{sec:results} presents the spatially resolved line measurements, excitation diagrams, and diagnostic ratios. Section~\ref{sec:discussion} discusses their physical interpretation, and Section~\ref{sec:conclusion} summarizes the main conclusions.  We adopt a flat $\Lambda$CDM cosmology with $H_0=70~{\rm km\,s^{-1}\,Mpc^{-1}}$, $\Omega_{\rm m}=0.3$, and $\Omega_\Lambda=0.7$. At $z=0.01003$ \citep{collaboration_bulk_2025} \footnote{We retain the redshift and distance conventions adopted in the companion analyses of the same JWST/NIRSpec dataset for consistency. At the low redshift of NGC~4696, redshift-independent distance estimates can differ from the Hubble-flow distance because peculiar velocities are non-negligible. The adopted values are therefore used as analysis conventions throughout this work rather than as a new distance determination.}, this corresponds to a physical scale of $206$\,pc\,arcsec$^{-1}$. Unless stated otherwise, quoted uncertainties are $1\sigma$.

\section{Observations} \label{sec:observations}

\begin{figure*}
    \centering
    \includegraphics[width=\linewidth]{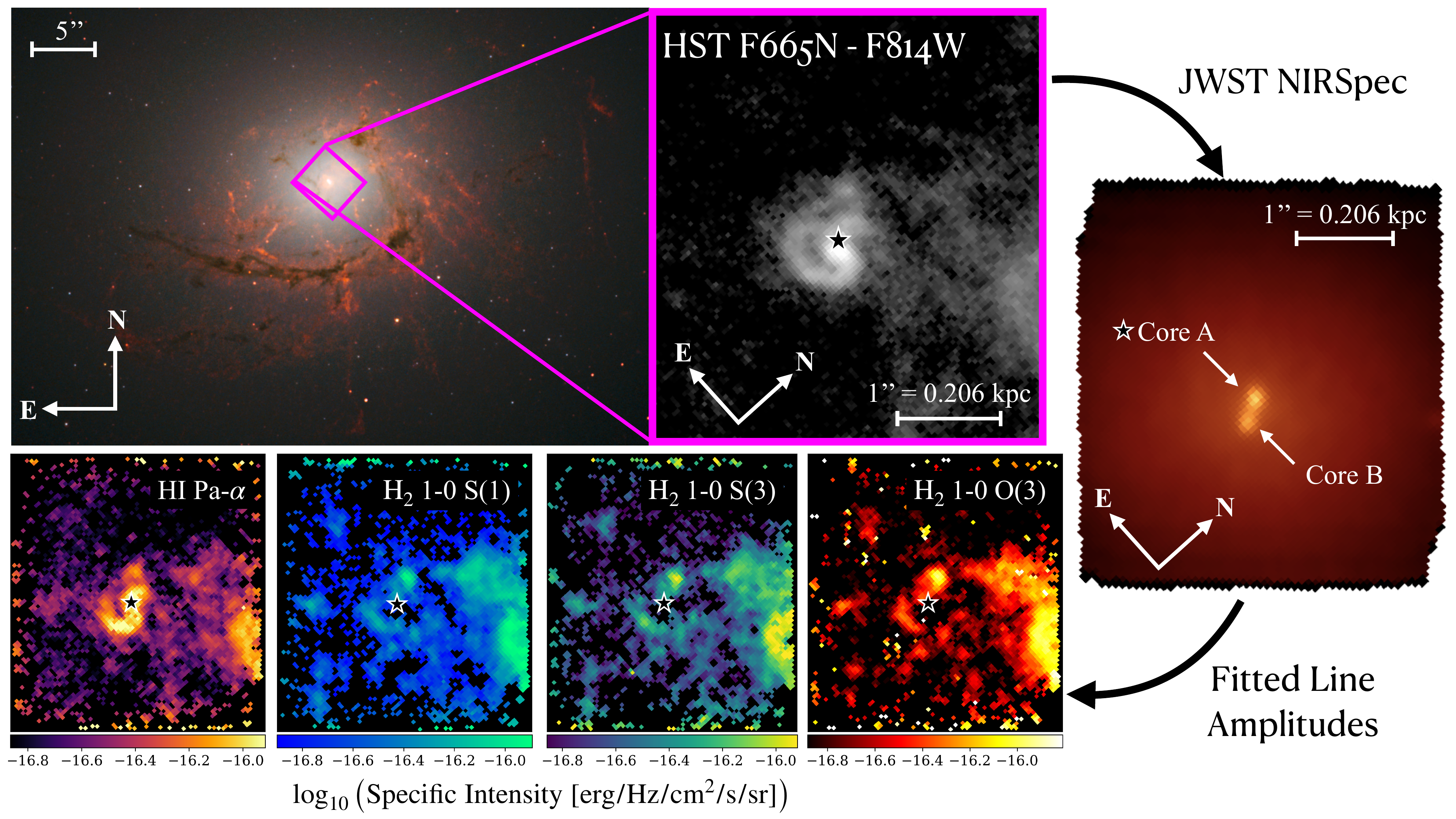}
    \caption{Multi-scale view of the BCG NGC~4696 in the Centaurus cluster and the JWST/NIRSpec IFU footprint. Upper left: Composite HST WFC3 image (F467M, F814W and H$\alpha$ F665N\footnote{DOI \url{10.17909/11gm-pb27}}, showing the extended dust lanes and outer ionized filamentary structures over several kiloparsecs. The 3''$\times$3'' NIRSpec IFU field of view is outlined with the magenta square, corresponding to the nucleus and associated circumnuclear swirl of gas. Upper middle: Zoomed continuum-subtraction HST H$\alpha$ (F665N - F814W) image of the central region, emphasizing the spiral pattern of ionized filaments and dust traced within the footprint as observed by \citet{fabian_hst_2016}. Right: NIRSpec G235H/F170LP IFU cube collapsed over the full 1.7--3.1$\mu$m range, revealing compact knots (labelled as \textit{Core A} and \textit{Core B}) of enhanced near-infrared emission. The IFU data are sampled at 0.05" ($\sim10$ pc) per spaxel after dithering. Lower panels: Logarithmic intensity maps of the Pa$\alpha$, $\rm H_2$ 1--0 S(1),  $\rm H_2$ 1--0 S(3) and $\rm H_2$ 1--0 O(3) lines, obtained using the \textsc{LOKI} spectral fitting pipeline (see Section \ref{sec:LOKI}). These maps trace the spatial distribution of ionized and warm molecular gas within the IFU field and reveal a strong structural correspondence between the warm Pa$\alpha$ $\rm H_2$-emitting filaments and the optically visible dust and H$\alpha$ swirl. North and east directions, as well as the angular-physical scale, are indicated in relevant panels. The location of \textit{Core A} is marked with a black star throughout the zoomed panels. All panels except the upper left share the same orientation and spatial aperture (3.0''$\times$3.0''; $\rm\sim 620\times620\rm ~pc^2$).}
    \label{fig:FOV}
\end{figure*} 

\subsection{Data Collection and Reduction}

\subsubsection{JWST NIRSpec}\label{sec: NIRSpec}

Observations of NGC~4696 were conducted with the \textit{JWST} NIRSpec IFU on 2025 June 5 (Program ID 05354, PI: Hlavacek-Larrondo\footnote{under the proposal titled ``Mapping a Black Hole Accretion Flow with JWST/NIRSpec''}). The observations used the G235H grating and F170LP filter, providing $1.7$--$3.1~\mu$m coverage at a spectral resolution of $R\simeq2700$ ($\sim110~{\rm km\,s^{-1}}$). The observations consist of a single visit with an effective exposure time of approximately 27,719 seconds ($\sim7.7$ hours) using the NRSIRS2 readout pattern. The IFU has a nominal field of view (FOV) of $3.0''\times3.0''$ ($\sim 620\times 620$ pc$^2$ at a luminosity distance 43.5 Mpc, see Figure~\ref{fig:FOV}). Ten-position medium cycling dithers were employed to improve spatial sampling and mitigate artifacts such as bad pixels and cosmic rays, extending the effective FOV through dither overlap, and enhancing the spatial sampling of the cube to 0.05'' spaxels through drizzle-weighted reconstruction. In addition, a micro-shutter assembly (MSA) leakage-calibration (\textit{leakcal}) exposure was obtained at one of the dither positions with the same exposure configuration to characterize and correct for flux leakage through the MSA. While border pixels exhibit reduced sensitivity due to edge effects, the primary science apertures (Figure~\ref{fig:spider}) are well-contained within the central $\sim3''\times3''$ region with median signal-to-noise ratio (SNR) $>30$ across the Pa$\alpha$ and $\rm H_2$ 1--0 $S(1)$, and $S(3)$ complexes. 

The raw data (\texttt{uncal} images) were retrieved from the archive and reduced with the standard JWST pipeline (Build 12.2, CRDS context 1464; \citealt{bushouse_2025}), following the same detector, calibration, and cube-building steps as \citealt{marquis_mapping_2026} and \citealt{hlavacek-centaurus-2026}, summarized as follows. Most of the reduction procedures were executed using the standard pipeline routines offered by STScI. During \texttt{calwebb\_detector1}, we utilized the \texttt{snowblind} software \citep{davies_2025_15791615} to enhance the identification and mitigation of snowball and shower artifacts. The removal of $1/f$ noise was carried out using the \texttt{nsclean} algorithm \citep{Rauscher2024}.
Finally, a customized outlier rejection step was performed to remove a small number of bad pixels with erroneous flux values.

In Figure~\ref{fig:FOV}, we show how the NIRSpec IFU footprint relates to the large-scale dust and filamentary structure of NGC~4696 previously observed. The full HST view\footnote{DOI \url{10.17909/11gm-pb27}} of the Centaurus BCG is shown in the top-left panel and highlights the extended dust lanes and outer filaments, while the zoomed H$\alpha$ image within the NIRSpec FOV is shown in the top-middle panel and isolates the swirl of line-emitting gas reported by \citet{fabian_hst_2016}. The integrated white-light image of the NIRSpec cube is shown in the right panel and reveals the compact bright knots used to define the \textit{Core A} and \textit{Core B} regions, further highlighted in Figure~\ref{fig:spider}. Four examples of intensity maps resulting from global \textsc{LOKI} fits (see Section \ref{sec:LOKI}) are shown for the Pa$\alpha$, $\rm H_2$ 1--0 O(3), $\rm H_2$ 1--0 S(1) and $\rm H_2$ 1--0 S(3) intensities. 

\subsubsection{MUSE}

Optical integral-field spectroscopy of NGC 4696 was obtained with the VLT/MUSE instrument in Wide Field Mode, and a detailed description of these observations and their reduction is given by \cite{collaboration_bulk_2025}. We use a combined dataset of twenty-two pointings taken between 2014 and 2023, totalling 5.66 hours of exposure under seeing conditions of $~\sim0.7''-1.2''$ and airmass 1.08-1.72. The resulting datacube covers an area of about $70''\times 70''$ ($\sim14\times14\rm\;kpc^2$) with $0.2''$ spatial sampling and includes the entire filamentary nebula as well as the region sampled by our JWST/NIRSpec IFU. 

Following the standard reduction, additional sky residuals were removed using the Zurich Atmosphere Purge (ZAP) software, and the combined cube was analyzed with the TARDIS pipeline, which uses pPXF \citep{cappellari_parametric_2004} and E-MILES templates \citep{vazdekis_uv-extended_2016} to model the stellar continuum and derive emission-line fluxes and kinematics \citep{collaboration_bulk_2025}. In Section~\ref{sec:line ratios}, we use these MUSE measurements to construct classical BPT diagrams \citep{baldwin_bpt_1981} and \ion{N}{2}/H$\alpha$ line ratios within apertures matched to our NIRSpec-defined regions, providing the optical excitation context for the near-infrared line-ratio diagnostics. The MUSE cube was not corrected for foreground Milky Way extinction, but since our optical BPT diagnostics rely on ratios of closely spaced emission lines (e.g., \ion{N}{2}/H$\alpha$, \ion{O}{2}/H$\beta$), the wavelength-dependent Galactic extinction is negligible over these narrow baselines and does not affect our line-ratio analysis or classification.

\subsection{Defining Regions}\label{sec:regions}
\begin{figure*}
    \centering
    \includegraphics[width=\linewidth]{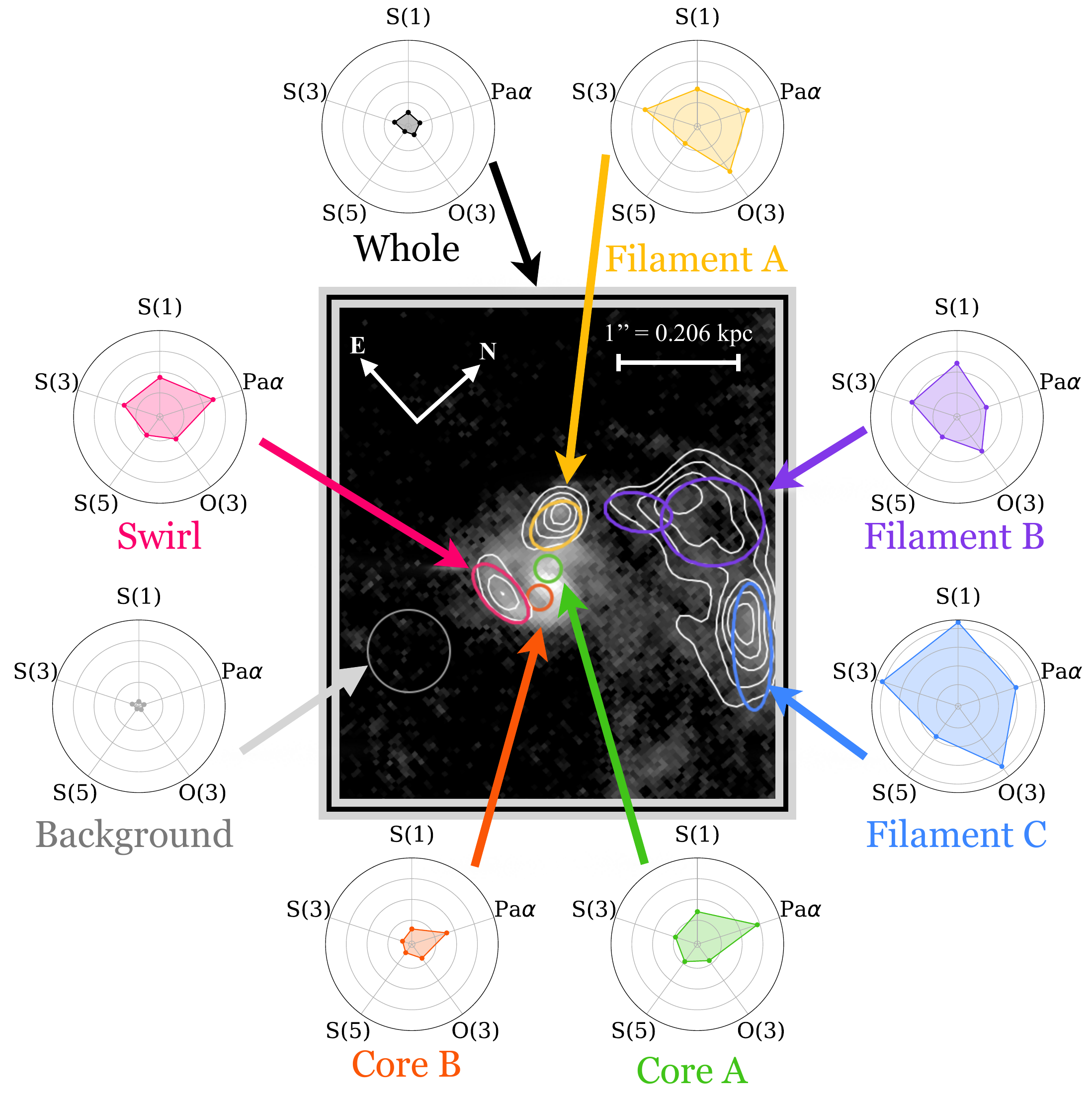}
    \caption{Definitions of regions and line-composition ``spider'' plots. Central panel: continuum-subtracted HST H$\alpha$ (F665N - F814W) image of NGC~4696 with NIRSpec bright-contour levels and the seven regions of interest (\textit{Core A}, \textit{Core B}, \textit{Swirl}, \textit{Filament A}, \textit{Filament B}, \textit{Filament C}, \textit{Background}) outlines. The \textit{Whole} region corresponds to the entire FOV. The \textit{Background} aperture denotes a faint, relatively quiescent continuum-reference region within the NIRSpec field; it is not a blank-sky region and may include diffuse projected line emission. Surrounding panels: Radial spider plots for each region, showing the relative specific intensity of brightest lines in the spectrum: Pa$\alpha$ and $\rm H_2$ 1--0 S(1), S(3), S(5), 1--0 O(3) lines. Grid lines are at intervals of $2\times10^{-17}\rm~erg~Hz^{-1}cm^{-2}s^{-1}sr^{-1}.$}
    \label{fig:spider}
\end{figure*}

Following \citet{marquis_mapping_2026}, we adopt seven spatial apertures that trace distinct morphological and spectroscopic components within the NIRSpec field of view: \textit{Core A}, \textit{Core B}, \textit{Swirl}, \textit{FilamentsA-C}, and \textit{Background}. We additionally define a \textit{Whole} aperture covering the full NIRSpec field of view.

Initial region boundaries were identified based on the morphology of Pa$\alpha$ and $\rm H_2$ 1–0 S(1) and S(3) emission, together with the dust and filamentary structures in archival HST imaging (Figure~\ref{fig:FOV}; \citealt{fabian_hst_2016}). To quantitatively assess whether these preliminary regions correspond to distinct spectral populations, we performed a weighted expectation-maximization principal component analysis (EMPCA; \citealt{bailey_principal_2012}) on the NIRSpec data cube, an algorithm that uses an expectation-maximization scheme to compute principal components from noisy data with missing or down-weighted pixels, ensuring that the eigenvectors capture true spectral variations rather than heteroskedastic noise. The analysis was restricted to spaxels with a median SNR of 30 or greater across the whole wavelength range, ensuring that robust spectral features drive the decomposition. Each spectrum was weighted by the inverse variance, $w=\sigma^{-2}$, while flagged or invalid pixels were excluded (see Section~\ref{sec: NIRSpec}).

Before decomposition, all spectra were mean-subtracted to isolate variance associated with emission lines rather than continuum shape. The EMPCA was computed using 10 components and 10 iterations, with mild spectral smoothing applied to suppress noise-driven fluctuations. We find that the first four principal components account for approximately 25\% of the total fractional variance and capture the dominant emission-line structure in the cube, particularly variations in Pa$\alpha$ and the $\rm H_2$ 1--0 S(1) and S(3) transitions. Including additional components increases the cumulative variance only marginally (reaching $\sim40\%$ at 10 components) and primarily captures noise or low-level spectral features not relevant to the present analysis. We therefore performed clustering in the space of the first four principal component coefficients. These coefficients were standardized before clustering, and a Gaussian mixture model (GMM) with 10 components was used to identify groups of spaxels with similar spectral properties. Because the PCA decomposition captures correlated variations between emission lines, clustering in this space naturally groups spaxels with similar line ratios, providing a physically meaningful segmentation of excitation conditions.

The resulting clusters were then compared to the physically motivated regions defined from imaging and line maps. We find strong agreement between the two approaches: spaxels grouped in PCA space form spatially contiguous structures that correspond to known morphological features such as the \textit{Core A} and \textit{Core B}, and filamentary emission. Minor adjustments to region boundaries were made to ensure spatial continuity and alignment with these structures. Maps of the principal components and GMM clustering are presented in Appendix~\ref{app:gmm}.

The final set of seven regions (\textit{Core A}, \textit{Core B}, \textit{Swirl}, \textit{Filament A}, \textit{Filament B}, \textit{Filament C}, and \textit{Background}, as well as the \textit{Whole} NIRSpec FOV) therefore, represents a hybrid segmentation that is both physically interpretable and statistically validated (see Figure~\ref{fig:spider}). \textit{Core A} and \textit{Core B} trace compact Pa$\alpha$ and $\rm H_2$ emission associated with the immediate AGN vicinity, following the radio nuclear peaks seen by VLBA \citep{taylor_radio_2006} and the bright knots in HST imaging \citep{fabian_hst_2016}, while the \textit{Swirl} region matches both a PCA-defined cluster and the HST H$\alpha$ spiral within the black hole's sphere of influence. \textit{Filament A}, \textit{Filament B}, and \textit{Filament C} capture coherent filamentary structures that connect the outer nebula and dust lanes to the central region. The \textit{Background} aperture samples a faint, relatively quiescent portion of the NIRSpec field, and a sub-section of it provided a representative measurement of the underlying stellar continuum. It is not intended as a blank-sky or emission-free subtraction region. Because the full NIRSpec field lies within the multiphase core of NGC~4696, this aperture can include low-surface-brightness diffuse ionized and molecular emission, as well as projected contributions from the extended filamentary network. We therefore retain it primarily as a continuum reference and show its line ratios for completeness, but interpret them cautiously given their comparatively low signal-to-noise. To visualize how the relative emission strengths (normalized by aperture size) differ among these regions, Figure~\ref{fig:spider} also shows radial `spider' plots, with axes corresponding to Pa$\alpha$ and the $\rm H_2$ 1--0 S(1), S(3), S(5) and 1--0 O(3) lines, and radial coordinates indicating line intensities. The plots are normalized to a common dynamic range, highlighting the strong contrast between core-dominated and filamentary environments in the molecular-to-recombination line balance. 

\subsection{Spectral Fitting with LOKI}\label{sec:LOKI}

\begin{figure*}
    \centering
    \includegraphics[width=\linewidth]{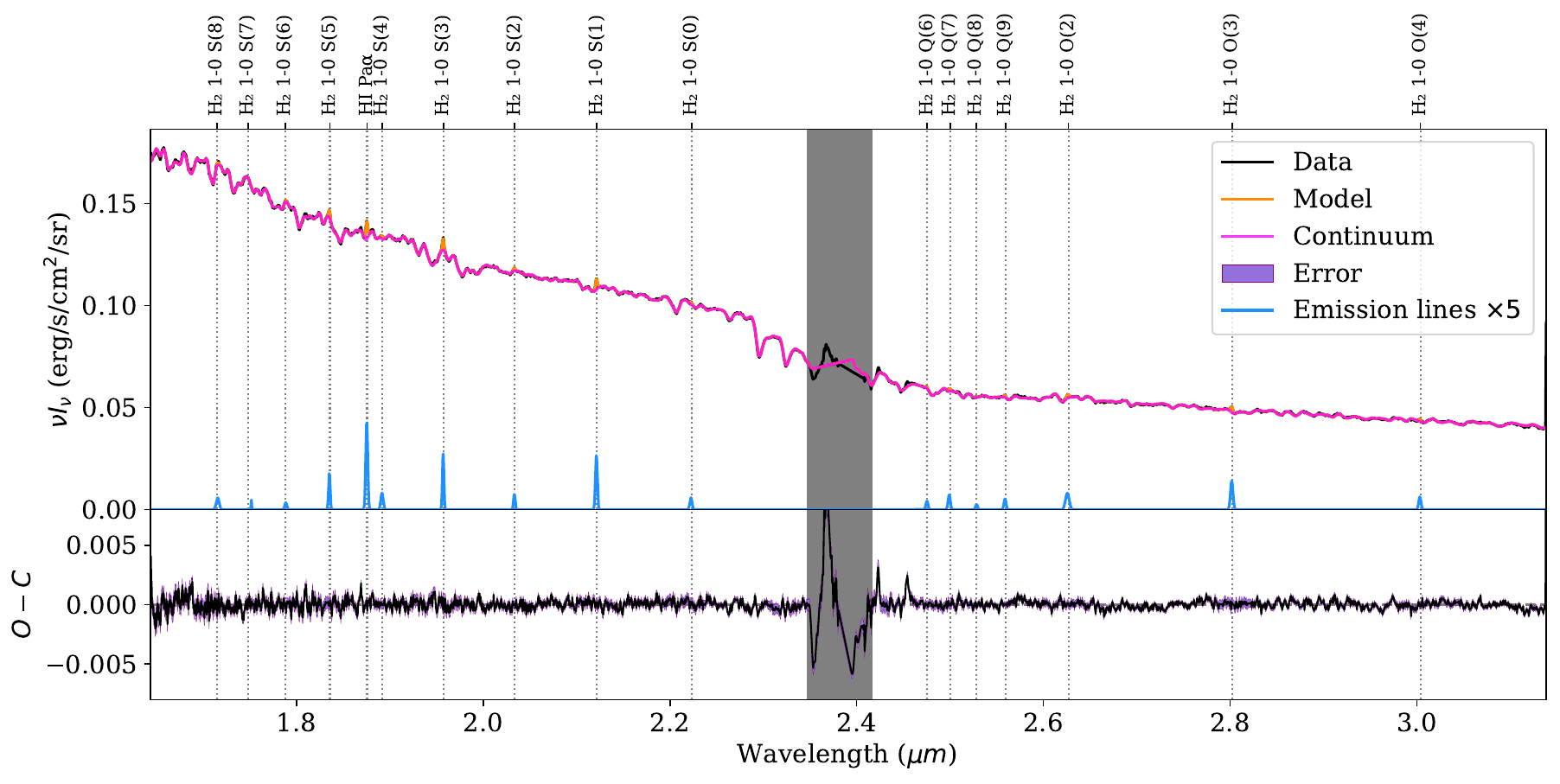}
    \caption{Representative NIRSpec spectrum and \textsc{LOKI} fit for the \textit{Swirl} region. Top panel: Observed spectrum (black) overplotted with the best-fitting continuum model in magenta, consisting of the data template, LOSVD, and multiplicative Chebyshev polynomials, and the Gaussian emission lines plotted in orange. The isolated emission-line models are plotted in blue, scaled up by a factor of 5 for clarity. The greyed-out wavelength interval marks channels excluded from the fit due to the NIRSpec chip gap. The masked interval appears broader than the instrumental gap because the stellar continuum template is taken from a subsection of the \textit{Background} region, which has its own chip gap at a slightly offset wavelength, so both ranges are combined into a single excluded band in the plotted model. Bottom panel: residuals expressed as $O-C$ (observed minus computed), i.e. the observed spectrum minus the full best-fitting continuum-plus-emission-line model. The greyed-out region near 2.4~$\mu$m represents the NIRSpec chip gap and is excluded from the fitting. }
    \label{fig:LOKI}
\end{figure*}


For each of the seven spatial apertures and the Whole-FOV extraction, we fit the spectra with the Likelihood Optimization of gas Kinematics in IFUs \citep[LOKI;][]{reefe_nature_2025}\footnote{\url{https://github.com/Michael-Reefe/Loki.jl}}.  This is a relatively new analysis software package, written in Julia, that enables flexible modelling, fitting, and plotting of IFU data.  We fit each spectrum with a combination of continuum templates (convolved with a line-of-sight velocity distribution, or LOSVD), a 10th-order multiplicative Chebyshev polynomial, and a series of emission lines modelled with Gaussian profiles. This configuration closely follows the \textsc{LOKI} setup used by \citealt{marquis_mapping_2026} for the kinematic analysis, but here our primary focus is on robust integrated line fluxes and ratios rather than detailed velocity-field reconstruction. In particular, we use a data-derived continuum template for the main emission-line fits, and only employ BT-Settl templates \citep{Allard2012} in a separate auxiliary fit to derive E(B–V), whereas \citealt{marquis_mapping_2026} use physically motivated stellar templates throughout their three-round fitting scheme.


As we are primarily concerned with modelling the emission lines, rather than attempting a full physical model of the spectrum, the continuum template, which primarily captures the stellar component, is extracted directly from an emission-line-free region of the data and provides a good match to the small-scale stellar absorption and emission features throughout the continuum.   The Chebyshev polynomials, meanwhile, allow this template some flexibility to account for large-scale differences observed in the stellar continuum across different spatial regions in the cube.  And the LOSVD allows the template to be shifted and broadened to account for differences in stellar kinematics. 

We constrain the emission lines to line-of-sight velocities within $\pm 800$ \kms and FWHMs of $\leqslant 1500$ \kms. 
We do not tie the kinematics of different species. This allows for small offsets between \ion{H}{1} and H$_2$, or among H$_2$ transitions, that could reflect distinct excitation or dynamical components; tying the lines would risk imposing a single velocity structure where the data permit modest differences. Moreover, detailed kinematic modelling is explored in \citealt{marquis_mapping_2026}; here we treat velocities and line widths primarily as supporting parameters for the excitation analysis. All line widths are corrected for the instrumental line spread function, for which we assume a width of $\sim 111$ \kms~ (corresponding to an $R$ of 2700).  Lines nominally have a single kinematic component, but for Pa$\alpha$ and the $\rm H_2$ 1--0 $S$ lines, we allow up to 2 kinematic components if they are statistically preferred over a single component.  We determine this preference using a statistical $F$-test, with a 99.7\% confidence threshold (corresponding to $3\sigma$).

The final spectral model may be summarized as:
\begin{equation}
    I_\nu(\lambda) = \sum_iC_i(\lambda)\bigg(S(v) \otimes G_{*}(v)\bigg) + \sum_{k}G_{k}(\lambda)
\end{equation}
where $C_i(\lambda)$ are the Chebyshev polynomials, $S(v)$ is the continuum template, $G_*$ is the continuum LOSVD, and $G_k$ are the Gaussian emission line profiles.

The model optimization is performed in two stages.  First, the emission lines are masked, and the continuum is fit using templates and polynomials.  The optimal model parameters are determined using a Levenberg-Marquardt local least squares algorithm \citep{Markwardt2009}.  Then, the continuum model is subtracted, and the emission lines are fit to the residuals.  This fit is performed with simulated annealing, a global optimization algorithm.  
There is a caveat to our approach of using continuum templates derived from the data themselves: we cannot determine an intrinsic dust reddening.  To address this, we perform a separate fit to the data with a set of physically motivated stellar continuum templates and a \citet{Calzetti2000} extinction law.  The templates are taken from the BT-Settl model atmosphere library \citep{Allard2012} due to its high resolution, wide coverage in stellar atmosphere parameters ($T_{\rm eff}$, $\log g$, and $\log (Z/Z_\odot)$), and specialization for low $T_{\rm eff}$ stars, which dominate the near infrared flux.  We constrain the parameter range of our templates to $T_{\rm eff} \in (2000,7000)~{\rm K}$, $\log g \in (-0.5, +5.0)$, and $\log (Z/Z_\odot) \in (-0.5, +0.5)$.  The Chebyshev polynomials are turned off during this fit, as they can be degenerate with the extinction law.  The $E(B-V)$ values obtained from this fit are then used to correct the emission line fluxes we measure in our primary fits, using the standard stellar-to-gas reddening conversion of $E(B-V)_*=0.44E(B-V)_{\rm gas}$ \citep{Calzetti2000}. The fitted reddening is applied as a total line-of-sight correction and therefore includes both foreground Milky Way and internal attenuation.

An example fit for the \textit{Swirl} region is shown in Figure~\ref{fig:LOKI}. The model reproduces both the stellar continuum and the suite of near-infrared emission lines, with residuals consistent with the noise level across the full spectral range. The greyed-out wavelength interval appears wider than the NIRSpec chip gap because the continuum template is drawn from the \textit{Background} region, which itself contains a chip gap at a slightly different wavelength, so both ranges are masked in the combined fit. Corresponding fits for all seven regions and the \textit{Whole} FOV are provided in Appendix~\ref{app:specfits} for direct comparison.

\subsection{CLOUDY photoionization models}
\label{sec:cloudy}

We compare the measured near-infrared line ratios with grids of photoionization models calculated using \textsc{CLOUDY} \citep{cloudy25}. We consider two incident spectral energy distributions (SEDs): the built-in \texttt{table AGN} continuum, representing illumination by a hard AGN-like ionizing source, and the built-in \texttt{table cooling flow} continuum, representing the EUV and soft-X-ray radiation associated with gas cooling from the X-ray-emitting intracluster medium (ICM). We refer to the latter as the ICM cooling-radiation model hereafter.

These grids are intended as physically motivated reference excitation sequences, rather than exhaustive or mutually exclusive descriptions of the multiphase gas.

For both SEDs, we assume a constant hydrogen density of $n_{\rm H}=10^{4}~{\rm cm}^{-3}$, ISM abundances, the \textsc{CLOUDY} background cosmic-ray field, and the molecular-hydrogen database. 

We adopt constant-density rather than constant-gas-pressure models because the pressure balance of the line-emitting gas is uncertain. Magnetic fields, turbulence, and bulk gas motions may contribute significantly to the total pressure in cool-core filaments, so imposing constant thermal gas pressure would introduce an additional poorly constrained assumption (e.g. \citealt{Fabian_Johnstone_Sanders_Conselice_Crawford_Iii_Zweibel_2008, hitomicollaborationQuiescentIntraclusterMedium2016}). The models are stopped at $A_V=10$ to ensure that the calculation includes the transition from ionized to molecular gas, and are iterated three times to convergence. We do not impose a temperature stopping criterion. For each SED, the ionization parameter is varied over $-7\leq\log U\leq-1$ in steps of $0.25$ dex, where $U$ is the dimensionless ratio of ionizing-photon to hydrogen density at the illuminated cloud face. Emergent line intensities are then used to construct the NIR H$_2$/Pa$\alpha$ and H$_2$-only ratios shown in Figures~\ref{fig:cornerplot} and \ref{fig:s3s5}.

The AGN grid tests whether variations in the intensity of a hard central radiation field can account for the observed line-ratio sequence. At fixed density and cloud properties, a decrease in $U$ could arise from a lower incident ionizing flux, for example, because of a larger distance from a central source, although density variations, attenuation, and cloud geometry can produce similar effects. The grid should therefore not be interpreted as a unique AGN-photoionization model.

Likewise, agreement with the ICM cooling-radiation model does not demonstrate that an aperture contains an isolated, steady, or classical cooling flow, nor that gas cooling radiatively through the observed near-infrared lines supplies the measured luminosity. Rather, it indicates that an ICM-related EUV/soft-X-ray ionizing spectrum can reproduce the relevant line ratios under the adopted cloud assumptions. The contribution of this channel depends on the soft-X-ray and EUV flux incident on the line-emitting clouds. \citet{polles_excitation_2021} found that ICM-powered photoionization can reproduce much of the optical-to-infrared filament spectrum when the radiation incident on the gas is substantially stronger than directly observed, as may occur if cooling is obscured, geometrically distributed, or embedded in unresolved multiphase structure.

Neither grid includes an explicit stellar \ion{H}{2}-region component, shocks, turbulent dissipation, mixing-layer physics, or enhanced heating of suprathermal particles. The model grids are therefore used to establish broad consistency with hard AGN or ICM-related radiative excitation and to identify departures from a single-component photoionization sequence. Quantitative constraints on the relative contributions of shocks, particle heating, and radiative processes require composite modelling and additional diagnostics, and are deferred to future work.

\section{Results} \label{sec:results}

\subsection{Integrated Line Fluxes} \label{sec:fluxes}

The dereddened emission-line fluxes and kinematic measurements for all fitted transitions in the seven spatial apertures and the Whole-FOV extraction defined in Section~\ref{sec:regions} are presented in Table~\ref{tab:linefluxes} in Appendix~\ref{app:linefluxes}. We detect Pa$\alpha$ and H$_2$ 1--0 S(1) and S(3) in all apertures. Detection rates decrease for weaker transitions, including H$_2$ 1--0 Q(6)--Q(9) and O(2)--O(4). The reported fluxes are integrated over each aperture and corrected using the \textsc{LOKI}-derived $E(B-V)_{\rm gas}$ and the \citet{Calzetti2000} attenuation law (Section~\ref{sec:LOKI}).

Table~\ref{tab:linefluxes} and the spider plots in Figure~\ref{fig:spider} together show a clear spatial segregation in line emission, with the contrast between nuclear and filamentary regions particularly evident in the spider plots. Although the nuclear apertures are compact, they are Pa$\alpha$-bright relative to the filamentary apertures. Their integrated Pa$\alpha$ fluxes are $\log_{10}[F({\rm Pa}\alpha)/({\rm erg\,s^{-1}\,cm^{-2}})] = -16.51\pm0.01$ and $-16.48\pm0.01$ for \textit{Core A} and \textit{Core B}, respectively. \textit{Filaments A} and \textit{B} cover areas of $0.13$ and $0.59~\mathrm{arcsec}^2$, respectively, compared with $0.03$--$0.04~\mathrm{arcsec}^2$ for the compact nuclear apertures. Despite their substantially larger areas, their integrated Pa$\alpha$ fluxes are only $\sim2$--$3$ times higher than those of the compact nuclear apertures, with $\log_{10}[F({\rm Pa}\alpha)/({\rm erg\,s^{-1}\,cm^{-2}})] = -16.27^{+0.03}_{-0.02}$ and $-16.32\pm0.02$ for \textit{Filaments A} and \textit{B}, respectively, demonstrating that their Pa$\alpha$ surface brightnesses are lower than in the nucleus. Table~\ref{tab:lineratios} quantifies key $\rm H_2$/Pa$\alpha$ ratios.

\begin{deluxetable}{lcccc}
\tabletypesize{\footnotesize}
\tablecaption{Key NIR line ratios. \label{tab:lineratios}}
\tablehead{
\colhead{Region} & \colhead{S(1)/Pa$\alpha$} & \colhead{S(3)/Pa$\alpha$} & \colhead{S(5)/Pa$\alpha$}}
\startdata
Core A     & $0.31^{+0.01}_{-0.02}$ & $0.08^{+0.01}_{-0.01}$ & $0.29^{+0.04}_{-0.02}$ \\
Core B     & $0.22^{+0.03}_{-0.01}$ & $0.05^{+0.02}_{-0.01}$ & $0.20^{+0.03}_{-0.02}$ \\
Swirl      & $0.49^{+0.03}_{-0.02}$ & $0.45^{+0.02}_{-0.02}$ & $0.34^{+0.02}_{-0.02}$ \\
Filament A & $0.89^{+0.05}_{-0.06}$ & $0.90^{+0.05}_{-0.06}$ & $0.51^{+0.04}_{-0.04}$ \\
Filament B & $1.43^{+0.08}_{-0.08}$ & $1.44^{+0.10}_{-0.10}$ & $0.81^{+0.06}_{-0.05}$ \\
Filament C & $1.24^{+0.06}_{-0.07}$ & $1.37^{+0.06}_{-0.08}$ & $0.74^{+0.04}_{-0.05}$ \\
Background & $2.13^{+0.73}_{-1.14}$ & $1.98^{+0.96}_{-1.17}$ & $1.07^{+0.51}_{-0.65}$ \\
Whole      & $1.16^{+0.35}_{-0.46}$ & $0.92^{+0.28}_{-0.37}$ & $0.37^{+0.13}_{-0.16}$ \\
\enddata
\tablecomments{Ratios are computed using asymmetric error propagation (Equation~\ref{eq:ratioerr}). The listed quantities are H$_2$ 1--0 S(1)/Pa$\alpha$, S(3)/Pa$\alpha$, and S(5)/Pa$\alpha$.}
\end{deluxetable}

\subsection{Line ratios and Optical-NIR comparison} \label{sec:line ratios}

All line fluxes and uncertainties in Table~\ref{tab:linefluxes} are measured from \textsc{LOKI} fits to spectra extracted over the seven spatial apertures defined in Section~\ref{sec:regions}, together with the Whole-FOV extraction. We compute line ratios that probe both the relative strength of warm molecular and ionized-gas emission, through H$_2$ 1--0 S$(J)$/Pa$\alpha$, and the internal excitation structure of the warm molecular gas, through ratios among H$_2$ transitions.

We use H$_2$ 1--0 S(1)/Pa$\alpha$, S(3)/Pa$\alpha$, and S(5)/Pa$\alpha$ to compare the relative strength of warm molecular and ionized-gas emission across the apertures (e.g., \citealt{ogle_jet-powered_2010,johnstone_discovery_2007}). We additionally use the pure-H$_2$ ratios S(3)/S(1) and S(5)/S(1) as excitation diagnostics \citep{turner_quadrupole_1977,black_fluorescent_1987}. For a ratio $R = F_1/F_2$, where $\sigma_{i,+}$ and $\sigma_{i,-}$ are the upper and lower $1\sigma$ uncertainties on flux $F_i$, respectively, we assume independent errors and propagate the upper and lower ratio uncertainties separately:
\begin{equation}
\sigma_{R,+} = R
\left[
\left(\frac{\sigma_{1,+}}{F_1}\right)^2 +
\left(\frac{\sigma_{2,-}}{F_2}\right)^2
\right]^{1/2},
\label{eq:ratioerr}
\end{equation}
\begin{equation*}
\sigma_{R,-} = R
\left[
\left(\frac{\sigma_{1,-}}{F_1}\right)^2 +
\left(\frac{\sigma_{2,+}}{F_2}\right)^2
\right]^{1/2}.
\end{equation*}

We apply this prescription to all H$_2$/Pa$\alpha$, pure-H$_2$, and optical line ratios discussed below. 

\begin{figure*}
    \centering
    \includegraphics[width=\linewidth]{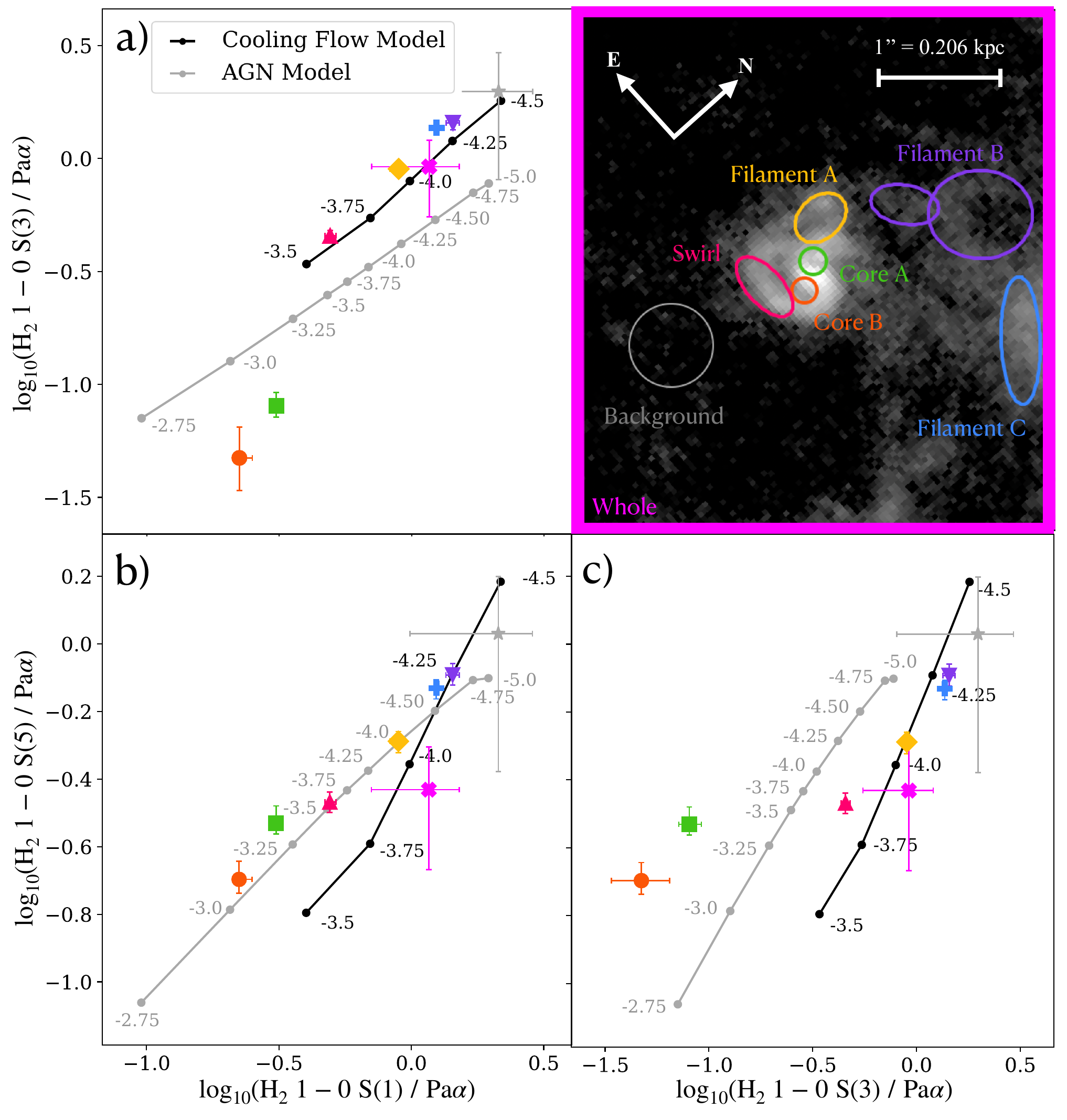}
    \caption{Near-infrared molecular and ionized-gas diagnostic diagrams for the seven spatial apertures and the Whole-FOV extraction in NGC~4696. The panels show H$_2$ 1--0 S(1)/Pa$\alpha$ versus S(3)/Pa$\alpha$ in (a), S(1)/Pa$\alpha$ versus S(5)/Pa$\alpha$ in (b), and S(3)/Pa$\alpha$ versus S(5)/Pa$\alpha$ in (c). Symbols identify \textit{Core A}, \textit{Core B}, the \textit{Swirl}, \textit{Filaments A--C}, the \textit{Background} aperture, and the \textit{Whole}-FOV extraction. The overplotted \textsc{CLOUDY} grids show reference photoionization sequences for an AGN-like ionizing continuum (gray) and an ICM-cooling-radiation continuum (black), evaluated over a range of ionization parameters. The accompanying map shows the aperture definitions overlaid on the continuum-subtracted HST H$\alpha$ image, F665N$-$F814W.}
    \label{fig:cornerplot}
\end{figure*}

\begin{figure*}
    \centering    \includegraphics[width=\linewidth]{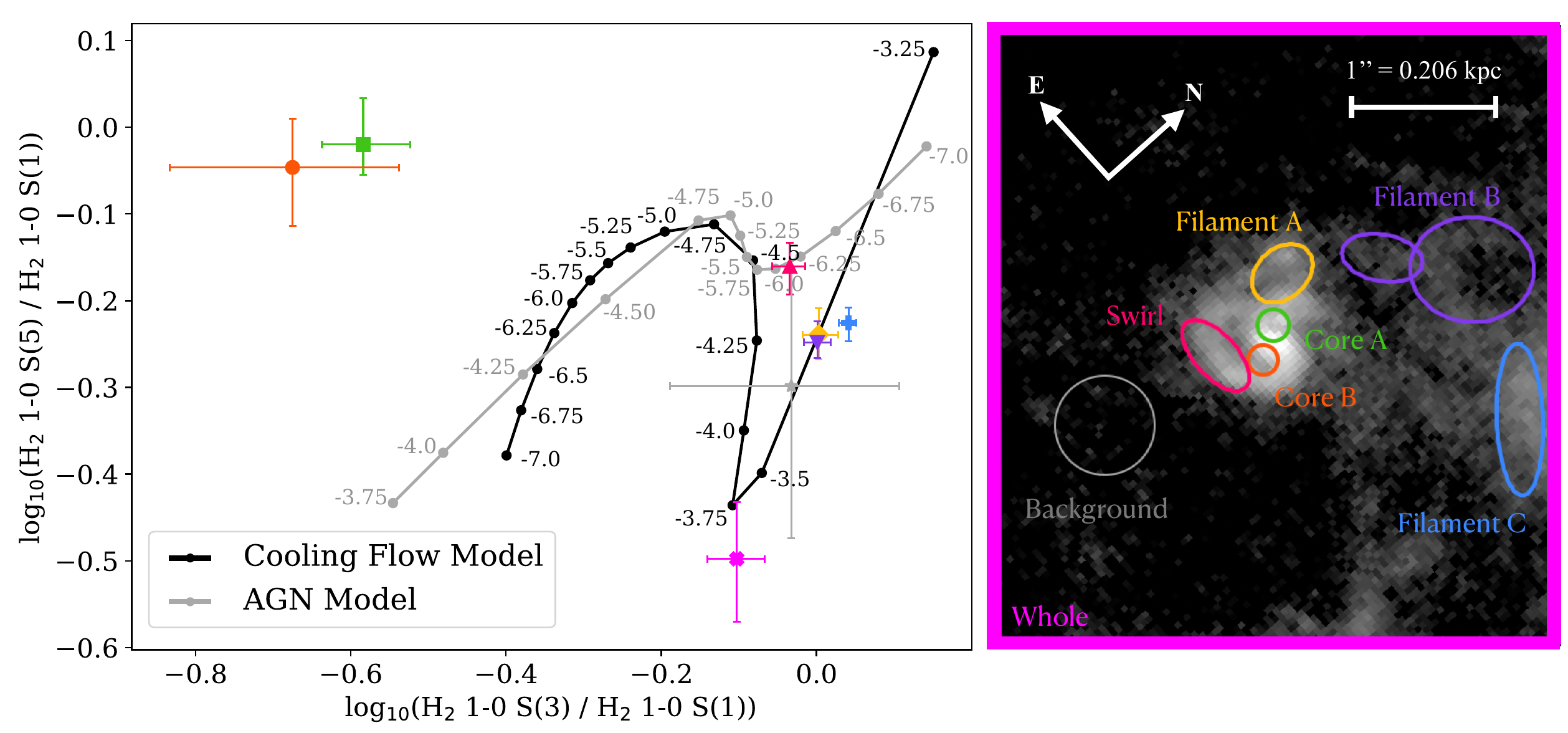}
    \caption{Pure-H$_2$ excitation diagnostic diagram for the seven spatial apertures and the Whole-FOV extraction in NGC~4696. The left panel shows H$_2$ 1--0 S(3)/S(1) versus H$_2$ 1--0 S(5)/S(1), calculated from aperture-integrated line fluxes. Symbols identify the NIRSpec apertures shown in the inset map. The \textsc{CLOUDY} grids show the AGN and ICM-cooling-radiation reference photoionization sequences over a range of ionization parameters. The inset map shows the aperture definitions overlaid on the HST continuum-subtracted H$\alpha$ image, F665N$-$F814W.}
    \label{fig:s3s5}
\end{figure*}

Figure~\ref{fig:cornerplot} presents the NIR H$_2$/Pa$\alpha$ diagnostic diagrams: S(1)/Pa$\alpha$ versus S(3)/Pa$\alpha$, S(1)/Pa$\alpha$ versus S(5)/Pa$\alpha$, and S(3)/Pa$\alpha$ versus S(5)/Pa$\alpha$. Figure~\ref{fig:s3s5} shows the pure-H$_2$ excitation diagram, S(3)/S(1) versus S(5)/S(1). Both figures include the \textsc{CLOUDY} AGN and ICM-cooling-radiation reference grids described in Section~\ref{sec:cloudy}.

\begin{figure*}
    \centering
    \includegraphics[width=\linewidth]{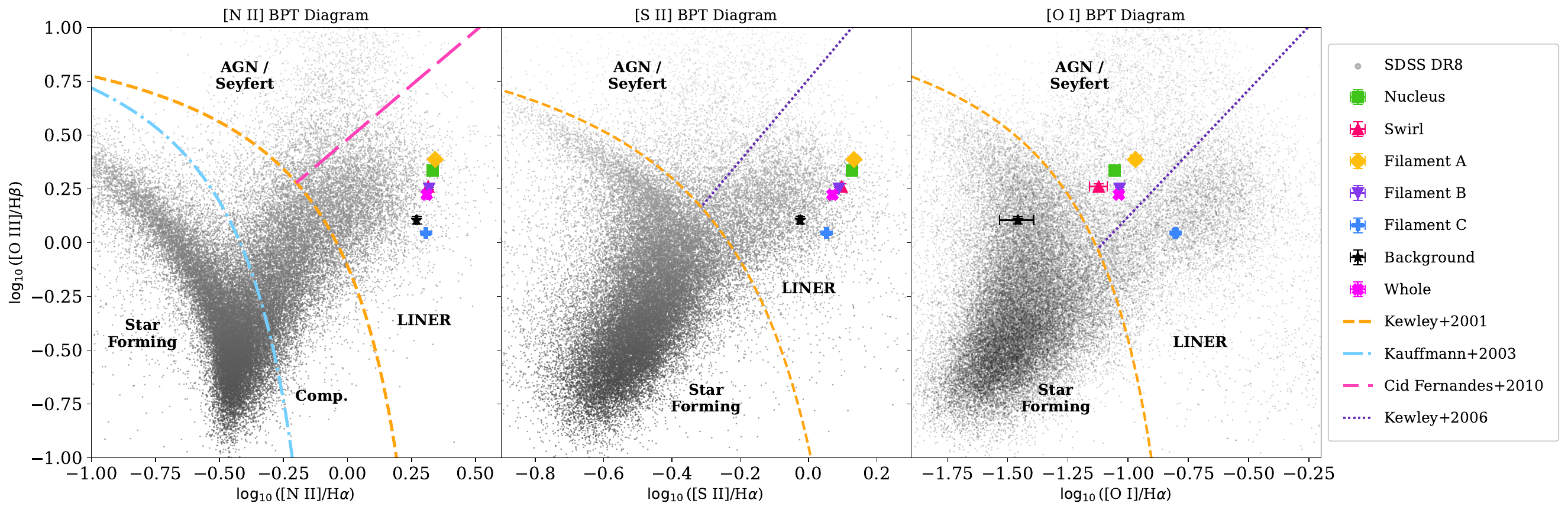}
    \caption{Classical optical BPT diagrams from matched-aperture MUSE spectroscopy of NGC~4696.
    \textit{Left:} $\log_{10}(\mbox{[O\,{\sc iii}]}\,\lambda5007/{\rm H}\beta)$ versus
    $\log_{10}(\mbox{[N\,{\sc ii}]}\,\lambda6583/{\rm H}\alpha)$.
    \textit{Middle:} $\log_{10}(\mbox{[O\,{\sc iii}]}\,\lambda5007/{\rm H}\beta)$ versus
    $\log_{10}(\mbox{[S\,{\sc ii}]}\,\lambda6716,6731/{\rm H}\alpha)$.
    \textit{Right:} $\log_{10}(\mbox{[O\,{\sc iii}]}\,\lambda5007/{\rm H}\beta)$ versus
    $\log_{10}(\mbox{[O\,{\sc i}]}\,\lambda6300/{\rm H}\alpha)$.
    The demarcation curves from \citet{kauffmann_host_2003}, \citet{kewley_theoretical_2001,kewley_host_2006}, and \citet{cid_fernandes_alternative_2010} delineate star-forming, composite, Seyfert-like, and LINER-like regions where applicable. Gray points show the SDSS DR8 comparison sample \citep{sdss_dr8}. The MUSE measurements comprise a \textit{Nucleus} aperture combining the NIRSpec \textit{Core A} and \textit{Core B} apertures, together with the \textit{Swirl}, \textit{Filaments A--C}, \textit{Background}, and \textit{Whole-FOV} apertures.}

    \label{fig:optical}
\end{figure*}

For comparison, Figure~\ref{fig:optical} shows matched-aperture MUSE BPT diagrams of [\ion{N}{2}]/H$\alpha$, [\ion{S}{2}]/H$\alpha$, and [\ion{O}{1}]/H$\alpha$ versus [\ion{O}{3}]/H$\beta$. The MUSE apertures match the NIRSpec definitions except in the nucleus, where \textit{Core A} and \textit{Core B} are combined into a single \textit{Nucleus} aperture to account for the lower spatial resolution of MUSE. The regions occupy a narrow range in the linear [\ion{N}{2}]/H$\alpha$ ratio, approximately $1.40$--$1.41$, consistent with the LINER-like optical spectra previously reported for the NGC~4696 nebula \citep{crawford_extended_2005,canning_deep_2011,hamer_discovery_2019}. For visual context, we plot SDSS DR8 galaxies using emission-line measurements from the \texttt{galSpecLine} catalogue. We require positive fluxes and uncertainties, with $\mathrm{S/N}>2$ in each of the four lines entering a given BPT diagnostic. We apply no redshift selection and randomly downsample each quality-selected sample to at most 50,000 galaxies for clarity.

The optical diagnostic diagrams show substantially less regional separation than the NIR ratios. In particular, the apertures span only a narrow range in [\ion{N}{2}]/H$\alpha$ ($\simeq1.40$--$1.41$) and occupy the LINER-like portion of the BPT diagrams. By contrast, H$_2$ 1--0 S(3)/Pa$\alpha$ increases from $0.08^{+0.01}_{-0.01}$ in \textit{Core A} to $0.90^{+0.05}_{-0.06}$ and $1.44\pm0.10$ in \textit{Filaments A} and \textit{B}, respectively. The filamentary apertures occupy the high-H$_2$/Pa$\alpha$ regime, with S(3)/Pa$\alpha$ of $0.90^{+0.05}_{-0.06}$ in \textit{Filament A} and $1.44\pm0.10$ in \textit{Filament B}. Thus, although the optical line ratios identify broadly similar low-ionization conditions, the JWST diagnostics resolve strong variations in the relative warm-molecular and ionized-gas emission across the central field.

To supplement the conventional BPT demarcations, we apply the fiducial four-component Gaussian mixture model of \citet{desouza2017} to the matched-aperture measurements of $\log_{10}([\mathrm{N\,II}]\lambda6583/\mathrm{H}\alpha)$, $\log_{10}([\mathrm{O\,III}]\lambda5007/\mathrm{H}\beta)$, and $\log_{10}[\mathrm{EW}(\mathrm{H}\alpha)/\mathrm{\AA}]$. All seven region measurements overwhelmingly favour GC2,
with point-estimate membership probabilities exceeding $0.9999999$. In the original calibration, this component is associated with the BPT AGN population and the WHAN weak-AGN/retired populations, rather than a distinct LINER subclass. Their fiducial four-component solution does not distinguish Seyferts and LINERs as separate statistical subclasses, although a LINER-associated component emerges when a fifth Gaussian component is included. We caution that such a subdivision may contribute only weakly to the global data variance and can therefore be overlooked by standard model-selection criteria. Our optical measurements thus share a broad component assignment under the adopted four-component model, despite their regional variation. This result neither excludes finer statistical subdivisions nor establishes identical excitation conditions across the apertures. In contrast, the NIR diagnostics distinguish the compact nuclear and filamentary environments through their molecular-to-recombination-line ratios and molecular excitation structure. Because the statistical model was calibrated on SDSS galaxy spectra rather than resolved BCG structures, we interpret its memberships as a reference classification conditional on the adopted model and measured inputs, not as probabilities for individual excitation mechanisms.

The nuclear apertures behave differently across the near-infrared diagnostics. In the H$_2$ 1--0 S(1)/Pa$\alpha$ versus H$_2$ 1--0 S(5)/Pa$\alpha$ plane (Figure~\ref{fig:cornerplot}), \textit{Core A} and \textit{Core B} lie near the AGN reference sequence at $\log U\sim-3.5$ to $-3.0$. However, they do not follow the AGN sequence in the diagnostics involving H$_2$ 1--0 S(3). This discrepancy is most evident in the pure-H$_2$ diagram (Figure~\ref{fig:s3s5}), where both cores occupy the upper-left portion of the plane, separated from the filamentary apertures and from the \textsc{CLOUDY} reference sequences. The \textit{Background} aperture is shown for completeness but should be interpreted cautiously: it traces low-surface-brightness diffuse emission within the NIRSpec field rather than blank sky, and its large uncertainties permit overlap with the filamentary loci.

For reference, the published H$_2$ 1--0 S(1)/Pa$\alpha$ ratio of the Orion Nebula, $6\times10^{-3}$ \citep{ferland_origin_2008}, lies well outside the range spanned by the NGC~4696 apertures. However, because corresponding measurements of the higher-excitation H$_2$ lines are unavailable, the Orion Bar provides only a one-dimensional comparison and cannot be placed directly on the full NIR diagnostic diagrams. Its offset nevertheless illustrates that the line-ratio regime sampled by the NGC~4696 warm molecular filaments differs strongly from that of this canonical UV-illuminated PDR.

\subsection{Excitation Diagrams}

Using fluxes from Table~\ref{tab:linefluxes}, we construct $\rm H_2$ excitation diagrams following the standard Boltzmann analysis \citep{turner_quadrupole_1977}. For each of the $\rm H_2$ 1--0 $S$-series transitions observed in our data, the measured flux can be converted to a column density via 
\begin{equation}
N_{u,i} = \frac{4\pi F_i\lambda_i}{A_i h c \Omega},
\label{eq:column}
\end{equation}
where $N_{u,i}$ is the column density of H$_2$ molecules in the upper ro-vibrational state of transition $i$, $F_i$ is the aperture-integrated line flux in ${\rm erg\,cm^{-2}\,s^{-1}}$, $\lambda_i$ is the rest wavelength, $A_i$ is the Einstein coefficient from \citet{turner_quadrupole_1977}, $h$ is Planck's constant, $c$ is the speed of light, and $\Omega$ is the solid angle of the extraction aperture in steradians.
We normalize the column densities by the statistical weight $g_u=(2I+1)(2J+1)$, where $I=1$ for ortho-H$_2$ (odd $J$) and $I=0$ for para-H$_2$ (even $J$).

\begin{figure*}
    \includegraphics[width=\linewidth]{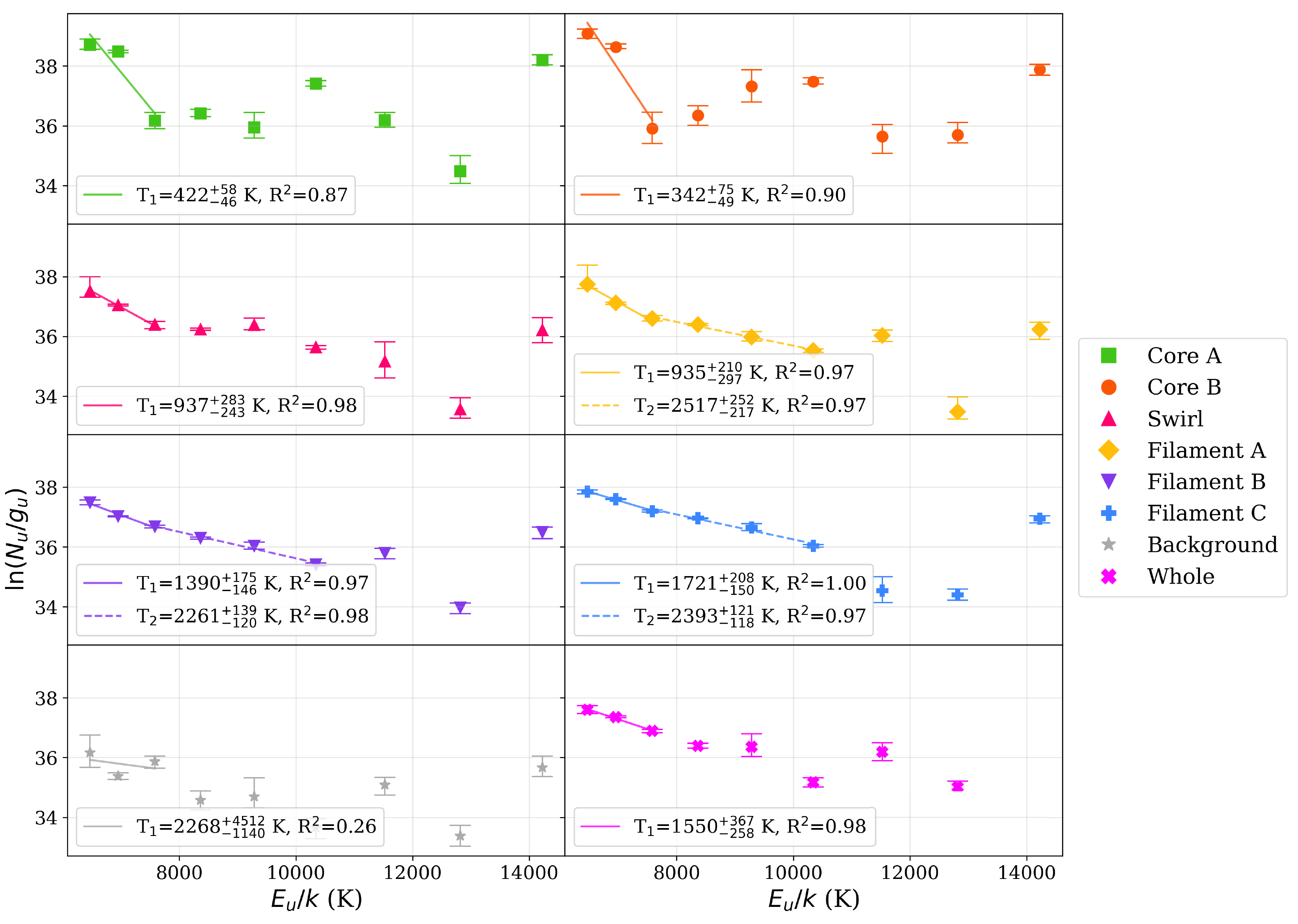}
    \caption{$\rm H_2$ 1--0 S(0)-S(8) excitation diagrams for the seven regions of interest. Each panel plots $\ln(N_u/g_u)$ versus $E_u/k$ with single-temperature fits (Equation \ref{eq:boltzmann}) yielding characteristic temperatures of $T_1=1390^{+175}_{-146}$ K (\textit{Filament A}), $1721^{+208}_{-150}$ K \textit{Filament B}; with a second component of $T_2=2261^{+139}_{-120}$ K and  $2393^{+121}_{-118}$ K where required for \textit{Filament A} and \textit{B} respectively. The upward flattening above $E_u/k_{\rm B}\sim10^4$~K shows an excess population in high-lying ro-vibrational levels relative to a single-temperature extrapolation, indicating multi-temperature and/or non-thermal excitation.
} 
    \label{fig:excitation-diagram}
\end{figure*}

Excitation diagrams were constructed by plotting $\ln\left(N_{u}/g_{u}\right)$ against the upper energy level $E_u/k_B$ for each transition, with $E_u$ values from \citet{dabrowski_lyman_1984}. We fit the resulting level populations with the Boltzmann relation:
\begin{equation}
\ln\left(\frac{N_{u}}{g_{u}}\right) = -\frac{1}{T}\frac{E_u}{k_B}+\ln\left(\frac{N_{\rm tot}}{Z(T)}\right),
\label{eq:boltzmann}
\end{equation}
where $T$ is the excitation temperature and $Z(T)$ is the rotational partition function. The fitted excitation diagrams are shown in Figure~\ref{fig:excitation-diagram}.

We first fit a single-temperature component to the low-excitation H$_2$ 1--0 S(0)--S(2) transitions, which sample upper-level energies of $E_u/k_{\rm B}=6471$--$7584$\,K. These transitions define the low-excitation component used to derive the formal warm-H$_2$ column densities and masses. In regions where the higher-excitation lines showed statistically significant curvature relative to this fit, we additionally fit a hotter component. \textit{Filaments A--C} require such a second component, fitted to the H$_2$ 1--0 S(2)--S(5) transitions, which span $E_u/k_{\rm B}=7584$--$10341$\,K. The higher-excitation transitions deviate from a single-temperature relation, indicating curvature in the excitation diagrams.

Most apertures show a clear flattening of the excitation diagram above $E_u/k_{\rm B}\sim8{,}000$--$10{,}000$~K, where the populations of the higher-lying ro-vibrational levels lie above the extrapolation of the low-excitation component. A single thermalized population would produce an approximately linear relation in $\ln(N_u/g_u)$ versus $E_u/k_{\rm B}$; the observed flattening therefore demonstrates an excess population of highly excited H$_2$ relative to a single-temperature description. In \textit{Filaments~A--C}, this behaviour is represented phenomenologically by an additional hotter component. Physically, such curvature can arise from an unresolved distribution of gas temperatures, collisional excitation in shocks or turbulent dissipation, excitation by suprathermal particles, and/or radiative pumping followed by collisional reprocessing \citep[][]{black_fluorescent_1987,sternberg_infrared_1989,hollenbach_molecule_1989,guillard_h2_2009,ferland_collisional_2009}. The available $v=1$ ro-vibrational transitions do not uniquely distinguish among these mechanisms; however, they robustly show that the warm molecular gas cannot be represented by a single LTE temperature at the sampled excitation energies.

We estimate uncertainties through 6000 Monte Carlo realizations that incorporate the asymmetric flux uncertainties from the \textsc{LOKI} Gaussian fits. Because odd- and even-$J$ transitions sample different upper-level energies, and because a single temperature distribution does not describe the level populations, the ratio of detected ortho- and para-state populations cannot be interpreted as the total H$_2$ ortho-to-para abundance ratio. The fitted intercept, $b=\ln[N_{\rm tot}/Z(T_{\rm ex})]$, yields the total warm-H$_2$ column density:

\begin{equation}
N_{\rm tot}=Z(T_{\rm ex})\,\exp(b),
\label{eq:Ntot}
\end{equation}
where $T_{\rm ex}$ is the fitted excitation temperature. The partition function was computed as
\begin{equation}
Z(T) = \sum_J g_J \exp\left(-\frac{E_J}{k_B T}\right),
\end{equation}
where $g_J=(2I+1)(2J+1)$ includes the rotational and nuclear-spin degeneracies, with $I=0$ for even-$J$ para-H$_2$ levels and $I=1$ for odd-$J$ ortho-H$_2$ levels; we sum over $J=0$--$10$, following \citet{black_fluorescent_1987}. Values at $T\sim300$--$2000$ K range from $Z\approx2.6$--$12$, consistent with \citet{johnstone_discovery_2007}.

We tested for systematic ortho--para offsets by fitting the odd- and even-$J$ ladders separately. The inferred relative normalizations depend on the selected transitions and do not yield a consistent constraint on a single ortho-to-para ratio, particularly in the nuclear apertures. We therefore do not infer a total H$_2$ ortho-to-para ratio from the present near-infrared ro-vibrational data.

\subsection{Warm molecular hydrogen mass estimates}\label{sec:mass}

Warm $\rm H_2$ column densities $N_{\rm tot}$ were derived from excitation diagram intercepts assuming LTE Boltzmann statistics (Equation~\ref{eq:boltzmann}), with the partition function $Z(T_{\rm ex})$ computed over $J=0$--10 following standard spin-weighting \citep{turner_quadrupole_1977}. Because these ro-vibrational lines trace only a small, vibrationally excited H$_2$ fraction, the inferred warm-H$_2$ masses depend sensitively on the assumed temperature distribution. They should be interpreted primarily as characterizations of the excited cooling/heating channel rather than of the total molecular reservoir. Mass estimates were obtained as 
\begin{equation}
M_{\rm H_2} = N_{\rm tot}\Omega D_A^2 m_{\rm H},
\label{eq:H2 mass}
\end{equation}
using the angular diameter distance $D_A=43.5$\,Mpc and individual aperture solid angles $\Omega$. At fixed measured fluxes, aperture solid angles, and fitted excitation parameters, these masses scale as $D_A^2$. A distance approximately $25\%$ smaller would therefore reduce all quoted warm-H$_2$ masses by approximately $44\%$. We retain the nominal mass estimates and report this common distance-scale systematic separately from their statistical uncertainties. For the nominal mass calculation, we used the low-redshift approximation $D_A\simeq D_L\simeq43.5$~Mpc and the individual aperture solid angles $\Omega$; the distinction between $D_A$ and $D_L$ is small compared with the distance-scale systematic considered here.

Monte Carlo uncertainties (6000 iterations) propagate $T_{\rm ex}$ and intercept errors from Table~\ref{tab:linefluxes}. Increasing the number of iterations further did not change the results within the quoted uncertainties.  Table~\ref{tab:warm_h2_masses} lists the resulting mass estimates, spanning 6.1 dex across regions from $10^3$ M$_\odot$ in \textit{Filament B} to $10^9$ M$_\odot$ in \textit{Core B}. 

However, these values must be interpreted cautiously because the LTE-based conversion extrapolates from a limited set of observed ro-vibrational levels to the full warm-$\rm H_2$ population. The excitation diagrams show curvature and, in several regions, require more than one characteristic temperature component; these effects can bias a single-temperature column-density extrapolation. The extreme range in the formal mass estimates is not driven by aperture area alone. Although the compact nuclear apertures have smaller projected areas (\textit{Core A}: $0.04''^2$, \textit{Core B}: $0.03''^2$), their low fitted excitation temperatures imply very large LTE-extrapolated warm-H$_2$ column densities, which dominate the mass calculation and produce the formally large \textit{Core A} and \textit{Core B} values. These nuclear masses should therefore not be interpreted as evidence for a correspondingly large physical concentration of warm molecular gas. Aperture area enters linearly into the conversion from column density to integrated mass. Still, in the present fits, its effect is secondary to the highly temperature-sensitive extrapolation of $N_{\rm tot}$ from the observed ro-vibrational levels.

In addition, they cannot be taken as robust evidence for true physical concentration without corrections for filling factor, beam dilution, and possible nuclear absorption, as further discussed in Section~\ref{sec:disc-caveats}. 

The warm-H$_2$ masses in Table~\ref{tab:warm_h2_masses} are formal LTE excitation-diagram estimates obtained by applying a common procedure to all apertures. The values for \textit{Core A} and \textit{Core B}, which encompass the immediate AGN and jet environment, require particular caution. Their low fitted excitation temperatures imply a large extrapolation from the observed H$_2$ 1--0 ro-vibrational upper levels to the total warm-H$_2$ column density. Because a unique single-temperature population does not adequately describe the nuclear excitation diagrams, this extrapolation can substantially bias the inferred column densities and, in particular, may overestimate the formal warm-H$_2$ masses \citep{tremblay_galaxy-scale_2018, ogle_jet-powered_2010}. Note that the \textit{Whole} panel is derived from a single spectrum extracted over the full NIRSpec footprint. It is not a sum of the plotted regional column densities: each excitation diagram is normalized by its own aperture solid angle and therefore represents an aperture-averaged H$_2$ level population distribution.

Several observational effects further compound this uncertainty. The bright nuclear continuum decreases H$_2$ equivalent widths and reduces the contrast of weak molecular features, increasing their measurement uncertainties \citep{davies_fueling_2014}. In addition, the small physical scales sampled by the nuclear apertures ($\sim3$--$5$\,pc) make the filling factor and the effects of beam dilution or possible self-absorption poorly constrained \citep{johnstone_discovery_2007, togi_lighting_2016}. We therefore retain the \textit{Core A} and \textit{Core B} values for completeness and methodological consistency, but regard them as model-dependent reference values rather than robust measurements of the nuclear warm-H$_2$ content. They should not be interpreted as evidence for a physical concentration of warm molecular gas, nor used for quantitative comparison with the more extended filamentary apertures. A more detailed discussion of these limitations is provided in Section~\ref{sec:disc-caveats}.

As for the extended regions (\textit{Swirl}: $0.14''^2$, \textit{Filament A}: $0.13''^2$, \textit{Filament B/C}: $0.25-0.59''^2$), they provide the most robust \textit{relative} comparisons because their aperture-integrated spectra have higher total signal and are less dominated by the uncertain geometry of a single compact emitting component. Their absolute warm-H$_2$ masses nevertheless remain order-of-magnitude estimates.

Continuous-temperature models, such as those proposed by \citet{togi_lighting_2016}, can, in principle, be used to estimate the total cold+warm $\rm H_2$ mass from suites of $\rm H_2$ lines. However, they are most robust when applied to pure rotational 0–-0 lines spanning a broad range of energies; with only the 1-–0 rovibrational lines available in our data, the inferred cold mass would be dominated by systematics. We therefore restrict the analysis to the warm H$_2$ component traced by the observed near-infrared ro-vibrational transitions. These data do not constrain the much cooler molecular gas expected to dominate the total H$_2$ reservoir, because the G235H spectrum lacks the low-energy pure-rotational H$_2$ lines required to characterize that population. The inferred masses should consequently be interpreted as model-dependent estimates of the warm, rovibrationally excited gas rather than measurements of the total molecular mass.

These excitation properties build on the NIR line ratios (see Section~\ref{sec:line ratios}), where filamentary enhancement in $\rm H_2$/Pa$\alpha$ becomes most apparent. 

\begin{table}[t]
\caption{Warm H$_2$ column-density and mass estimates for each region. $N_{\rm tot}$ is the aperture-averaged warm-H$_2$ column density inferred from the excitation-diagram fit. \textit{Caution:} \textit{Core A} and \textit{Core B} warm-H$_2$ column densities and masses are formal, highly model-dependent LTE extrapolations. Owing to their low fitted temperatures, limited line coverage, compact apertures, and non-single-temperature excitation, these values may be substantially biased high and should not be used for quantitative comparison with the filamentary values; see Section~\ref{sec:disc-caveats}.}
\label{tab:warm_h2_masses}
\begin{tabular}{l c c}
\hline\hline
Region &
$N_{\rm tot}$ ($\mathrm{cm}^{-2}$) &
Mass ($M_\odot$) \\
\hline
Core A &
$2.94^{+17.9}_{-2.51}\times10^{24}$ &
$8.39^{+51.0}_{-7.16}\times10^{7}$ \\
Core B &
$1.42^{+40.1}_{-1.37}\times10^{26}$ &
$2.53^{+71.3}_{-2.43}\times10^{9}$ \\
Swirl &
$1.70^{+13.7}_{-1.51}\times10^{20}$ &
$1.67^{+13.4}_{-1.48}\times10^{4}$ \\
Filament A &
$2.01^{+17.7}_{-1.81}\times10^{20}$ &
$1.79^{+15.8}_{-1.61}\times10^{4}$ \\
Filament B &
$1.80^{+1.44}_{-0.789}\times10^{19}$ &
$7.48^{+6.01}_{-3.29}\times10^{3}$ \\
Filament C &
$1.22^{+0.648}_{-0.419}\times10^{19}$ &
$2.18^{+1.15}_{-0.746}\times10^{3}$ \\
Whole &
$1.39^{+2.07}_{-0.824}\times10^{19}$ &
$2.94^{+4.38}_{-1.74}\times10^{5}$ \\
\hline
\end{tabular}
\end{table}

\section{Discussion} \label{sec:discussion}

Filamentary emission-line nebulae in cool-core BCGs are highly multiphase structures whose optical and infrared emission cannot generally be explained by stellar H~{\sc ii}-region photoionization alone. Previous optical, near-infrared, and far-infrared studies have instead pointed to a combination of cooling-ICM radiation, AGN-related photoionization, mechanical dissipation, and particle heating, although their relative contributions remain uncertain \citep[e.g.,][]{Crawford_1999,ferland_collisional_2009,Donahue_2011,russell_driving_2019,polles_excitation_2021}. In particular, the strong H$_2$ emission observed in BCG filaments and its close association with optical and far-infrared line emission demonstrate that the warm molecular phase is a sensitive tracer of the local energy balance within the feedback-regulated ICM \citep{Salome_2008,Donahue_2011,mittal_herschel_2011}.

Against this backdrop, the JWST/NIRSpec observations of NGC 4696 resolve how warm ionized and molecular-gas excitation changes between the compact nuclear environment (\textit{Core A} and \textit{B}), the circumnuclear \textit{Swirl}, and the surrounding filamentary network (\textit{Filaments A, B} and \textit{C}). The matched-aperture optical diagnostics show relatively little regional separation: all apertures occupy a narrow, LINER-like locus in the BPT diagrams. In contrast, the near-infrared H$_2$/Pa$\alpha$ and pure-H$_2$ ratios strongly distinguish the nuclear and filamentary regions, showing that a common low-ionization optical classification does not imply a common warm-molecular excitation state.

The \textit{Filament} apertures are H$_2$-bright relative to Pa$\alpha$ and show curvature in their H$_2$ excitation diagrams, including an additional hotter component in \textit{Filaments A--C}, whereas the compact nuclear apertures \textit{Core A} and \textit{B} are Pa$\alpha$-bright and have lower H$_2$/Pa$\alpha$ ratios. These spatial contrasts motivate comparison with both AGN-like and ICM-related cooling-radiation fields, while departures from either single-component reference sequence leave room for additional mechanical or particle heating. The following sections examine the resolved warm-gas structure, the new near-infrared diagnostic diagrams, and the physical limitations on assigning a unique heating channel to each region.

\subsection{Spatially resolved warm-gas structures}
\label{sec:disc-structure}

The JWST/NIRSpec observations resolve the central $3\arcsec\times3\arcsec$ ($\sim620\times620$~pc$^2$) of NGC~4696 into compact nuclear regions, the circumnuclear \textit{Swirl}, and surrounding filamentary structures (Figures~\ref{fig:FOV} and  \ref{fig:spider}).  The nuclear apertures, \textit{Core A} and \textit{Core B}, are relatively Pa$\alpha$-bright, whereas \textit{Filaments A--C} exhibit H$_2$/Pa$\alpha$ ratios that are approximately $4$--$8$ times higher (Table~\ref{tab:lineratios}; see also Figure~\ref{fig:spider}). The resolved ratio maps in Appendix~\ref{fig:ratio-maps} show that this aperture-level contrast traces genuine spatial structure: H$_2$ 1--0 S(1)/Pa$\alpha$ is lowest in the compact nuclear region and increases along the  urrounding filaments. The molecular-to-recombination-line contrast is therefore not solely an effect of aperture selection.  In contrast, the H$_2$ 1--0 S(5)/S(1) map shows localized enhancements, indicating that the high-excitation molecular component is spatially non-uniform across the defined regions.  This strong NIR separation is substantially greater than the regional variation in the matched optical BPT diagnostics (Figure~\ref{fig:optical}).

The H$_2$ excitation diagrams provide complementary evidence that a single common excitation state does not describe molecular gas
(Figure~\ref{fig:excitation-diagram}).  The filamentary apertures show curvature at high upper-level energies and require an additional hotter component under our fitting prescription, whereas the compact nuclear lower-temperature, low-excitation components characterize apertures. The Whole-FOV spectrum instead provides an emission-weighted description of the mixed central field: because each regional excitation diagram represents an aperture-averaged level-population distribution, the integrated diagram is not expected to reproduce the slopes or intercepts of the individual regions. Its low-excitation fitted temperature of approximately $1550$~K should therefore not be interpreted as evidence that all components share a common thermal structure, nor as evidence that a single spatial component dominates the integrated molecular emission.  Rather, the morphology and line-ratio maps indicate that the extended filaments make an important contribution to the molecular emission, while the compact nuclear regions remain relatively Pa$\alpha$-bright.  The spatial connection between the extended filaments, dust lanes, and the circumnuclear \textit{Swirl} is consistent with the rotating inflow structure described by \citet{hlavacek-centaurus-2026} and 
\citet{marquis_mapping_2026}, and with a broader feedback-regulated circulation cycle involving condensation, uplift, and return of gas toward the nucleus \citep{voit_precipitation-regulated_2015,
tremblay_galaxy-scale_2018}.

\subsection{New Infrared Diagnostic Diagrams}\label{sec:disc-excitation}

A central result of this work is that the new near-infrared diagnostics reveal regional excitation differences that remain largely unresolved in the matched-aperture optical classifications. Using lines available within the JWST/NIRSpec G235H bandpass, these diagrams compare molecular-to-ionized-gas emission and internal H$_2$ excitation on spatial scales inaccessible to earlier \textit{Spitzer} and \textit{Herschel} observations.

In Figure~\ref{fig:optical}, we show the classical BPT diagrams from matched-aperture MUSE spectroscopy. All regions fall within the LINER/AGN regime, with [\ion{N}{2}] and [\ion{S}{2}] consistently classifying the gas as LINER-like and [\ion{O}{1}] indicating AGN-like excitation in most regions. This optical uniformity masks the strong spatial variation revealed by the near-infrared diagnostics. The warm molecular phase is therefore a more sensitive tracer of local excitation differences in this field than the optical line ratios alone.

In Figure~\ref{fig:cornerplot}, we show the $\rm H_2$/Pa$\alpha$ diagnostic diagrams, plotting S(1)/Pa$\alpha$ versus S(3)/Pa$\alpha$, S(1)/Pa$\alpha$ versus S(5)/Pa$\alpha$, and S(3)/Pa$\alpha$ versus S(5)/Pa$\alpha$ for each spatial region, overlaid with \textsc{CLOUDY} AGN-powered and ICM-cooling-radiation-powered photoionization grids (see Section~\ref{sec:cloudy}). We use these two reference continua because the matched-aperture MUSE spectra occupy the AGN/LINER-like portion of the BPT diagrams and do not show a distinct star-forming sequence. An explicit stellar \ion{H}{2}-region grid is not included because the optical spectra do not show a distinct star-forming branch. The low star-formation rate of NGC~4696 relative to its extended multiphase filamentary nebula, as inferred from its far-infrared emission \citep{mittal_herschel_2011}, suggests that stellar photoionization is unlikely to dominate the central NIRSpec field. Instead, the AGN and ICM cooling-radiation grids bracket the range of hard versus softer ionizing continua expected in cluster cores, and highlight loci where additional non-radiative processes (e.g., shocks or particle heating) may be required.

The nuclear apertures lie near the AGN reference sequence in some diagnostic projections. In contrast, the filamentary apertures occupy higher-H$_2$/Pa$\alpha$ loci that are broadly consistent with portions of the ICM-cooling-radiation sequence. However, no single model family reproduces every observed line-ratio combination, and the two reference grids overlap substantially in several planes.

The ICM cooling-radiation-labelled \textsc{CLOUDY} grid should be interpreted specifically as a photoionization calculation using an ICM-related cooling radiation field, rather than as a unique dynamical model for the origin of the emitting gas. Its approximate correspondence with the filamentary line ratios indicates that photoionization by cooling ICM plasma can contribute under the adopted spectral and geometric assumptions (see Section~\ref{sec:cloudy}). It does not, by itself, establish that the individual filaments are undergoing steady, spatially local cooling flows. This possibility is particularly relevant in light of \citet{polles_excitation_2021}, who showed that the optical and infrared filament spectra can be reproduced by photoionization when the soft-X-ray radiation field incident on the clouds is approximately 2.5 dex stronger than that inferred from the directly observed hot-gas emission. Such an enhancement could arise if cooling radiation is partially absorbed, so that the line-emitting clouds receive a larger EUV and soft-X-ray flux than is directly observed. More generally, the discrepancy may indicate that the ionizing spectrum differs from that of unobscured X-ray cooling; for example, turbulent mixing layers at hot--cold gas interfaces can produce strong EUV emission and thereby lower the observed X-ray-to-line-emission ratio \citep{slavin_1993, begelman_1990}. Additional heating beyond stellar photoionization has long been invoked to explain the strong optical and infrared line emission in cool-core central-galaxy nebulae (e.g. \citealt{voit_donahue_1997,Crawford_1999,Edwards_2009}). Recent hidden-cooling-flow scenarios, therefore, motivate treating ICM-related photoionization as a serious alternative to, or partner of, particle and mechanical heating \citep{fabian_hidden_2022, fabian_hidden_2023b, fabian_hidden_2024, fabian_hidden_2025}.

It is worth noting that the ICM cooling-radiation grid does not include an explicit \ion{H}{2}-region component. However, the low star-formation rate of Centaurus \citep[$<1~M_\odot~{\rm yr}^{-1}$]{mittal_herschel_2011}, together with the absence of a star-forming branch in the matched-aperture optical BPT diagrams, indicates that stellar photoionization is unlikely to dominate the warm H$_2$ excitation within the central NIRSpec field. This conclusion does not require star formation in cluster-core filaments to be globally inefficient: molecular-gas depletion times in cluster centres can be comparable to those in normal disk galaxies, while young stars, cold molecular gas, and line-emitting filaments may be spatially offset by feedback and environmental effects such as ram pressure \citep{2011ApJ...738L..24V, kruijssenFastInefficientStar2019, li_effects_2018}.

The correspondence between the filamentary regions and the ICM cooling-radiation photoionization tracks indicates that an ICM-related ionizing radiation field could substantially contribute to their observed emission. Additional non-radiative heating, including mechanical dissipation and excitation by high-energy particles, remains physically motivated by the multiphase, feedback-dominated environment and may be required to explain regions or transitions that depart from the simple photoionization sequences. However, the present line-ratio comparison does not uniquely determine the relative contributions of these channels.

A possible interpretation of the AGN reference sequence is that part of the observed range in H$_2$/H~I ratios reflects a low-ionization-parameter medium. In the adopted \textsc{CLOUDY} grid, the approximately $0.8$-dex range in the diagnostic ratios can be reproduced over a modest range in \(\log U\), primarily at \(\log U\lesssim-3\). If the cloud densities are similar and the gas is illuminated predominantly by a central source, then \(U\propto r^{-2}\), so gas at larger projected distances could occupy lower-\(U\) portions of the sequence. This interpretation is also qualitatively consistent with the LINER-like optical BPT classifications, which are commonly associated with low-ionization conditions \citep{1983ApJ...264..105F}. The approximate correspondence of \textit{Core A} and \textit{Core B} with the low-$U$ AGN sequence in the H$_2$ 1--0 S(1)/Pa$\alpha$ versus H$_2$ 1--0 S(5)/Pa$\alpha$ diagram therefore shows that AGN photoionization remains a plausible contributor in the inner regions. A low ionization parameter can arise either from a weak ionizing continuum, a relatively large cloud distance, a high gas density, or a combination of these factors. However, the infrared diagnostics do not support a single, purely radial AGN-photoionization sequence as a complete description of the system. The filamentary regions also lie near the ICM cooling-radiation model tracks, while \textit{Core A} and \textit{Core B} do not follow a single AGN locus in all diagnostic planes. Thus, the molecular excitation is more complex than a one-parameter variation in $U$, but the observed offsets do not by themselves exclude a significant photoionized component.

The differing locations of \textit{Core A} and \textit{Core B} among the infrared diagnostic diagrams indicate that their molecular excitation is not reproduced by a single value of the ionization parameter within either of the adopted \textsc{CLOUDY} reference sequences. Their approximate agreement with the AGN sequence in the H$_2$ 1--0 S(1)/Pa$\alpha$ versus H$_2$ 1--0 S(5)/Pa$\alpha$ diagram suggests that a hard nuclear radiation field contributes to the excitation. However, their offsets from the AGN locus in diagrams involving H$_2$ 1--0 S(3), and particularly their isolated position in the H$_2$ 1--0 S(5)/S(1) versus H$_2$ 1--0 S(3)/S(1) plane, show that the relative populations of the rovibrational levels differ from those predicted by the single-component AGN and ICM cooling-radiation models. This behaviour is consistent with a composite and spatially complex nuclear medium, in which AGN-associated photoionization may coexist with non-radiative molecular heating, such as shocks, turbulent dissipation, or excitation by suprathermal particles. Such processes can alter the relative populations of the higher-excitation H$_2$ levels and produce line-ratio combinations not captured by the adopted photoionization grids. We therefore interpret the nuclear offsets as evidence that the \textsc{CLOUDY} grids provide useful reference excitation sequences, rather than unique physical descriptions of the compact nuclear regions.

This contrast has a clear physical interpretation. As shown in Figure~\ref{fig:spider}, the emission in the nuclear regions is dominated by strong, compact Pa$\alpha$, indicating a significant ionized-gas contribution. At the same time, the comparatively lower molecular excitation temperatures found in Figure~\ref{fig:excitation-diagram} are consistent with warm molecular gas that contributes a smaller fraction of the total emission in that environment. In contrast, \textit{Filaments A} and \textit{B} have elevated H$_2$/Pa$\alpha$ ratios: S(1)/Pa$\alpha$ is $0.89^{+0.05}_{-0.06}$ and $1.43\pm0.08$, respectively, while S(3)/Pa$\alpha$ is $0.90^{+0.05}_{-0.06}$ and $1.44\pm0.10$ (Table~\ref{tab:lineratios}). These ratios exceed those of the compact nuclear apertures and are not uniformly reproduced by the adopted single-component photoionization sequences, suggesting that additional excitation channels may contribute.

The observed $\rm H_2$ rovibrational transitions trace a range of excitation conditions: lower-J lines probe collisionally excited gas at $T\sim500$--2000~K, while higher-J transitions are sensitive to additional heating mechanisms such as shocks, UV fluorescence, or excitation by suprathermal particles \citep{black_fluorescent_1987, hollenbach_molecule_1989, sternberg_infrared_1989}. A detailed comparison of our high-$J$ H$_2$ line ratios with grids of shock and cosmic-ray heating models (e.g., \citealt{ferland_collisional_2009, guillard_h2_2009, reefe_apj_2025}) would be valuable for quantitatively constraining the relative contributions of non-radiative excitation, but lies beyond the scope of this initial line-ratio study and will be explored in future work. Similar elevated $\rm H_2$/Pa$\alpha$ ratios are observed in molecular hydrogen emission galaxies (MOHEGs), where shocks and mechanical energy deposition dominate $\rm H_2$ excitation \citep{ogle_jet-powered_2010, guillard_h2_2009, appleton_powerful_2017}, lending support to a mechanically heated filamentary component in NGC~4696.

The excitation diagrams provide complementary evidence that the warm molecular gas is not thermodynamically uniform. Their curvature, especially at high excitation energies and in the filamentary regions, indicates that a single-temperature population does not fully describe the detected ro-vibrational levels. Together with the elevated $\rm H_2$/Pa$\alpha$ ratios and the placement of the filaments in the infrared diagnostic diagrams, this supports spatially varying molecular excitation and a contribution from non-radiative heating processes, making NGC~4696 an especially useful target for future comparison with simulations of AGN-driven turbulence, uplift, and condensation in cluster cores.

\subsection{Spatially Varying Molecular Excitation}\label{sec:disc-ratios}

Another key result from this paper is the pronounced variation in near-infrared line ratios between the nucleus and the filaments. The near-infrared ratios show a strong contrast between the compact nuclear apertures and the filamentary structures. In Table~\ref{tab:lineratios}, H$_2$ 1--0 S(1)/Pa$\alpha$ increases from $0.22$--$0.31$ in \textit{Core A} and \textit{Core B} to $0.89$--$1.43$ in \textit{Filament A} and \textit{Filament B}, respectively. This factor of approximately $4$--$8$ change establishes a marked increase in the relative strength of warm molecular emission outside the compact nuclear region. In combination with the higher $\rm H_2$ excitation inferred from the excitation diagram analyses, this large increase in $\rm H_2$ 1--0 S(1)/Pa$\alpha$ could indicate that non-stellar heating (e.g. low-velocity shocks, particle heating, or ICM-driven excitation) becomes progressively more important at larger radii, while direct photoionization and star formation play a comparatively smaller role in powering the filament emission. Thus, although the global star formation efficiency in cluster centres can be similar to that of disk galaxies, the observed offsets between cold gas and young stars, combined with strong environmental effects, imply that stellar photoionization is unlikely to dominate the excitation of warm H₂ in systems such as NGC 4696. This behaviour is qualitatively similar to the elevated $\rm H_2$/IR and low-ionization line ratios seen in cool-core BCGs with \textit{Spitzer} and \textit{Herschel}, and NGC~4696 thus extends that picture by resolving the same trends on tens-of-parsecs scales.

In Figure~\ref{fig:s3s5}, we show the pure $\rm H_2$ excitation diagnostic diagram, plotting $\rm H_2$ 1--0 S(3)/S(1) versus S(5)/S(1) for each region overlaid with \textsc{CLOUDY} grids. The non-core regions (\textit{Swirl} and \textit{Filaments A, B} and \textit{C}) occupy a distinct regime from the \textit{Core A} and \textit{Core B} in this pure-$\rm H_2$ ratio space. This separation shows that the internal excitation structure of the warm molecular gas differs between nuclear and non-nuclear environments, independent of recombination-line strength. The filaments also require higher excitation temperatures (see Figure~\ref{fig:excitation-diagram}) and, in several cases, a multi-temperature description. In contrast, \textit{Core A} and \textit{Core B} are better described by a single, cooler component, as shown by the excitation diagram fits in Section~\ref{sec:disc-structure}. 

The elevated $\rm H_2$/Pa$\alpha$ ratios, the distinct pure-$\rm H_2$ line-ratio loci, and the multi-temperature excitation structure (Figure~\ref{fig:excitation-diagram}) collectively indicate that a single uniform excitation state does not describe the filamentary gas. These diagnostics are consistent with a mixture of mechanical heating, particle heating, shocks, and radiative processes, although the present data do not uniquely determine the relative contributions of these processes. Similar non-equilibrium, multi-phase conditions have been inferred in other BCG filaments and shocked systems: \citet{ferland_collisional_2009} used collisional-heating models for cool-core filaments to show that extra heating or cosmic rays can produce strong $\rm H_2$ emission with mixed H/H$^+$/$\rm H_2$ gas and rapidly evolving ortho--para ratios, while \citet{guillard_h2_2009} modelled the Stephan's Quintet galaxy-wide shock with low-velocity MHD shocks in dense molecular gas that reproduce curved $\rm H_2$ excitation diagrams and sustained non-thermal excitation. In this context, the combination of high $\rm H_2$/Pa$\alpha$ ratios and multi-temperature excitation in NGC~4696 points to a dynamically evolving, continuously perturbed molecular component, rather than a steady-state, thermally relaxed molecular phase. A more quantitative comparison with shock and PDR grids, along the lines of the Phoenix analysis in \citealt{reefe_apj_2025}, would be an important next step for constraining the non-radiative contribution to the $\rm H_2$ emission. Still, the present work focuses on establishing the spatially resolved excitation diagnostics and their broad physical interpretation.

\subsection{Caveats and model degeneracies}
\label{sec:disc-caveats}

An important limitation of the thermal analysis is that it relies exclusively on near-infrared ro-vibrational H$_2$ lines. The low-energy pure-rotational $v=0-0$ transitions, which trace cooler molecular gas, lie outside the $1.7$--$3.1~\mu$m coverage of the NIRSpec G235H data: H$_2$ S(0), S(1), and S(2)--S(5) occur at $28.22~\mu$m, $17.03~\mu$m, and $12.28$--$6.91~\mu$m, respectively. Their absence limits our sensitivity to the cooler warm-H$_2$ component and biases the measured excitation distribution toward the warmer gas traced by the observed ro-vibrational transitions.

The line-ratio comparisons also do not yield a unique excitation decomposition. In Figure~\ref{fig:s3s5}, the apertures do not follow either the AGN-like or ICM-cooling-radiation \textsc{CLOUDY} reference sequence consistently across all diagnostic planes, and the two model families overlap substantially in several projections. Shocks, UV fluorescence, and particle heating can produce similar H$_2$ ro-vibrational ratios, so the observed partial agreement with different reference sequences is consistent with a mixed excitation environment rather than a unique identification of the dominant heating mechanism. More detailed composite photoionization and shock modelling will be required to quantitatively separate these contributions.

The warm-H$_2$ masses in Table~\ref{tab:warm_h2_masses} should therefore be treated as order-of-magnitude, method-dependent estimates. They depend on the adopted apertures, fitted excitation temperatures, and LTE/Boltzmann extrapolation from the observed ro-vibrational levels to the total warm-gas column; the latter is especially uncertain because the unobserved low-energy rotational population may dominate the total column. Similar limitations have been emphasized in previous studies: \citet{johnstone_discovery_2007} inferred a warm-H$_2$ mass of order $10^5\,M_\odot$ for NGC~4696 from \textit{Spitzer} data, \citet{ferland_collisional_2009} noted that the warm molecular component in cooling-flow filaments is a small fraction of the total molecular reservoir, and LTE-based estimates in other BCGs span approximately $10^5$--$10^8\,M_\odot$ depending on the source, aperture, and adopted temperature range \citep[e.g.,][]{Egami_2006,leftley_jwstnirspec_2024,ghodsi_molecular_2024}. The formal $10^3$--$10^9\,M_\odot$ range in Table~\ref{tab:warm_h2_masses} consequently requires particular caution: the extreme nuclear values are driven by the temperature-sensitive LTE extrapolation and should not be compared directly with the more extended apertures.

The \textit{Whole}-FOV warm-H$_2$ mass inferred from the near-infrared excitation diagram is approximately $3\times10^5\,M_\odot$ (Table~\ref{tab:warm_h2_masses}), substantially below the CO-traced cold molecular mass associated with the central filamentary system \citep{marquis_mapping_2026}. Although the warm- and cold-gas measurements are not matched in aperture or angular resolution, and the cold-gas mass depends on the adopted CO-to-H$_2$ conversion factor, their large contrast is consistent with a multiphase molecular medium in which a small, continuously heated warm-H$_2$ component is associated with a much more massive cold reservoir. 

An additional common systematic arises from the adopted distance: the warm-H$_2$ masses could be approximately $44\%$ smaller if NGC~4696 lies approximately $25\%$ closer. This uncertainty affects the absolute mass normalization, but not the relative masses of apertures analyzed at the same distance, and is distinct from the excitation-model uncertainties discussed above.

\section{Conclusion}
\label{sec:conclusion}

We use JWST/NIRSpec integral-field spectroscopy to study warm ionized and molecular gas in the central $3\arcsec\times3\arcsec$ ($\sim620\times620$~pc$^2$) of NGC~4696. Using physically motivated, data-driven spatial apertures, we isolate compact nuclear regions, the circumnuclear \textit{Swirl}, and surrounding filamentary structures and measure their spatially resolved hydrogen recombination and H$_2$ ro-vibrational emission.

Our main findings are as follows:
\begin{enumerate}
    \item The NIRSpec field contains compact Pa$\alpha$-bright nuclear emission embedded within a circumnuclear \textit{Swirl} and extended filamentary H$_2$ emission. The nuclear apertures have lower H$_2$/Pa$\alpha$ ratios than the \textit{Swirl} and filaments, demonstrating strong spatial variation in the relative strength of warm molecular and ionized-gas emission.

    \item The H$_2$ ro-vibrational line ratios and excitation diagrams vary substantially across the field. The filamentary apertures show curved excitation diagrams and, in several cases, require an additional hotter component under the adopted fitting procedure. In contrast, the nuclear apertures are characterized by lower-temperature fitted components. The detected ro-vibrational lines do not robustly constrain the total H$_2$ ortho-to-para ratio.

    \item Matched-aperture optical MUSE diagnostics show comparatively limited regional separation: all apertures occupy the AGN/LINER-like region of the BPT diagrams and span a narrow range in [\ion{N}{2}]/H$\alpha$. In contrast, the near-infrared H$_2$/Pa$\alpha$ ratios clearly distinguish compact nuclear and filamentary environments.

    \item The \textsc{CLOUDY} reference sequences do not uniquely describe every aperture or diagnostic plane. Filamentary regions occupy high-H$_2$/Pa$\alpha$ loci broadly consistent with portions of the ICM-cooling-radiation sequence, while the \textit{Swirl} and compact nuclear apertures lie closer to the AGN reference sequence in selected projections. These regional differences and departures from both simple model families imply a spatially varying mixture of radiative and non-radiative excitation processes rather than a single common heating mechanism.
\end{enumerate}

Taken together, the central region of NGC~4696 contains thermally and spatially structured warm gas whose excitation differs between compact nuclear and extended filamentary environments. The elevated H$_2$/Pa$\alpha$ ratios and non-uniform excitation in the filaments are consistent with a contribution from non-radiative heating, as seen in mechanically disturbed warm molecular gas in other JWST-observed radio galaxies and cool-core systems \citep{leftley_jwstnirspec_2024,perna_ga-nifs_2023,ubler_ga-nifs_2024}. In contrast, the nuclear apertures remain broadly consistent with an AGN-related radiative component in selected diagnostic planes. NGC~4696, therefore, demonstrates how JWST/NIRSpec can resolve the multiphase excitation structure of the feedback cycle in cool-core BCGs on tens-of-parsecs scales.

Future JWST/MIRI spectroscopy of the pure-rotational H$_2$ lines will extend the temperature baseline to cooler warm molecular gas, better constrain the level populations and ortho-to-para ratio, and enable more discriminating composite models of AGN photoionization, ICM-related radiation, shocks, turbulent dissipation, and particle heating.

\begin{acknowledgments}

OP acknowledges the support of the Natural Sciences and Engineering Research Council of Canada (NSERC). JHL acknowledges funding support from the Canada Research Chairs Program, as well as NSERC through the Discovery Grant, Accelerator Supplement programs and the Arthur B. McDonald Fellowship. In addition, this research was made possible through funding from the Canadian Space Agency (GRANT 24JWGO3A06). MM acknowledges the support of NSERC and of the Fonds de recherche du Qu\'ebec (FRQ) (\href{https://doi.org/10.69777/365532}{doi: 10.69777/365532}). MR acknowledges support from the National Science Foundation Graduate Research Fellowship under Grant No. 2141064. VO acknowledges support from the DICYT ESO-Chile Comite Mixto PS 1757, Fondecyt Regular 1251702, and CIRAS-AI Project, code FIUF137139-USACH. J. B-B acknowledges support from project UNAM DGAPA-PAPIIT AG 101025, Mexico, as well as support from the PASPA 2025 grant. LA acknowledges support from the Canadian Space Agency through grant 22EXPJWST. MP acknowledges funding from the Physics Department of the University of Montreal (UdeM) and the Centre for Research in Astrophysics of Qu\'ebec (AstroQu\'ebec). MLGM acknowledges financial support from NSERC through the Discovery grant program and the Canada Research Chair program. YL acknowledges support from NASA grant 80NSSC22K0668, Chandra X-ray Observatory grant TM3-24005X, NSF grant AST-2510198, and CAREER award AST-2516092. 

This work was supported by a grant \url{https://doi.org/10.69777/377645} from the Fonds de recherche du Québec and by the Center for Research in Astrophysics of Québec (AstroQuébec).

This work is based in part on observations made with the NASA/ESA/CSA James Webb Space Telescope. The data were obtained from the Mikulski Archive for Space Telescopes at the Space Telescope Science Institute, which is operated by the Association of Universities for Research in Astronomy, Inc., under NASA contract NAS 5-03127 for JWST. These observations are associated with program \#5354. The specific observations analyzed can be accessed via \dataset[doi: 10.17909/yqjb-gg89]{https://doi.org/10.17909/yqjb-gg89}.

Funding for SDSS-III has been provided by the Alfred P. Sloan Foundation, the Participating Institutions, the National Science Foundation, and the U.S. Department of Energy Office of Science. The SDSS-III website is http://www.sdss3.org/.

SDSS-III is managed by the Astrophysical Research Consortium for the Participating Institutions of the SDSS-III Collaboration including the University of Arizona, the Brazilian Participation Group, Brookhaven National Laboratory, Carnegie Mellon University, University of Florida, the French Participation Group, the German Participation Group, Harvard University, the Instituto de Astrofisica de Canarias, the Michigan State/Notre Dame/JINA Participation Group, Johns Hopkins University, Lawrence Berkeley National Laboratory, Max Planck Institute for Astrophysics, Max Planck Institute for Extraterrestrial Physics, New Mexico State University, New York University, Ohio State University, Pennsylvania State University, University of Portsmouth, Princeton University, the Spanish Participation Group, University of Tokyo, University of Utah, Vanderbilt University, University of Virginia, University of Washington, and Yale University.

The Universit\'e de Montr\'eal recognizes that it is located on unceded (no treaty) Indigenous territory, and wishes to salute those who, since time immemorial, have been its traditional custodians. The University expresses its respect for the contribution of Indigenous peoples to the culture of societies here and around the world. The Universit\'e de Montr\'eal is located where, long before French settlement, various Indigenous peoples interacted with one another. We wish to pay tribute to these Indigenous peoples, to their descendants, and to the spirit of fraternity that presided over the signing in 1701 of the Great Peace of Montr\'eal, a peace treaty that founded lasting peaceful relations between France, its Indigenous allies, and the Haudenosauni Confederacy (pronounced: O-di-no-sho-ni). The spirit of fraternity that inspired this treaty is a model for our academic community.
\end{acknowledgments}

\facilities{JWST/NIRSpec, HST(STIS), MUSE (WFM)}


\software{
    NumPy \citep{harris_array_2020}, 
    Astropy \citep{theastropycollaboration_astropy_2022}, 
    Matplotlib \citep{hunter_matplotlib_2007},
}

\appendix \label{sec:appendix}

\section{GMM Clustering and Region Definition}
\label{app:gmm}

\begin{figure*}
    \centering
    \includegraphics[width=\linewidth]{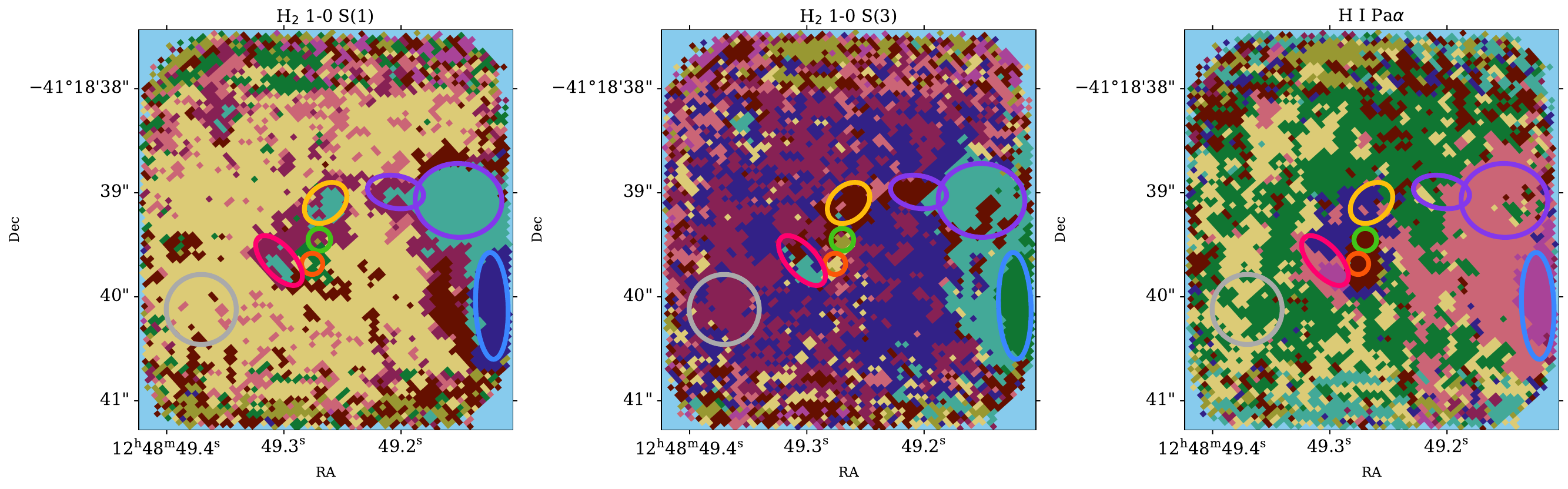}
    \caption{Gaussian-mixture-model (GMM) clustering maps used to inform the spatial apertures adopted in this work. The panels show cluster assignments derived from the continuum-subtracted H$_2$ 1--0 S(1) (left), H$_2$ 1--0 S(3) (middle), and Pa$\alpha$ (right) emission-line maps. Each spatial pixel is coloured according to its maximum-posterior-probability GMM cluster assignment, and equatorial coordinates are shown in the ICRS frame. Black contours mark the final extraction regions used in this work. These regions were defined by considering the data-driven GMM groupings, the H$_2$ and Pa$\alpha$ morphologies, and the dust and filamentary structures visible in archival HST imaging.}
    \vspace{-10pt}
    \label{fig:gmm-clustering}
\end{figure*}

To complement the morphology-based region selection described in
Section~\ref{sec:regions}, we applied Gaussian-mixture-model (GMM)
clustering to continuum-subtracted H$_2$ 1--0 S(1), H$_2$ 1--0 S(3),
and Pa$\alpha$ maps. Figure~\ref{fig:gmm-clustering} shows the resulting
pixel-level cluster assignments. These data-driven groupings were considered
alongside the emission-line morphology and the dust and filamentary
structures visible in archival HST imaging when defining the final
extraction apertures.

\section{Spatial Line-Ratio Maps}
\label{app:ratiomaps}

\begin{figure*}
    \centering
    \includegraphics[width=0.49\linewidth]{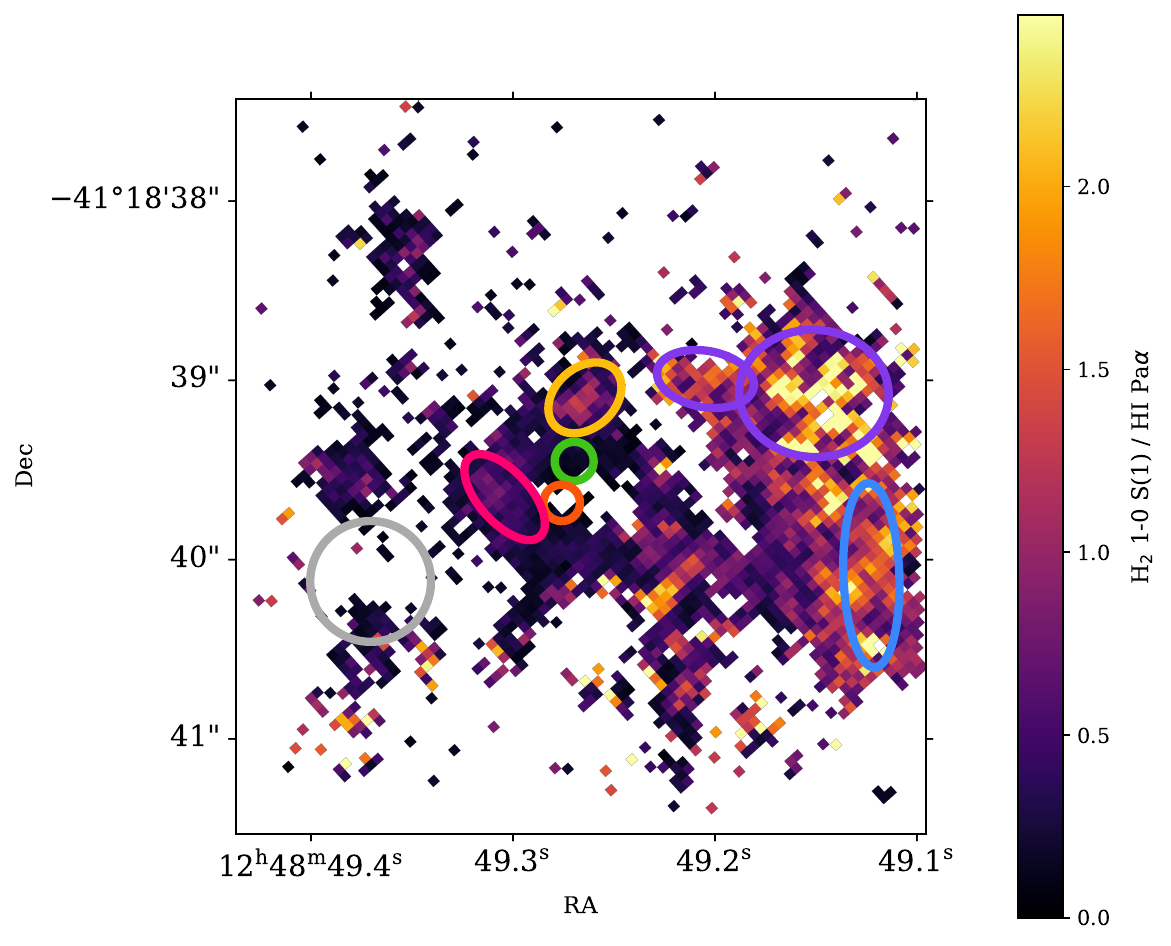}
    \includegraphics[width=0.49\linewidth]{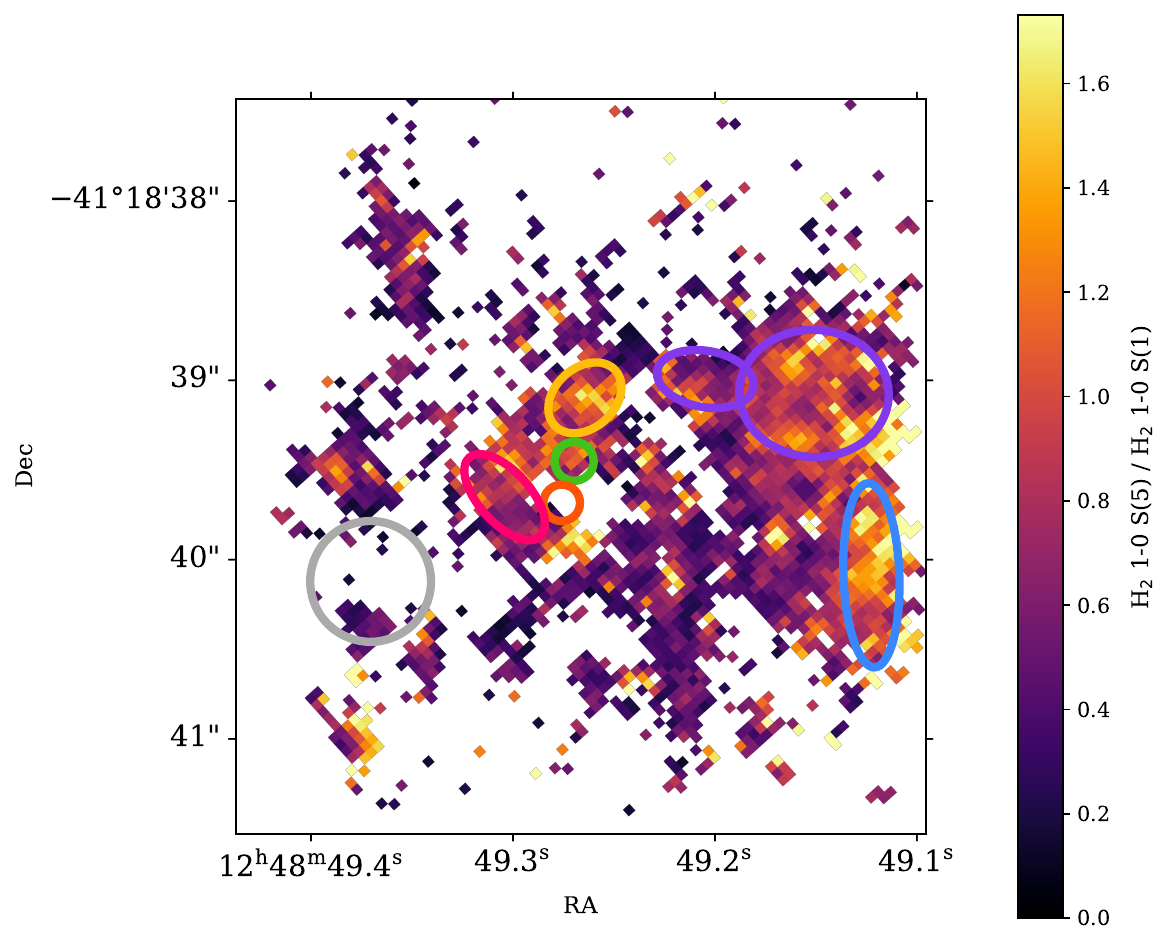}
    \caption{\textit{Left:} H$_2$ 1--0 S(1)/Pa$\alpha$ line-ratio map. The ratio is lowest near the compact nuclear regions and increases along the surrounding filamentary structures, confirming the aperture-integrated trend shown in Figure~\ref{fig:s3s5}. \textit{Right:} H$_2$ 1--0 S(5)/S(1) line-ratio map. Unlike the smoother large-scale behaviour of H$_2$ 1--0 S(1)/Pa$\alpha$, this high-excitation molecular-line ratio is spatially patchy, with localized regions of enhanced S(5)/S(1) within several defined apertures. Both maps are shown only where the two lines entering the relevant ratio have S/N greater than 3. Region boundaries from Figure~\ref{fig:spider} are overlaid for reference.}
    \label{fig:ratio-maps}
\end{figure*}

Figure~\ref{fig:ratio-maps} shows spatially resolved molecular-to-ionized and molecular-excitation line-ratio maps. These maps provide a spatial context for the aperture-integrated ratios discussed in the main text and demonstrate that the nuclear-to-filament trends are not driven solely by the adopted extraction regions.

\clearpage

\section{Spectral Fits for Individual Regions}
\label{app:specfits}

\begin{figure*}
    \centering
    \includegraphics[width=0.48\linewidth]{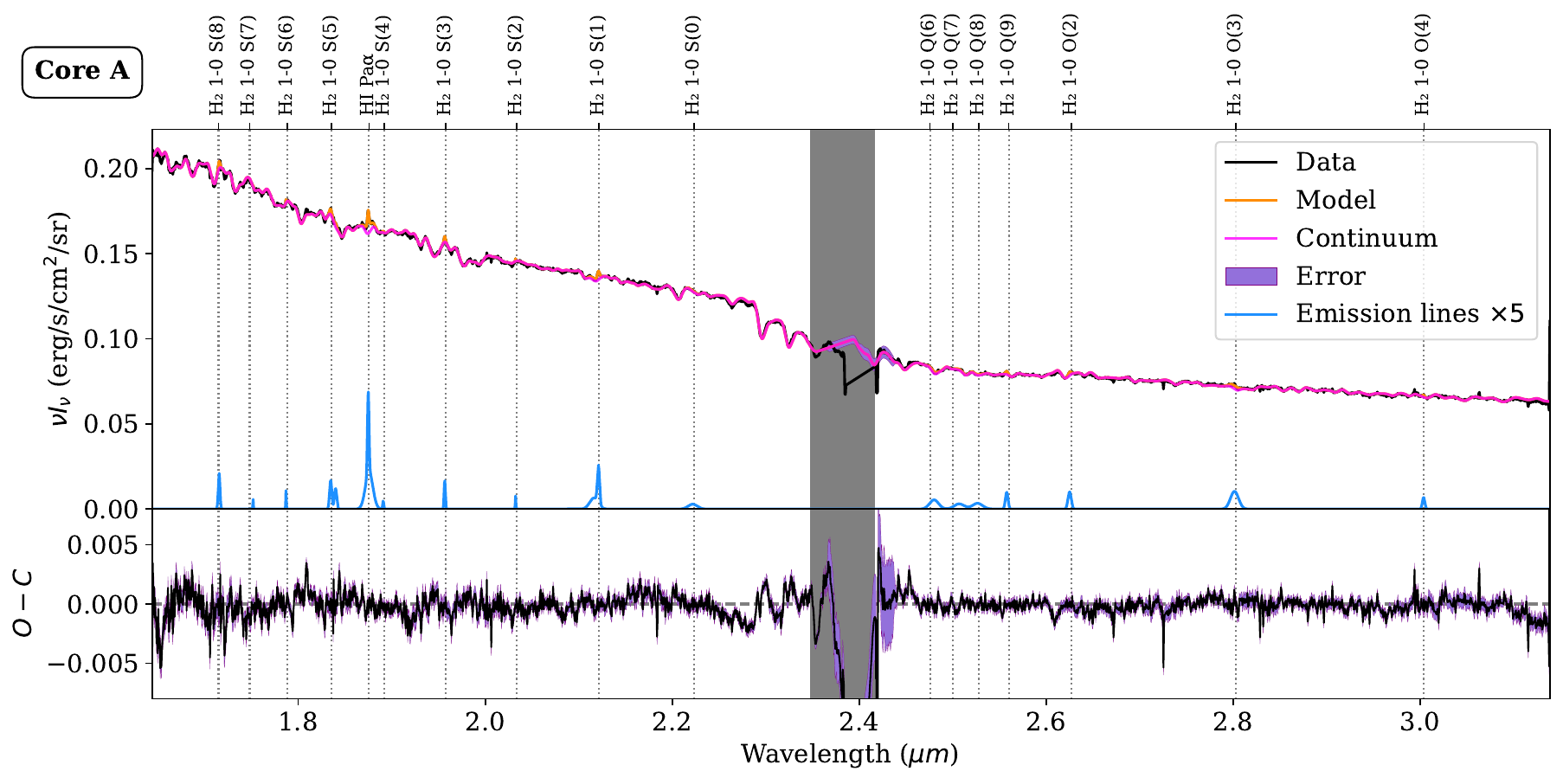}
    \includegraphics[width=0.48\linewidth]{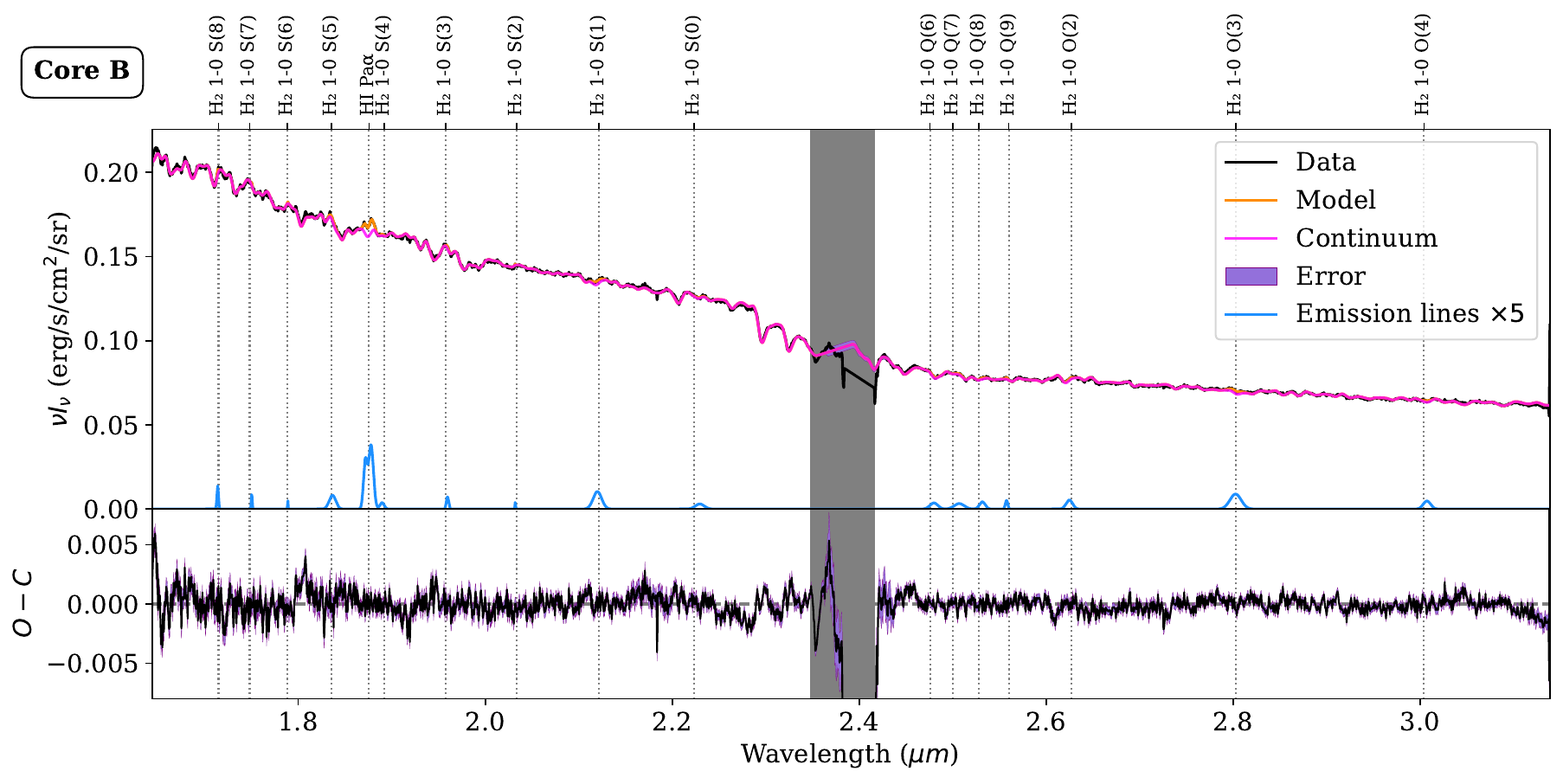}
    \includegraphics[width=0.48\linewidth]{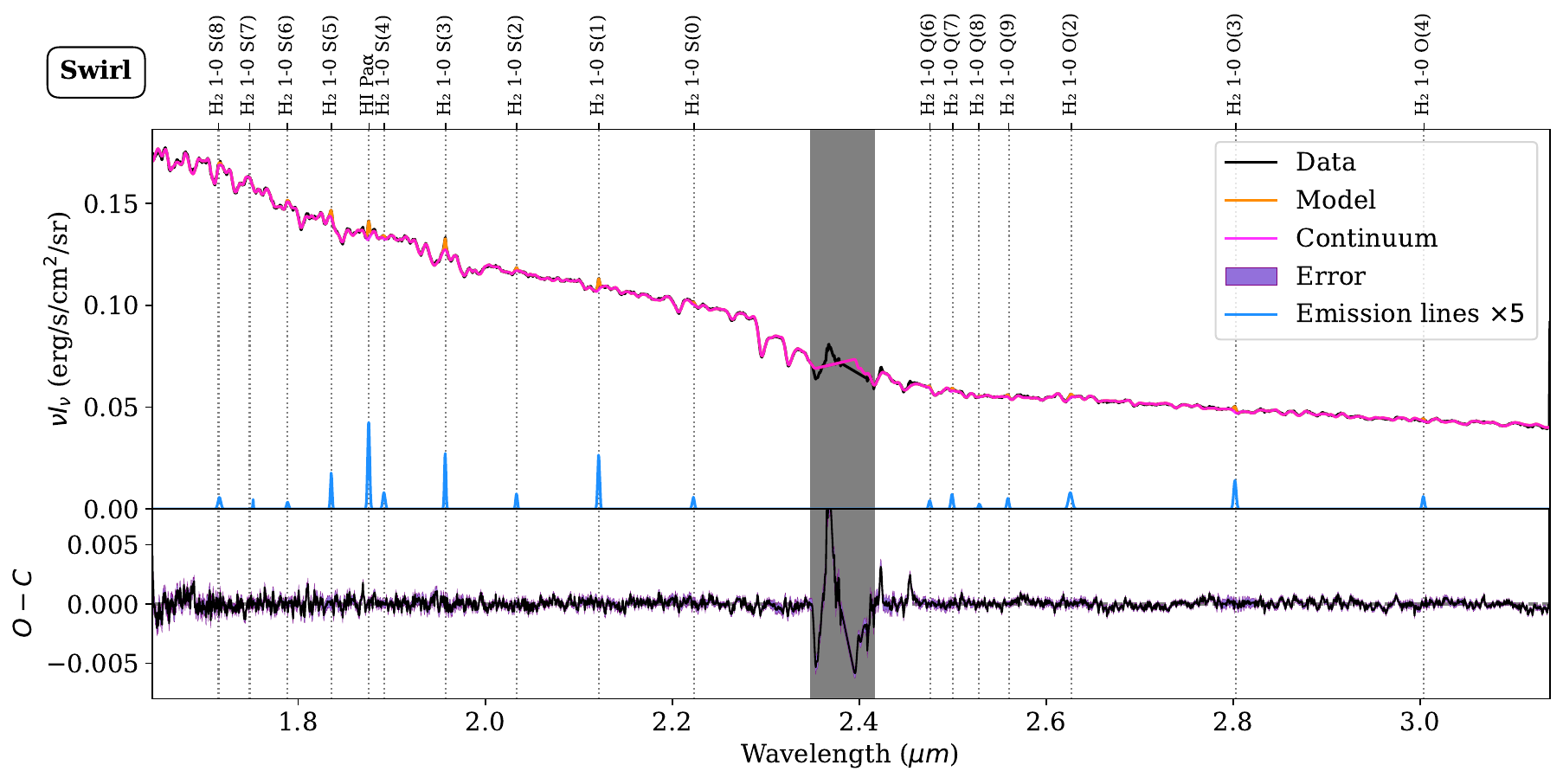}
    \includegraphics[width=0.48\linewidth]{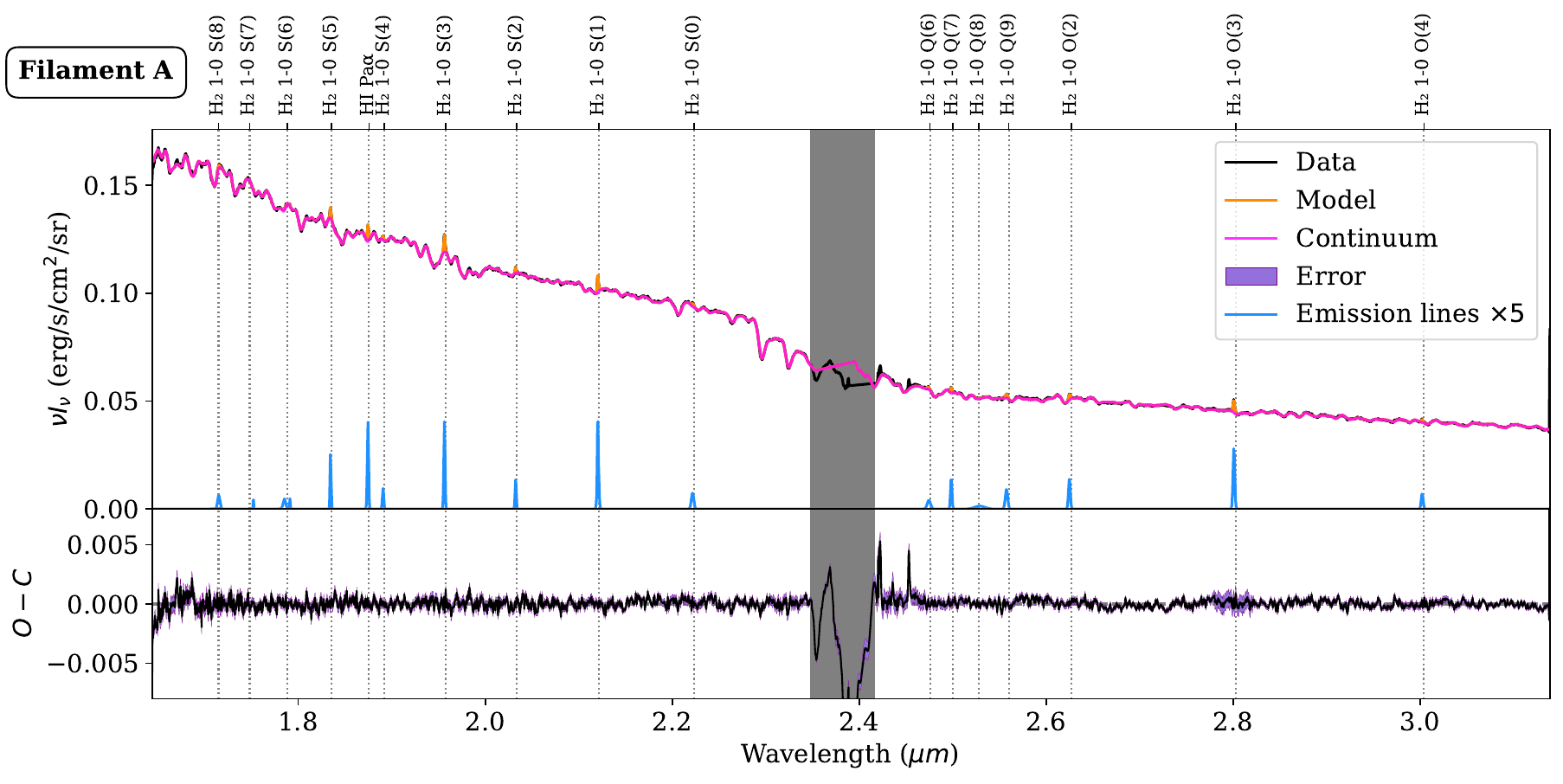}
    \includegraphics[width=0.48\linewidth]{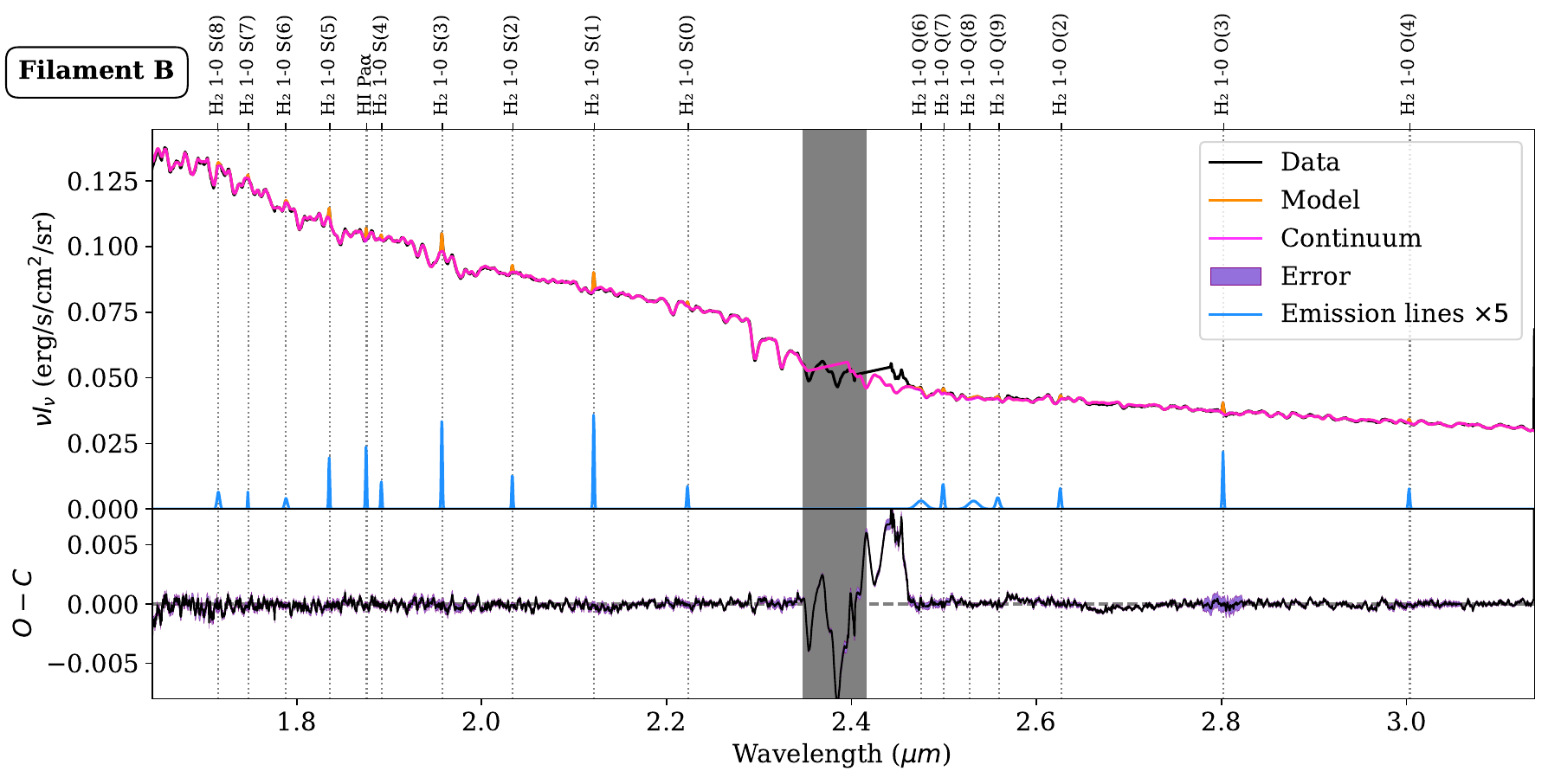}
    \includegraphics[width=0.48\linewidth]{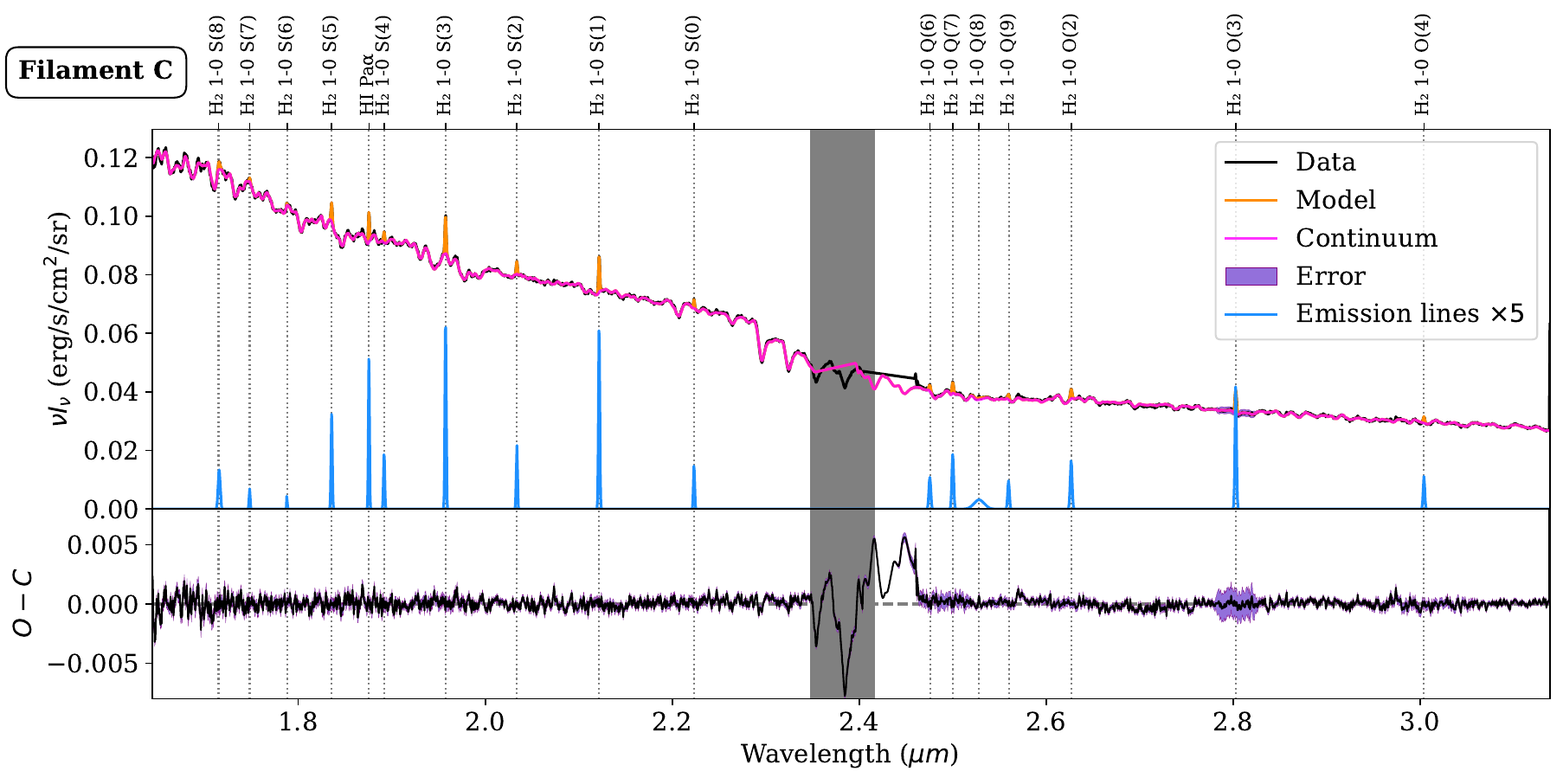}
    \includegraphics[width=0.48\linewidth]{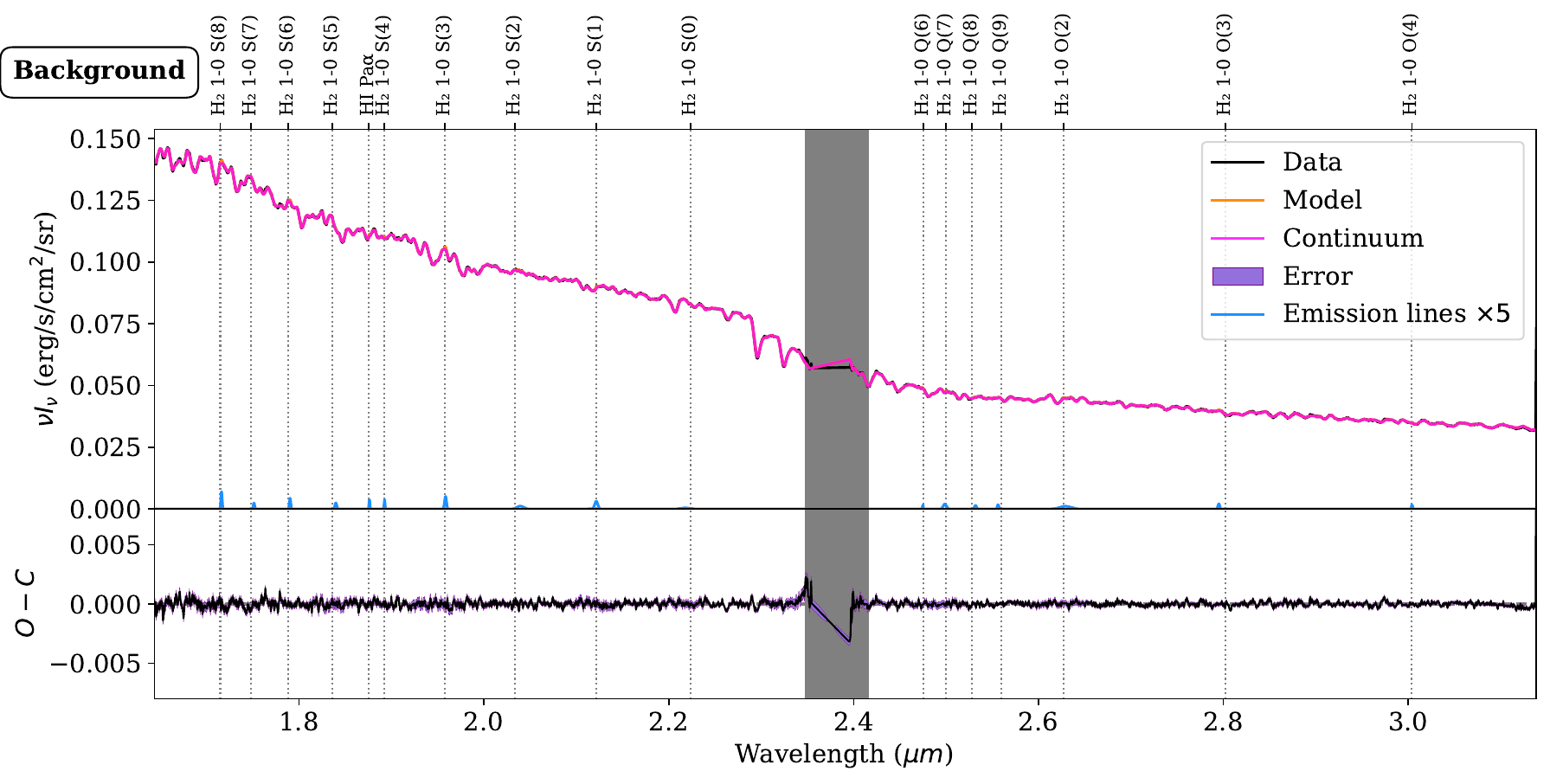}
    \includegraphics[width=0.48\linewidth]{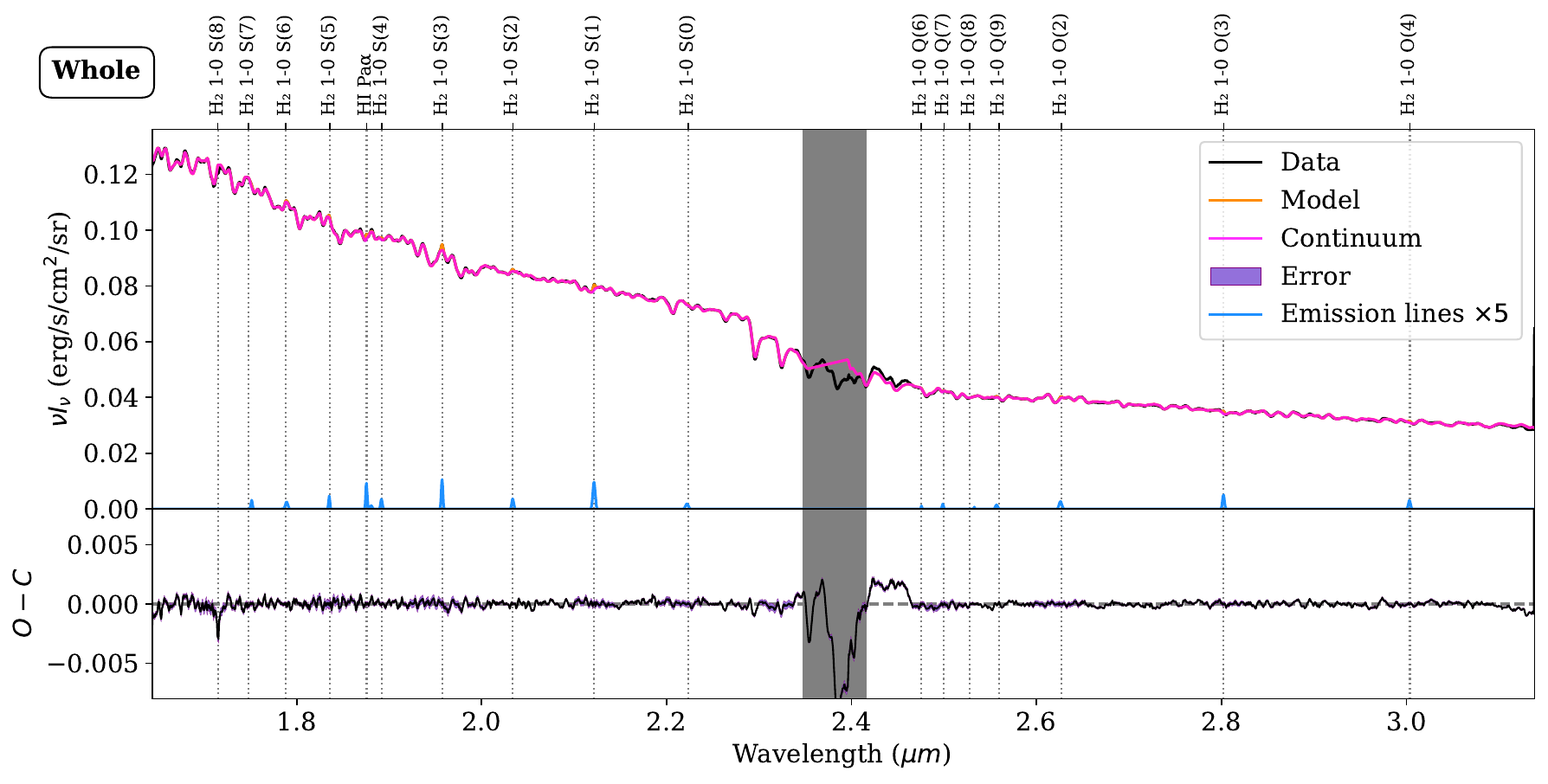}
    \caption{LOKI spectral fits for the eight spatial apertures analyzed in this work. Each panel follows the conventions of Figure~\ref{fig:LOKI}: the observed spectrum, best-fitting model, continuum model, and emission-line components are shown as defined there. The aperture represented in each panel is identified in the upper-left corner. These fits provide the region-by-region measurements used to derive the line fluxes, ratios, and excitation diagnostics presented in the main text.}
    \label{fig:all-spec-fits}
\end{figure*}

Figure~\ref{fig:all-spec-fits} presents the best-fitting \textsc{LOKI}
models for the eight spatial apertures used throughout this work. The
spectra and fit components are displayed using the same layout, wavelength
range, and conventions as Figure~\ref{fig:LOKI}. The aperture name is given
in the upper-left corner of each panel.

\clearpage

\section{Integrated Line Fluxes}
\label{app:linefluxes}

Table~\ref{tab:linefluxes} presents the integrated emission-line flux
measurements for the spatial apertures used throughout this work. Aperture
areas and reddening estimates are included in the table headers, while the
Full line-flux measurements are provided here to keep the main text focused
on the spatial trends and diagnostic ratios.

\begin{center}
\begin{longtable}{@{}cccccccc@{}}
\caption{Integrated line fluxes and kinematics for the eight regions. (1) Line name, (2) Rest wavelength $\lambda_{\rm rest}$, (3) $\log_{10}$(Line Flux), dereddened using LOKI-derived $E(B-V)_{\rm gas}$ values, (4) Equivalent Width (EW), (5) Signal-to-Noise Ratio (SNR), (6) Median Velocity Offset $v_{\rm med}$, (7) $W_{80}$, the velocity range holding 80\% of the total light. ``---'' denotes lines for which the parameters could not be constrained due to limited SNR. Aperture size and extinction $E(B-V)$ values for each region are included in their respective header row.}
\label{tab:linefluxes}\\

\hline
\shortstack{(1) Line} &
\shortstack{(2) $\lambda_{\rm rest}$\\($\mu$m)} &
\shortstack{(3) $\log_{10}$(Flux)\\($\rm erg~cm^{-2}~s^{-1}$)} &
\shortstack{(4) EW/$10^{-5}$\\($\mu$m)} &
\shortstack{(5) SNR} &
\shortstack{(6) $v_{\rm med}$\\($\rm km~s^{-1}$)} &
\shortstack{(7) $W_{80}$\\($\rm km~s^{-1}$)} \\ 
\hline
\endfirsthead

\caption[]{Integrated line fluxes and kinematics for the eight regions (continued).}\\
\hline
\shortstack{(1) Line} &
\shortstack{(2) $\lambda_{\rm rest}$\\($\mu$m)} &
\shortstack{(3) $\log_{10}$(Flux)\\($\rm erg~cm^{-2}~s^{-1}$)} &
\shortstack{(4) EW/$10^{-5}$\\($\mu$m)} &
\shortstack{(5) SNR} &
\shortstack{(6) $v_{\rm med}$\\($\rm km~s^{-1}$)} &
\shortstack{(7) $W_{80}$\\($\rm km~s^{-1}$)} \\
\hline
\endhead

\hline
\endfoot

\multicolumn{8}{c}{\textbf{Core A:} $0.04''^2$, $E(B-V) = 1.07$} \\ \hline
H$_2$ 1--0 S(8) &
  $1.7147$ &
  ${-17.29}^{+0.07}_{-0.08}$ &
  ${4.77}^{+0.80}_{-0.82}$ &
  2.6 &
  ${76}^{+31}_{-36}$ &
  ${384}^{+60}_{-70}$ \\
H$_2$ 1--0 S(7) &
  $1.7480$ &
  ${-18.36}^{+0.18}_{-0.22}$ &
  ${0.44}^{+0.18}_{-0.23}$ &
  $<$2 &
  ${579}^{+11}_{-64}$ &
  --- \\
H$_2$ 1--0 S(6) &
  $1.7880$ &
  ${-18.06}^{+0.10}_{-0.11}$ &
  ${0.93}^{+0.22}_{-0.24}$ &
  $<$2 &
  ${-198}^{+34}_{-8}$ &
  --- \\
H$_2$ 1--0 S(5) &
  $1.8358$ &
  ${-17.05}^{+0.03}_{-0.05}$ &
  ${10.35}^{+0.70}_{-1.20}$ &
  5.1 &
  ${9}^{+66}_{-82}$ &
  ${1181}^{+52}_{-88}$ \\
\ion{H}{1} Pa$\alpha$ &
  $1.8756$ &
  ${-16.51}^{+0.01}_{-0.01}$ &
  ${38.85}^{+0.96}_{-0.82}$ &
  14 &
  ${-115}^{+8}_{-10}$ &
  ${1312}^{+47}_{-36}$ \\
H$_2$ 1--0 S(4) &
  $1.8920$ &
  ${-18.19}^{+0.15}_{-0.22}$ &
  ${0.81}^{+0.29}_{-0.41}$ &
  $<$2 &
  ${-215}^{+81}_{-102}$ &
  ${151}^{+151}_{-159}$ \\
H$_2$ 1--0 S(3) &
  $1.9576$ &
  ${-17.57}^{+0.05}_{-0.06}$ &
  ${3.68}^{+0.42}_{-0.49}$ &
  3.4 &
  ${-188}^{+9}_{-19}$ &
  ${255}^{+65}_{-39}$ \\
H$_2$ 1--0 S(2) &
  $2.0338$ &
  ${-18.25}^{+0.11}_{-0.12}$ &
  ${0.86}^{+0.23}_{-0.23}$ &
  $<$2 &
  ${-232}^{+43}_{-22}$ &
  --- \\
H$_2$ 1--0 S(1) &
  $2.1218$ &
  ${-16.91}^{+0.02}_{-0.02}$ &
  ${20.74}^{+0.94}_{-0.80}$ &
  7.2 &
  ${-308}^{+19}_{-22}$ &
  ${1447}^{+30}_{-27}$ \\
H$_2$ 1--0 S(0) &
  $2.2235$ &
  ${-17.56}^{+0.07}_{-0.08}$ &
  ${5.24}^{+0.81}_{-1.01}$ &
  $<$2 &
  ${-264}^{+574}_{-254}$ &
  ${1633}^{+1}_{-3}$ \\
H$_2$ 1--0 Q(6) &
  $2.4756$ &
  ${-17.32}^{+0.04}_{-0.05}$ &
  ${15.88}^{+1.56}_{-1.70}$ &
  $<$2 &
  ${492}^{+81}_{-97}$ &
  ${1482}^{+125}_{-112}$ \\
H$_2$ 1--0 Q(7) &
  $2.5001$ &
  ${-17.52}^{+0.09}_{-0.08}$ &
  ${10.08}^{+2.02}_{-1.79}$ &
  $<$2 &
  --- &
  --- \\
H$_2$ 1--0 Q(8) &
  $2.5280$ &
  ${-17.49}^{+0.08}_{-0.07}$ &
  ${11.11}^{+2.11}_{-1.82}$ &
  $<$2 &
  ${-229}^{+125}_{-177}$ &
  --- \\
H$_2$ 1--0 Q(9) &
  $2.5598$ &
  ${-17.59}^{+0.05}_{-0.04}$ &
  ${9.13}^{+1.07}_{-0.80}$ &
  3.8 &
  ${-295}^{+19}_{-25}$ &
  ${396}^{+36}_{-47}$ \\
H$_2$ 1--0 O(2) $+$ Br$\beta$ &
  $2.6269$ &
  ${-17.51}^{+0.05}_{-0.05}$ &
  ${11.27}^{+1.21}_{-1.30}$ &
  3.1 &
  ${-263}^{+21}_{-23}$ &
  ${479}^{+54}_{-68}$ \\
H$_2$ 1--0 O(3) &
  $2.8025$ &
  ${-17.13}^{+0.04}_{-0.06}$ &
  ${32.01}^{+3.12}_{-4.50}$ &
  2.7 &
  ${-170}^{+33}_{-48}$ &
  ${1175}^{+217}_{-334}$ \\
H$_2$ 1--0 O(4) &
  $3.0039$ &
  ${-17.91}^{+0.11}_{-0.17}$ &
  ${6.15}^{+1.57}_{-2.46}$ &
  $<$2 &
  ${-76}^{+42}_{-37}$ &
  ${311}^{+208}_{-58}$ \\ \hline
\multicolumn{8}{c}{\textbf{Core B:} $0.03''^2$, $E(B-V) = 0.98$ } \\ \hline
H$_2$ 1--0 S(8) &
  $1.7147$ &
  ${-17.60}^{+0.08}_{-0.07}$ &
  ${2.69}^{+0.52}_{-0.46}$ &
  2 &
  ${-191}^{+32}_{-45}$ &
  ${337}^{+20}_{-44}$ \\
H$_2$ 1--0 S(7) &
  $1.7480$ &
  ${-18.01}^{+0.12}_{-0.18}$ &
  ${1.11}^{+0.30}_{-0.46}$ &
  $<$2 &
  ${343}^{+33}_{-81}$ &
  ${129}^{+129}_{-244}$ \\
H$_2$ 1--0 S(6) &
  $1.7880$ &
  ${-18.47}^{+0.25}_{-0.17}$ &
  ${0.41}^{+0.24}_{-0.16}$ &
  $<$2 &
  ${76}^{+76}_{-40}$ &
  --- \\
H$_2$ 1--0 S(5) &
  $1.8358$ &
  ${-17.19}^{+0.04}_{-0.05}$ &
  ${8.59}^{+0.81}_{-1.01}$ &
  $<$2 &
  ${103}^{+94}_{-73}$ &
  ${1496}^{+105}_{-108}$ \\
\ion{H}{1} Pa$\alpha$ &
  $1.8756$ &
  ${-16.48}^{+0.01}_{-0.01}$ &
  ${47.02}^{+1.17}_{-1.25}$ &
  12 &
  ${2}^{+24}_{-24}$ &
  ${1572}^{+33}_{-33}$ \\
H$_2$ 1--0 S(4) &
  $1.8920$ &
  ${-17.77}^{+0.22}_{-0.24}$ &
  ${2.44}^{+1.26}_{-1.37}$ &
  $<$2 &
  ${-389}^{+252}_{-355}$ &
  ${907}^{+861}_{-588}$ \\
H$_2$ 1--0 S(3) &
  $1.9576$ &
  ${-17.77}^{+0.14}_{-0.14}$ &
  ${2.67}^{+0.88}_{-0.82}$ &
  $<$2 &
  ${268}^{+158}_{-82}$ &
  ${460}^{+333}_{-268}$ \\
H$_2$ 1--0 S(2) &
  $2.0338$ &
  ${-18.54}^{+0.22}_{-0.24}$ &
  ${0.50}^{+0.25}_{-0.27}$ &
  $<$2 &
  ${-270}^{+22}_{-270}$ &
  --- \\
H$_2$ 1--0 S(1) &
  $2.1218$ &
  ${-17.03}^{+0.02}_{-0.05}$ &
  ${18.19}^{+0.93}_{-1.98}$ &
  2.6 &
  ${-301}^{+99}_{-68}$ &
  --- \\
H$_2$ 1--0 S(0) &
  $2.2235$ &
  ${-17.58}^{+0.07}_{-0.06}$ &
  ${5.77}^{+0.92}_{-0.88}$ &
  $<$2 &
  --- &
  --- \\
H$_2$ 1--0 Q(6) &
  $2.4756$ &
  ${-17.60}^{+0.07}_{-0.05}$ &
  ${9.68}^{+1.58}_{-1.23}$ &
  $<$2 &
  ${467}^{+107}_{-130}$ &
  ${1287}^{+106}_{-122}$ \\
H$_2$ 1--0 Q(7) &
  $2.5001$ &
  ${-17.59}^{+0.07}_{-0.10}$ &
  ${10.18}^{+1.69}_{-2.29}$ &
  $<$2 &
  --- &
  --- \\
H$_2$ 1--0 Q(8) &
  $2.5280$ &
  ${-17.70}^{+0.11}_{-0.12}$ &
  ${8.03}^{+1.95}_{-2.29}$ &
  $<$2 &
  ${377}^{+120}_{-126}$ &
  ${922}^{+411}_{-260}$ \\
H$_2$ 1--0 Q(9) &
  $2.5598$ &
  ${-18.01}^{+0.14}_{-0.14}$ &
  ${4.03}^{+1.34}_{-1.32}$ &
  $<$2 &
  ${-309}^{+66}_{-85}$ &
  ${362}^{+329}_{-108}$ \\
H$_2$ 1--0 O(2) $+$ Br$\beta$ &
  $2.6269$ &
  ${-17.54}^{+0.06}_{-0.08}$ &
  ${12.26}^{+1.55}_{-2.30}$ &
  $<$2 &
  ${-295}^{+70}_{-88}$ &
  ${1002}^{+79}_{-85}$ \\
H$_2$ 1--0 O(3) &
  $2.8025$ &
  ${-17.11}^{+0.02}_{-0.02}$ &
  ${38.94}^{+1.76}_{-1.80}$ &
  2.5 &
  ${-52}^{+55}_{-64}$ &
  --- \\
H$_2$ 1--0 O(4) &
  $3.0039$ &
  ${-17.61}^{+0.10}_{-0.08}$ &
  ${14.23}^{+3.18}_{-2.76}$ &
  $<$2 &
  ${278}^{+142}_{-82}$ &
  ${982}^{+112}_{-117}$ \\ \hline
\multicolumn{8}{c}{\textbf{Swirl:} $0.14''^2$, $E(B-V) = 0.14$ } \\ \hline
H$_2$ 1--0 S(8) &
  $1.7147$ &
  ${-17.20}^{+0.19}_{-0.18}$ &
  ${2.22}^{+0.97}_{-0.91}$ &
  $<$2 &
  ${116}^{+116}_{-145}$ &
  ${671}^{+533}_{-223}$ \\
H$_2$ 1--0 S(7) &
  $1.7480$ &
  ${-17.81}^{+0.13}_{-0.16}$ &
  ${0.58}^{+0.18}_{-0.22}$ &
  $<$2 &
  ${584}^{+13}_{-69}$ &
  ${40}^{+40}_{-97}$ \\
H$_2$ 1--0 S(6) &
  $1.7880$ &
  ${-17.57}^{+0.24}_{-0.28}$ &
  ${1.08}^{+0.60}_{-0.71}$ &
  $<$2 &
  ${39}^{+112}_{-248}$ &
  ${423}^{+423}_{-353}$ \\
H$_2$ 1--0 S(5) &
  $1.8358$ &
  ${-16.91}^{+0.03}_{-0.03}$ &
  ${5.42}^{+0.34}_{-0.31}$ &
  5.1 &
  ${-130}^{+12}_{-9}$ &
  ${336}^{+26}_{-25}$ \\
\ion{H}{1} Pa$\alpha$ &
  $1.8756$ &
  ${-16.43}^{+0.02}_{-0.01}$ &
  ${17.67}^{+0.67}_{-0.58}$ &
  12.7 &
  ${-89}^{+8}_{-7}$ &
  ${418}^{+19}_{-14}$ \\
H$_2$ 1--0 S(4) &
  $1.8920$ &
  ${-17.10}^{+0.07}_{-0.09}$ &
  ${3.84}^{+0.62}_{-0.83}$ &
  2.5 &
  ${-100}^{+44}_{-61}$ &
  ${487}^{+109}_{-149}$ \\
H$_2$ 1--0 S(3) &
  $1.9576$ &
  ${-16.77}^{+0.01}_{-0.02}$ &
  ${8.92}^{+0.30}_{-0.33}$ &
  10.7 &
  ${-111}^{+6}_{-5}$ &
  ${301}^{+9}_{-20}$ \\
H$_2$ 1--0 S(2) &
  $2.0338$ &
  ${-17.30}^{+0.06}_{-0.05}$ &
  ${2.98}^{+0.43}_{-0.31}$ &
  3.5 &
  ${-98}^{+19}_{-24}$ &
  ${321}^{+65}_{-57}$ \\
H$_2$ 1--0 S(1) &
  $2.1218$ &
  ${-16.72}^{+0.01}_{-0.02}$ &
  ${12.73}^{+0.35}_{-0.48}$ &
  12 &
  ${-119}^{+5}_{-4}$ &
  ${338}^{+13}_{-11}$ \\
H$_2$ 1--0 S(0) &
  $2.2235$ &
  ${-17.29}^{+0.09}_{-0.21}$ &
  ${3.88}^{+0.80}_{-1.81}$ &
  3.2 &
  ${-175}^{+130}_{-28}$ &
  ${388}^{+87}_{-411}$ \\
H$_2$ 1--0 Q(6) &
  $2.4756$ &
  ${-17.56}^{+0.11}_{-0.11}$ &
  ${3.93}^{+0.97}_{-1.03}$ &
  $<$2 &
  ${-74}^{+43}_{-58}$ &
  ${323}^{+116}_{-125}$  \\
H$_2$ 1--0 Q(7) &
  $2.5001$ &
  ${-17.28}^{+0.05}_{-0.06}$ &
  ${7.69}^{+0.80}_{-1.09}$ &
  3.8 &
  ${-155}^{+19}_{-24}$ &
  ${337}^{+39}_{-82}$ \\
H$_2$ 1--0 Q(8) &
  $2.5280$ &
  ${-17.84}^{+0.13}_{-0.14}$ &
  ${2.28}^{+0.71}_{-0.71}$ &
  $<$2 &
  ${1}^{+106}_{-115}$ &
  ${305}^{+273}_{-154}$ \\
H$_2$ 1--0 Q(9) &
  $2.5598$ &
  ${-17.45}^{+0.06}_{-0.09}$ &
  ${5.65}^{+0.81}_{-1.16}$ &
  3 &
  ${-144}^{+25}_{-22}$ &
  ${316}^{+58}_{-94}$ \\
H$_2$ 1--0 O(2) $+$ Br$\beta$ &
  $2.6269$ &
  ${-17.05}^{+0.03}_{-0.03}$ &
  ${14.56}^{+0.97}_{-1.05}$ &
  4.3 &
  ${-175}^{+17}_{-14}$ &
  ${553}^{+44}_{-57}$ \\
H$_2$ 1--0 O(3) &
  $2.8025$ &
  ${-17.00}^{+0.02}_{-0.02}$ &
  ${19.81}^{+0.96}_{-0.94}$ &
  8.3 &
  ${-123}^{+8}_{-7}$ &
  ${340}^{+28}_{-20}$ \\
H$_2$ 1--0 O(4) &
  $3.0039$ &
  ${-17.44}^{+0.06}_{-0.07}$ &
  ${8.62}^{+1.17}_{-1.40}$ &
  3.4 &
  ${-100}^{+27}_{-21}$ &
  ${297}^{+41}_{-32}$ \\ \hline
\multicolumn{8}{c}{\textbf{Filament A:} $0.13''^2$, $E(B-V) = 0.10$ } \\ \hline
H$_2$ 1--0 S(8) &
  $1.7147$ &
  ${-17.20}^{+0.14}_{-0.10}$ &
  ${2.52}^{+0.84}_{-0.59}$ &
  $<$2 &
  ${7}^{+60}_{-81}$ &
  ${547}^{+154}_{-180}$ \\
H$_2$ 1--0 S(7) &
  $1.7480$ &
  ${-17.87}^{+0.11}_{-0.21}$ &
  ${0.60}^{+0.15}_{-0.29}$ &
  $<$2 &
  ${645}^{+40}_{-41}$ &
  ${31}^{+31}_{-75}$ \\
H$_2$ 1--0 S(6) &
  $1.7880$ &
  ${-17.22}^{+0.09}_{-0.08}$ &
  ${2.87}^{+0.58}_{-0.52}$ &
  3.3 &
  ${-447}^{+96}_{-66}$ &
  ${1151}^{+629}_{-347}$ \\
H$_2$ 1--0 S(5) &
  $1.8358$ &
  ${-16.97}^{+0.02}_{-0.02}$ &
  ${5.35}^{+0.30}_{-0.28}$ &
  9.5 &
  ${-257}^{+11}_{-7}$ &
  ${219}^{+47}_{-41}$ \\
\ion{H}{1} Pa$\alpha$ &
  $1.8756$ &
  ${-16.68}^{+0.02}_{-0.02}$ &
  ${11.60}^{+0.51}_{-0.59}$ &
  13.6 &
  ${-222}^{+7}_{-8}$ &
  ${290}^{+34}_{-6}$ \\
H$_2$ 1--0 S(4) &
  $1.8920$ &
  ${-17.30}^{+0.06}_{-0.08}$ &
  ${2.81}^{+0.39}_{-0.51}$ &
  3.3 &
  ${-250}^{+26}_{-24}$ &
  ${295}^{+90}_{-42}$ \\
H$_2$ 1--0 S(3) &
  $1.9576$ &
  ${-16.72}^{+0.02}_{-0.01}$ &
  ${11.50}^{+0.42}_{-0.39}$ &
  15.7 &
  ${-241}^{+8}_{-5}$ &
  ${202}^{+10}_{-55}$ \\
H$_2$ 1--0 S(2) &
  $2.0338$ &
  ${-17.23}^{+0.04}_{-0.04}$ &
  ${4.00}^{+0.36}_{-0.38}$ &
  5.2 &
  ${-233}^{+12}_{-18}$ &
  ${182}^{+49}_{-76}$ \\
H$_2$ 1--0 S(1) &
  $2.1218$ &
  ${-16.72}^{+0.02}_{-0.01}$ &
  ${15.03}^{+0.75}_{-0.45}$ &
  21.6 &
  ${-239}^{+6}_{-4}$ &
  ${259}^{+4}_{-12}$ \\
H$_2$ 1--0 S(0) &
  $2.2235$ &
  ${-17.22}^{+0.06}_{-0.28}$ &
  ${5.28}^{+0.75}_{-3.32}$ &
  3.8 &
  ${-280}^{+15}_{-26}$ &
  ${422}^{+106}_{-44}$ \\
H$_2$ 1--0 Q(6) &
  $2.4756$ &
  ${-17.42}^{+0.16}_{-0.18}$ &
  ${6.36}^{+2.30}_{-2.70}$ &
  $<$2 &
  ${-166}^{+92}_{-99}$ &
  ${528}^{+341}_{-320}$ \\
H$_2$ 1--0 Q(7) &
  $2.5001$ &
  ${-17.21}^{+0.04}_{-0.04}$ &
  ${10.55}^{+0.96}_{-0.96}$ &
  6.4 &
  ${-248}^{+13}_{-7}$ &
  ${204}^{+80}_{-56}$ \\
H$_2$ 1--0 Q(8) &
  $2.5280$ &
  ${-17.43}^{+0.09}_{-0.11}$ &
  ${6.72}^{+1.45}_{-1.76}$ &
  $<$2 &
  ${44}^{+882}_{-338}$ &
  ${1631}^{+2}_{-5}$ \\
H$_2$ 1--0 Q(9) &
  $2.5598$ &
  ${-17.19}^{+0.04}_{-0.03}$ &
  ${11.94}^{+1.06}_{-0.79}$ &
  4.8 &
  ${-323}^{+16}_{-20}$ &
  ${357}^{+30}_{-38}$ \\
H$_2$ 1--0 O(2) $+$ Br$\beta$ &
  $2.6269$ &
  ${-17.14}^{+0.03}_{-0.03}$ &
  ${13.90}^{+0.81}_{-1.03}$ &
  7 &
  ${-286}^{+8}_{-6}$ &
  ${262}^{+7}_{-42}$ \\
H$_2$ 1--0 O(3) &
  $2.8025$ &
  ${-16.89}^{+0.02}_{-0.01}$ &
  ${29.81}^{+1.09}_{-0.97}$ &
  16.1 &
  ${-240}^{+4}_{-5}$ &
  ${208}^{+11}_{-50}$ \\
H$_2$ 1--0 O(4) &
  $3.0039$ &
  ${-17.47}^{+0.10}_{-0.07}$ &
  ${9.34}^{+2.15}_{-1.45}$ &
  4.4 &
  ${-224}^{+30}_{-25}$ &
  ${256}^{+167}_{-48}$ \\ \hline
\multicolumn{8}{c}{\textbf{Filament B:} $0.59''^2$, $E(B-V) = 0.12$ } \\ \hline
H$_2$ 1--0 S(8) &
  $1.7147$ &
  ${-16.43}^{+0.09}_{-0.07}$ &
  ${3.58}^{+0.78}_{-0.62}$ &
  $<$2 &
  ${53}^{+52}_{-61}$ &
  ${609}^{+66}_{-88}$ \\
H$_2$ 1--0 S(7) &
  $1.7480$ &
  ${-16.99}^{+0.09}_{-0.06}$ &
  ${1.04}^{+0.22}_{-0.16}$ &
  2.7 &
  ${-171}^{+26}_{-13}$ &
  ${89}^{+89}_{-62}$ \\
H$_2$ 1--0 S(6) &
  $1.7880$ &
  ${-16.66}^{+0.08}_{-0.07}$ &
  ${2.49}^{+0.48}_{-0.38}$ &
  $<$2 &
  ${22}^{+55}_{-56}$ &
  ${582}^{+94}_{-129}$ \\
H$_2$ 1--0 S(5) &
  $1.8358$ &
  ${-16.36}^{+0.02}_{-0.02}$ &
  ${5.29}^{+0.21}_{-0.28}$ &
  9.1 &
  ${-147}^{+7}_{-7}$ &
  ${187}^{+10}_{-10}$ \\
\ion{H}{1} Pa$\alpha$ &
  $1.8756$ &
  ${-16.27}^{+0.03}_{-0.02}$ &
  ${7.28}^{+0.43}_{-0.40}$ &
  10.4 &
  ${-124}^{+11}_{-8}$ &
  ${198}^{+16}_{-60}$ \\
H$_2$ 1--0 S(4) &
  $1.8920$ &
  ${-16.62}^{+0.05}_{-0.05}$ &
  ${3.29}^{+0.36}_{-0.41}$ &
  4.8 &
  ${-125}^{+12}_{-19}$ &
  ${217}^{+101}_{-78}$ \\
H$_2$ 1--0 S(3) &
  $1.9576$ &
  ${-16.10}^{+0.02}_{-0.02}$ &
  ${11.55}^{+0.43}_{-0.42}$ &
  16 &
  ${-130}^{+5}_{-2}$ &
  ${257}^{+43}_{-9}$\\
H$_2$ 1--0 S(2) &
  $2.0338$ &
  ${-16.54}^{+0.02}_{-0.02}$ &
  ${4.81}^{+0.27}_{-0.21}$ &
  8 &
  ${-136}^{+6}_{-5}$ &
  ${215}^{+65}_{-47}$ \\
H$_2$ 1--0 S(1) &
  $2.1218$ &
  ${-16.09}^{+0.01}_{-0.01}$ &
  ${15.14}^{+0.21}_{-0.20}$ &
  22.6 &
  ${-121}^{+1}_{-2}$ &
  ${202}^{+4}_{-3}$ \\
H$_2$ 1--0 S(0) &
  $2.2235$ &
  ${-16.67}^{+0.03}_{-0.04}$ &
  ${4.55}^{+0.34}_{-0.40}$ &
  6.5 &
  ${-160}^{+9}_{-11}$ &
  ${259}^{+66}_{-41}$ \\
H$_2$ 1--0 Q(6) &
  $2.4756$ &
  ${-16.36}^{+0.08}_{-0.08}$ &
  ${17.74}^{+3.16}_{-3.06}$ &
  $<$2 &
  ${-101}^{+107}_{-130}$ &
  --- \\
H$_2$ 1--0 Q(7) &
  $2.5001$ &
  ${-16.50}^{+0.03}_{-0.04}$ &
  ${13.21}^{+0.82}_{-1.36}$ &
  3.2 &
  ${-166}^{+10}_{-11}$ &
  ${335}^{+33}_{-63}$ \\
H$_2$ 1--0 Q(8) &
  $2.5280$ &
  ${-16.35}^{+0.02}_{-0.02}$ &
  ${19.93}^{+1.00}_{-1.09}$ &
  2.1 &
  ${422}^{+64}_{-111}$ &
  ${1629}^{+1}_{-0}$ \\
H$_2$ 1--0 Q(9) &
  $2.5598$ &
  ${-16.62}^{+0.04}_{-0.04}$ &
  ${10.91}^{+1.11}_{-1.04}$ &
  3.3 &
  ${-229}^{+29}_{-37}$ &
  ${600}^{+51}_{-63}$ \\
H$_2$ 1--0 O(2) $+$ Br$\beta$ &
  $2.6269$ &
  ${-16.66}^{+0.04}_{-0.03}$ &
  ${10.13}^{+0.95}_{-0.72}$ &
  4.5 &
  ${-172}^{+13}_{-10}$ &
  ${298}^{+5}_{-11}$ \\
H$_2$ 1--0 O(3) &
  $2.8025$ &
  ${-16.40}^{+0.01}_{-0.02}$ &
  ${22.65}^{+0.69}_{-0.86}$ &
  17.5 &
  ${-119}^{+2}_{-2}$ &
  ${141}^{+13}_{-16}$  \\
H$_2$ 1--0 O(4) &
  $3.0039$ &
  ${-16.81}^{+0.03}_{-0.03}$ &
  ${10.39}^{+0.64}_{-0.75}$ &
  6.2 &
  ${-121}^{+9}_{-11}$ &
  ${157}^{+29}_{-27}$ \\ \hline
\multicolumn{8}{c}{\textbf{Filament C:} $0.25''^2$, $E(B-V) = 0.11$ } \\ \hline
H$_2$ 1--0 S(8) &
  $1.7147$ &
  ${-16.60}^{+0.06}_{-0.04}$ &
  ${6.43}^{+0.93}_{-0.60}$ &
  2.9 &
  ${82}^{+23}_{-25}$ &
  ${476}^{+50}_{-61}$ \\
H$_2$ 1--0 S(7) &
  $1.7480$ &
  ${-17.17}^{+0.08}_{-0.08}$ &
  ${1.83}^{+0.33}_{-0.36}$ &
  2.2 &
  ${-50}^{+19}_{-29}$ &
  ${201}^{+69}_{-96}$ \\
H$_2$ 1--0 S(6) &
  $1.7880$ &
  ${-17.57}^{+0.17}_{-0.21}$ &
  ${0.80}^{+0.31}_{-0.38}$ &
  $<$2 &
  ${-60}^{+47}_{-60}$ &
  ${40}^{+40}_{-47}$ \\
H$_2$ 1--0 S(5) &
  $1.8358$ &
  ${-16.46}^{+0.02}_{-0.02}$ &
  ${11.38}^{+0.51}_{-0.44}$ &
  9.6 &
  ${-51}^{+9}_{-8}$ &
  ${288}^{+32}_{-8}$ \\
\ion{H}{1} Pa$\alpha$ &
  $1.8756$ &
  ${-16.32}^{+0.02}_{-0.02}$ &
  ${17.04}^{+0.80}_{-0.92}$ &
  12.5 &
  ${-46}^{+6}_{-6}$ &
  ${193}^{+11}_{-10}$ \\
H$_2$ 1--0 S(4) &
  $1.8920$ &
  ${-16.71}^{+0.05}_{-0.06}$ &
  ${7.00}^{+0.73}_{-0.90}$ &
  5.1 &
  ${-69}^{+19}_{-16}$ &
  ${256}^{+85}_{-40}$ \\
H$_2$ 1--0 S(3) &
  $1.9576$ &
  ${-16.18}^{+0.01}_{-0.01}$ &
  ${25.77}^{+0.56}_{-0.31}$ &
  25.4 &
  ${-54}^{+1}_{-1}$ &
  ${254}^{+1}_{-2}$ \\
H$_2$ 1--0 S(2) &
  $2.0338$ &
  ${-16.68}^{+0.01}_{-0.02}$ &
  ${9.25}^{+0.31}_{-0.34}$ &
  9.9 &
  ${-56}^{+4}_{-5}$ &
  ${211}^{+8}_{-48}$ \\
H$_2$ 1--0 S(1) &
  $2.1218$ &
  ${-16.21}^{+0.01}_{-0.01}$ &
  ${30.65}^{+0.55}_{-0.57}$ &
  27.8 &
  ${-39}^{+1}_{-1}$ &
  ${260}^{+3}_{-5}$ \\
H$_2$ 1--0 S(0) &
  $2.2235$ &
  ${-16.88}^{+0.03}_{-0.02}$ &
  ${7.48}^{+0.49}_{-0.38}$ &
  7 &
  ${-74}^{+10}_{-10}$ &
  ${187}^{+70}_{-44}$ \\
H$_2$ 1--0 Q(6) &
  $2.4756$ &
  ${-16.89}^{+0.08}_{-0.17}$ &
  ${13.68}^{+2.66}_{-5.59}$ &
  $<$2 &
  ${-57}^{+20}_{-28}$ &
  ${297}^{+141}_{-131}$\\
H$_2$ 1--0 Q(7) &
  $2.5001$ &
  ${-16.67}^{+0.03}_{-0.04}$ &
  ${23.31}^{+1.49}_{-2.08}$ &
  5 &
  ${-56}^{+8}_{-7}$ &
  ${296}^{+30}_{-19}$ \\
H$_2$ 1--0 Q(8) &
  $2.5280$ &
  ${-16.71}^{+0.03}_{-0.03}$ &
  ${22.65}^{+1.58}_{-1.52}$ &
  $<$2 &
  ${-35}^{+67}_{-82}$ &
  --- \\
H$_2$ 1--0 Q(9) &
  $2.5598$ &
  ${-16.99}^{+0.04}_{-0.04}$ &
  ${12.24}^{+1.23}_{-1.20}$ &
  5.7 &
  ${-68}^{+14}_{-12}$ &
  ${257}^{+73}_{-44}$ \\
H$_2$ 1--0 O(2) $+$ Br$\beta$ &
  $2.6269$ &
  ${-16.72}^{+0.03}_{-0.03}$ &
  ${23.06}^{+1.63}_{-1.86}$ &
  8.1 &
  ${-73}^{+13}_{-8}$ &
  ${292}^{+36}_{-14}$ \\
H$_2$ 1--0 O(3) &
  $2.8025$ &
  ${-16.44}^{+0.01}_{-0.01}$ &
  ${54.92}^{+1.22}_{-1.31}$ &
  24 &
  ${-35}^{+7}_{-9}$ &
  ${168}^{+24}_{-26}$  \\
H$_2$ 1--0 O(4) &
  $3.0039$ &
  ${-17.04}^{+0.03}_{-0.04}$ &
  ${16.14}^{+1.32}_{-1.47}$ &
  5.1 &
  ${-40}^{+10}_{-10}$ &
  ${158}^{+41}_{-22}$ \\ \hline
\multicolumn{8}{c}{\textbf{Background:} $0.36''^2$, $E(B-V) = 0.08$ } \\ \hline
H$_2$ 1--0 S(8) &
  $1.7147$ &
  ${-16.98}^{+0.13}_{-0.17}$ &
  ${1.56}^{+0.46}_{-0.60}$ &
  2.9 &
  ${200}^{+74}_{-30}$ &
  ${240}^{+171}_{-181}$\\
H$_2$ 1--0 S(7) &
  $1.7480$ &
  ${-17.44}^{+0.15}_{-0.15}$ &
  ${0.59}^{+0.20}_{-0.21}$ &
  $<$2 &
  ${499}^{+68}_{-68}$ &
  ${258}^{+226}_{-76}$ \\
H$_2$ 1--0 S(6) &
  $1.7880$ &
  ${-17.16}^{+0.15}_{-0.11}$ &
  ${1.23}^{+0.42}_{-0.30}$ &
  2 &
  ${336}^{+78}_{-59}$ &
  ${294}^{+220}_{-125}$ \\
H$_2$ 1--0 S(5) &
  $1.8358$ &
  ${-17.32}^{+0.16}_{-0.13}$ &
  ${0.96}^{+0.35}_{-0.29}$ &
  $<$2 &
  ${599}^{+77}_{-70}$ &
  ${340}^{+226}_{-138}$ \\
\ion{H}{1} Pa$\alpha$ &
  $1.8756$ &
  ${-17.34}^{+0.17}_{-0.18}$ &
  ${0.94}^{+0.36}_{-0.40}$ &
  $<$2 &
  ${31}^{+43}_{-51}$ &
  ${80}^{+80}_{-224}$  \\
H$_2$ 1--0 S(4) &
  $1.8920$ &
  ${-17.39}^{+0.16}_{-0.27}$ &
  ${0.86}^{+0.32}_{-0.54}$ &
  $<$2 &
  ${11}^{+38}_{-70}$ &
  ${71}^{+71}_{-226}$\\
H$_2$ 1--0 S(3) &
  $1.9576$ &
  ${-17.04}^{+0.14}_{-0.13}$ &
  ${2.07}^{+0.67}_{-0.64}$ &
  2.3 &
  ${88}^{+68}_{-62}$ &
  ${359}^{+260}_{-128}$ \\
H$_2$ 1--0 S(2) &
  $2.0338$ &
  ${-17.09}^{+0.10}_{-0.08}$ &
  ${2.12}^{+0.47}_{-0.39}$ &
  $<$2 &
  --- &
  ${1273}^{+130}_{-177}$ \\
H$_2$ 1--0 S(1) &
  $2.1218$ &
  ${-17.00}^{+0.05}_{-0.05}$ &
  ${2.88}^{+0.33}_{-0.34}$ &
  2.2 &
  ${-81}^{+41}_{-62}$ &
  ${573}^{+95}_{-81}$ \\
H$_2$ 1--0 S(0) &
  $2.2235$ &
  ${-17.44}^{+0.21}_{-0.26}$ &
  ${1.17}^{+0.57}_{-0.69}$ &
  $<$2 &
  --- &
  --- \\
H$_2$ 1--0 Q(6) &
  $2.4756$ &
  ${-17.94}^{+0.23}_{-0.23}$ &
  ${0.72}^{+0.38}_{-0.39}$ &
  $<$2 &
  ${-102}^{+60}_{-102}$ &
  --- \\
H$_2$ 1--0 Q(7) &
  $2.5001$ &
  ${-17.29}^{+0.13}_{-0.18}$ &
  ${3.31}^{+1.01}_{-1.38}$ &
  $<$2 &
  ${-223}^{+136}_{-114}$ &
  ${500}^{+191}_{-263}$  \\
H$_2$ 1--0 Q(8) &
  $2.5280$ &
  ${-17.61}^{+0.16}_{-0.23}$ &
  ${1.68}^{+0.62}_{-0.88}$ &
  $<$2 &
  ${398}^{+170}_{-217}$ &
  ${299}^{+297}_{-299}$ \\
H$_2$ 1--0 Q(9) &
  $2.5598$ &
  ${-17.61}^{+0.18}_{-0.16}$ &
  ${1.71}^{+0.71}_{-0.65}$ &
  $<$2 &
  ${-466}^{+62}_{-59}$ &
  ${180}^{+180}_{-160}$ \\
H$_2$ 1--0 O(2) $+$ Br$\beta$ &
  $2.6269$ &
  ${-17.00}^{+0.10}_{-0.06}$ &
  ${7.28}^{+1.61}_{-1.04}$ &
  $<$2 &
  ${211}^{+312}_{-550}$ &
  ${1630}^{+3}_{-2}$ \\
H$_2$ 1--0 O(3) &
  $2.8025$ &
  ${-17.54}^{+0.14}_{-0.20}$ &
  ${2.47}^{+0.79}_{-1.15}$ &
  $<$2 &
  ${-835}^{+2}_{-1}$ &
  ${196}^{+154}_{-235}$ \\
H$_2$ 1--0 O(4) &
  $3.0039$ &
  ${-17.82}^{+0.14}_{-0.20}$ &
  ${1.59}^{+0.52}_{-0.71}$ &
  $<$2 &
  ${-9}^{+34}_{-16}$ &
  ${48}^{+48}_{-132}$ \\ \hline
\multicolumn{8}{c}{\textbf{Whole:} $15.49''^2$, $E(B-V) = 0.08$ } \\ \hline
H$_2$ 1--0 S(8) &
  $1.7147$ &
  --- &
  --- &
  $<$2 &
  --- &
  --- \\
H$_2$ 1--0 S(7) &
  $1.7480$ &
  ${-15.28}^{+0.07}_{-0.09}$ &
  ${1.15}^{+0.18}_{-0.23}$ &
  $<$2 &
  ${516}^{+34}_{-36}$ &
  ${337}^{+35}_{-49}$ \\
H$_2$ 1--0 S(6) &
  $1.7880$ &
  ${-15.23}^{+0.12}_{-0.13}$ &
  ${1.40}^{+0.40}_{-0.43}$ &
  $<$2 &
  ${143}^{+100}_{-103}$ &
  ${547}^{+226}_{-96}$ \\
H$_2$ 1--0 S(5) &
  $1.8358$ &
  ${-15.21}^{+0.07}_{-0.07}$ &
  ${1.58}^{+0.25}_{-0.24}$ &
  2.5 &
  ${-143}^{+29}_{-26}$ &
  ${275}^{+91}_{-35}$ \\
\ion{H}{1} Pa$\alpha$ &
  $1.8756$ &
  ${-14.78}^{+0.15}_{-0.14}$ &
  ${4.68}^{+1.64}_{-1.52}$ &
  5.5 &
  ${-54}^{+42}_{-111}$ &
  ${385}^{+131}_{-695}$\\
H$_2$ 1--0 S(4) &
  $1.8920$ &
  ${-15.22}^{+0.14}_{-0.19}$ &
  ${1.71}^{+0.55}_{-0.76}$ &
  $<$2 &
  ${-92}^{+89}_{-88}$ &
  ${337}^{+200}_{-238}$ \\
H$_2$ 1--0 S(3) &
  $1.9576$ &
  ${-14.81}^{+0.03}_{-0.03}$ &
  ${4.73}^{+0.36}_{-0.38}$ &
  5.9 &
  ${-101}^{+13}_{-12}$ &
  ${303}^{+7}_{-22}$  \\
H$_2$ 1--0 S(2) &
  $2.0338$ &
  ${-15.20}^{+0.03}_{-0.02}$ &
  ${2.21}^{+0.16}_{-0.12}$ &
  3 &
  ${-68}^{+13}_{-12}$ &
  ${342}^{+18}_{-28}$ \\
H$_2$ 1--0 S(1) &
  $2.1218$ &
  ${-14.71}^{+0.01}_{-0.02}$ &
  ${7.74}^{+0.19}_{-0.30}$ &
  8.5 &
  ${-81}^{+5}_{-5}$ &
  ${430}^{+28}_{-27}$ \\
H$_2$ 1--0 S(0) &
  $2.2235$ &
  ${-15.38}^{+0.05}_{-0.06}$ &
  ${1.84}^{+0.22}_{-0.26}$ &
  $<$2 &
  ${-234}^{+52}_{-45}$ &
  ${514}^{+92}_{-68}$ \\
H$_2$ 1--0 Q(6) &
  $2.4756$ &
  ${-16.13}^{+0.34}_{-0.30}$ &
  ${0.62}^{+0.48}_{-0.42}$ &
  $<$2 &
  ${0}^{+43}_{-10}$ &
  --- \\
H$_2$ 1--0 Q(7) &
  $2.5001$ &
  ${-15.76}^{+0.19}_{-0.25}$ &
  ${1.52}^{+0.68}_{-0.89}$ &
  $<$2 &
  ${-180}^{+95}_{-139}$ &
  ${164}^{+164}_{-150}$ \\
H$_2$ 1--0 Q(8) &
  $2.5280$ &
  ${-16.40}^{+0.43}_{-0.22}$ &
  ${0.37}^{+0.37}_{-0.18}$ &
  $<$2 &
  ${467}^{+467}_{-97}$ &
  --- \\
H$_2$ 1--0 Q(9) &
  $2.5598$ &
  ${-15.67}^{+0.10}_{-0.12}$ &
  ${2.00}^{+0.45}_{-0.55}$ &
  $<$2 &
  ${-381}^{+84}_{-74}$ &
  ${330}^{+217}_{-81}$ \\
H$_2$ 1--0 O(2) $+$ Br$\beta$ &
  $2.6269$ &
  ${-15.30}^{+0.05}_{-0.05}$ &
  ${4.88}^{+0.55}_{-0.54}$ &
  3.4 &
  ${-141}^{+21}_{-16}$ &
  ${381}^{+45}_{-54}$ \\
H$_2$ 1--0 O(3) &
  $2.8025$ &
  ${-15.20}^{+0.04}_{-0.05}$ &
  ${7.57}^{+0.74}_{-0.88}$ &
  5.3 &
  ${-71}^{+13}_{-14}$ &
  ${257}^{+65}_{-40}$ \\
H$_2$ 1--0 O(4) &
  $3.0039$ &
  ${-15.40}^{+0.03}_{-0.05}$ &
  ${5.59}^{+0.40}_{-0.62}$ &
  3 &
  ${-85}^{+15}_{-8}$ &
  ${258}^{+3}_{-9}$\\ \hline
\end{longtable}
\end{center}

\clearpage
\bibliography{biblio}{}

@ARTICLE{hlavacek-centaurus-2026,
       author = {{Hlavacek-Larrondo}, Julie and {Choi}, Hyunseop and {Guo}, Minghao and {Marquis}, Mathieu and {Pereira}, Olivia and {Voit}, G. Mark and {Lo{\"\i}c}, Albert and {Barrera-Ballesteros}, Jorge and {Canning}, Rebecca E.~A. and {D'Eugenio}, Francesco and {Donahue}, Megan and {Fabian}, Andrew C. and {Ferland}, Gary J. and {Gallagher}, John S. and {Gendron-Marsolais}, Marie-Lou and {Guillard}, Pierre and {Hatch}, Nina and {Kotulla}, Ralf and {Li}, Yuan and {Maiolino}, Roberto and {Man}, Allison and {McDonald}, Michael A. and {McNamara}, B.~R. and {Olivares}, Valeria and {Prunier}, Marine and {Reefe}, Michael and {Reynolds}, Christopher S. and {Rhea}, Carter and {Richard-Laferri{\`e}re}, Annabelle and {Russell}, Helen R. and {Salom{\'e}}, Philippe and {Sun}, Ming and {Tamhane}, Prathamesh and {Taylor}, Gregory and {Thilloy}, Auriane and {Tremblay}, Grant R. and {Vigneron}, Benjamin and {Walker}, Stephen A.},
        title = "{JWST Reveals How Black Holes are Fed: Kiloparsec-scale Multiphase Filaments Feed Subkiloparsec Circumnuclear Disks}",
      journal = {\apjl},
         year = 2026,
        month = {july},
       volume = {1006},
       number = {1},
          eid = {L9},
        pages = {L9},
          doi = {10.3847/2041-8213/ae81ae},
archivePrefix = {arXiv},
       eprint = {2606.06620},
 primaryClass = {astro-ph.GA},
       adsurl = {https://ui.adsabs.harvard.edu/abs/2026ApJ..1006L...9H}
}

@ARTICLE{baldwin_bpt_1981,
       author = {{Baldwin}, J.~A. and {Phillips}, M.~M. and {Terlevich}, R.},
        title = "{Classification parameters for the emission-line spectra of extragalactic objects.}",
      journal = {\pasp},
         year = 1981,
        month = feb,
       volume = {93},
        pages = {5-19},
          doi = {10.1086/130766},
       adsurl = {https://ui.adsabs.harvard.edu/abs/1981PASP...93....5B}
}

@ARTICLE{2011ApJ...738L..24V,
       author = {{Voit}, G. Mark and {Donahue}, Megan},
        title = "{The Fate of Stellar Mass Loss in Central Cluster Galaxies}",
      journal = {\apjl},
         year = 2011,
        month = sep,
       volume = {738},
       number = {2},
          eid = {L24},
        pages = {L24},
          doi = {10.1088/2041-8205/738/2/L24},
archivePrefix = {arXiv},
       eprint = {1108.2068},
 primaryClass = {astro-ph.CO},
       adsurl = {https://ui.adsabs.harvard.edu/abs/2011ApJ...738L..24V}
}

@article{kruijssenFastInefficientStar2019,
  title = {Fast and Inefficient Star Formation Due to Short-Lived Molecular Clouds and Rapid Feedback},
  author = {Kruijssen, J. M. Diederik and Schruba, Andreas and Chevance, M{\'e}lanie and Longmore, Steven N. and Hygate, Alexander P. S. and Haydon, Daniel T. and McLeod, Anna F. and Dalcanton, Julianne J. and Tacconi, Linda J. and Van Dishoeck, Ewine F.},
  year = 2019,
  month = may,
  journal = {Nature},
  volume = {569},
  number = {7757},
  pages = {519--522},
  issn = {0028-0836, 1476-4687},
  doi = {10.1038/s41586-019-1194-3},
  urldate = {2026-08-13},
  langid = {english}
}

@ARTICLE{fabian_hidden_2022,
       author = {{Fabian}, A.~C. and {Ferland}, G.~J. and {Sanders}, J.~S. and {McNamara}, B.~R. and {Pinto}, C. and {Walker}, S.~A.},
        title = "{Hidden cooling flows in clusters of galaxies}",
      journal = {\mnras},
         year = 2022,
        month = sep,
       volume = {515},
       number = {3},
        pages = {3336-3345},
          doi = {10.1093/mnras/stac2003},
archivePrefix = {arXiv},
       eprint = {2207.04951},
 primaryClass = {astro-ph.GA},
       adsurl = {https://ui.adsabs.harvard.edu/abs/2022MNRAS.515.3336F}
}

@ARTICLE{fabian_hidden_2023b,
       author = {{Fabian}, A.~C. and {Sanders}, J.~S. and {Ferland}, G.~J. and {McNamara}, B.~R. and {Pinto}, C. and {Walker}, S.~A.},
        title = "{Hidden cooling flows in clusters of galaxies - III. Accretion on to the central black hole}",
      journal = {\mnras},
         year = 2023,
        month = sep,
       volume = {524},
       number = {1},
        pages = {716-730},
          doi = {10.1093/mnras/stad1870},
archivePrefix = {arXiv},
       eprint = {2306.11077},
 primaryClass = {astro-ph.GA},
       adsurl = {https://ui.adsabs.harvard.edu/abs/2023MNRAS.524..716F}
}

@ARTICLE{fabian_hidden_2024,
       author = {{Fabian}, A.~C. and {Ferland}, G.~J. and {Sanders}, J.~S. and {Russell}, H.~R. and {McNamara}, B.~R. and {Pinto}, C. and {Hlavacek-Larrondo}, J. and {Walker}, S.~A. and {Ivey}, L.~R. and {McDonald}, M.},
        title = "{Hidden cooling flows - IV. More details on Centaurus and the efficiency of AGN feedback in clusters}",
      journal = {\mnras},
         year = 2024,
        month = dec,
       volume = {535},
       number = {3},
        pages = {2173-2188},
          doi = {10.1093/mnras/stae2414},
archivePrefix = {arXiv},
       eprint = {2410.10675},
 primaryClass = {astro-ph.GA},
       adsurl = {https://ui.adsabs.harvard.edu/abs/2024MNRAS.535.2173F}
}

@ARTICLE{fabian_hidden_2025,
       author = {{Fabian}, A.~C. and {Sanders}, J.~S. and {Ferland}, G.~J. and {Russell}, H.~R. and {McNamara}, B.~R. and {Pinto}, C. and {Walker}, S.~A.},
        title = "{Hidden (absorbed) Cooling Flows V: Groups and Galaxies including Spirals}",
      journal = {arXiv e-prints},
         year = 2025,
        month = aug,
          eid = {arXiv:2508.14785},
        pages = {arXiv:2508.14785},
          doi = {10.48550/arXiv.2508.14785},
archivePrefix = {arXiv},
       eprint = {2508.14785},
 primaryClass = {astro-ph.GA},
       adsurl = {https://ui.adsabs.harvard.edu/abs/2025arXiv250814785F}
}

@ARTICLE{ferland_origin_2008,
       author = {{Ferland}, G.~J. and {Fabian}, A.~C. and {Hatch}, N.~A. and {Johnstone}, R.~M. and {Porter}, R.~L. and {van Hoof}, P.~A.~M. and {Williams}, R.~J.~R.},
        title = "{The origin of molecular hydrogen emission in cooling-flow filaments}",
      journal = {\mnras},
         year = 2008,
        month = may,
       volume = {386},
       number = {1},
        pages = {L72-L76},
          doi = {10.1111/j.1745-3933.2008.00463.x},
archivePrefix = {arXiv},
       eprint = {0802.2535},
 primaryClass = {astro-ph},
       adsurl = {https://ui.adsabs.harvard.edu/abs/2008MNRAS.386L..72F}
}

@ARTICLE{rodriguez-ardila_2011,
       author = {{Rodr{\'\i}guez-Ardila}, A. and {Prieto}, M.~A. and {Portilla}, J.~G. and {Tejeiro}, J.~M.},
        title = "{The Near-infrared Coronal Line Spectrum of 54 nearby Active Galactic Nuclei}",
      journal = {\apj},
         year = 2011,
        month = dec,
       volume = {743},
       number = {2},
          eid = {100},
        pages = {100},
          doi = {10.1088/0004-637X/743/2/100},
archivePrefix = {arXiv},
       eprint = {1109.0984},
 primaryClass = {astro-ph.CO},
       adsurl = {https://ui.adsabs.harvard.edu/abs/2011ApJ...743..100R}
}

@ARTICLE{li_effects_2018,
       author = {{Li}, Yuan and {Ruszkowski}, Mateusz and {Tremblay}, Grant},
        title = "{The Effects of Ram Pressure on the Cold Clouds in the Centers of Galaxy Clusters}",
      journal = {\apj},
         year = 2018,
        month = feb,
       volume = {854},
       number = {2},
          eid = {91},
        pages = {91},
          doi = {10.3847/1538-4357/aaa843},
archivePrefix = {arXiv},
       eprint = {1703.06954},
 primaryClass = {astro-ph.GA},
       adsurl = {https://ui.adsabs.harvard.edu/abs/2018ApJ...854...91L}
}

@ARTICLE{1983ApJ...264..105F,
       author = {{Ferland}, G.~J. and {Netzer}, H.},
        title = "{Are there any shock-heated galaxies ?}",
      journal = {\apj},
         year = 1983,
        month = jan,
       volume = {264},
        pages = {105-113},
          doi = {10.1086/160577},
       adsurl = {https://ui.adsabs.harvard.edu/abs/1983ApJ...264..105F}
}

@ARTICLE{marquis_mapping_2026,
       author = {{Marquis}, Mathieu and {Hlavacek-Larrondo}, Julie and {Pereira}, Olivia and {Reefe}, Michael and {Choi}, Hyunseop and {Barrera-Ballesteros}, Jorge and {Vigneron}, Benjamin and {Sun}, Ming and {Canning}, Rebecca E.~A. and {Taylor}, Gregory and {Albert}, Lo{\"\i}c and {D'Eugenio}, Francesco and {Donahue}, Megan and {Fabian}, Andrew C. and {Ferland}, Gary J. and {Gallagher}, John S. and {Gendron-Marsolais}, Marie-Lou and {Guillard}, Pierre and {Guo}, Minghao and {Hatch}, Nina and {Kotulla}, Ralf and {Li}, Yuan and {Maiolino}, Roberto and {Man}, Allison and {McDonald}, Michael A. and {McNamara}, Brian R. and {Olivares}, Valeria and {Prunier}, Marine and {Reynolds}, Christopher S. and {Rhea}, Carter and {Richard-Laferri{\`e}re}, Annabelle and {Russell}, Helen R. and {Salom{\'e}}, Philippe and {Tamhane}, Prathamesh and {Thilloy}, Auriane and {Tremblay}, Grant R. and {Voit}, G. Mark and {Walker}, Stephen A.},
        title = "{Mapping gas accretion and stellar kinematics to sub-kiloparsec scales in NGC 4696 with JWST/NIRSpec}",
      journal = {arXiv e-prints},
         year = 2026,
        month = aug,
          eid = {arXiv:2608.07732},
        pages = {arXiv:2608.07732},
archivePrefix = {arXiv},
       eprint = {2608.07732},
 primaryClass = {astro-ph.GA},
       adsurl = {https://ui.adsabs.harvard.edu/abs/2026arXiv260807732M}
}

@ARTICLE{Calzetti2000,
       author = {{Calzetti}, Daniela and {Armus}, Lee and {Bohlin}, Ralph C. and {Kinney}, Anne L. and {Koornneef}, Jan and {Storchi-Bergmann}, Thaisa},
        title = "{The Dust Content and Opacity of Actively Star-forming Galaxies}",
      journal = {\apj},
         year = 2000,
        month = apr,
       volume = {533},
       number = {2},
        pages = {682-695},
          doi = {10.1086/308692},
archivePrefix = {arXiv},
       eprint = {astro-ph/9911459},
 primaryClass = {astro-ph},
       adsurl = {https://ui.adsabs.harvard.edu/abs/2000ApJ...533..682C}
}

@article{Salome_2008, title={Cold gas in the Perseus cluster core: excitation of molecular gas in filaments}, volume={484}, ISSN={0004-6361, 1432-0746}, url={http://www.aanda.org/10.1051/0004-6361:200809493}, DOI={10.1051/0004-6361:200809493}, number={2}, journal={Astronomy \& Astrophysics}, author={Salom{\'e}, P. and Combes, F. and Revaz, Y. and Edge, A. C. and Hatch, N. A. and Fabian, A. C. and Johnstone, R. M.}, year={2008}, pages={317–325} }

@ARTICLE{Olivarex_metalicity_2026,
       author = {{Olivares}, Valeria and {Su}, Yuanyuan and {Temi}, Pasquale and {Eskenasy}, Ryan and {Russell}, Helen and {Gaspari}, Massimo and {Salome}, Philippe and {Combes}, Francoise and {Sun}, Ming and {Treister}, Ezequiel and {Fogarty}, Kevin and {Jimenez-Gallardo}, Ana and {Lagos}, Patricio},
        title = "{The relationship between warm and hot gas-phase metallicity in massive elliptical galaxies and the influence of active galactic nucleus feedback}",
      journal = {\aap},
         year = 2026,
        month = jan,
       volume = {705},
          eid = {A249},
        pages = {A249},
          doi = {10.1051/0004-6361/202555070},
archivePrefix = {arXiv},
       eprint = {2511.10587},
 primaryClass = {astro-ph.GA},
       adsurl = {https://ui.adsabs.harvard.edu/abs/2026A&A...705A.249O}
}

@ARTICLE{Mateusz_2018_Halpha,
       author = {{Ruszkowski}, Mateusz and {Yang}, H.-Y. Karen and {Reynolds}, Christopher S.},
        title = "{Powering of H{\ensuremath{\alpha}} Filaments by Cosmic Rays}",
      journal = {\apj},
         year = 2018,
        month = may,
       volume = {858},
       number = {1},
          eid = {64},
        pages = {64},
          doi = {10.3847/1538-4357/aaba72},
archivePrefix = {arXiv},
       eprint = {1802.04298},
 primaryClass = {astro-ph.HE},
       adsurl = {https://ui.adsabs.harvard.edu/abs/2018ApJ...858...64R}
}

@ARTICLE{cloudy25,
       author = {{Gunasekera}, C.~M. and {van Hoof}, P.~A.~M. and {Dehghanian}, M. and {Chakraborty}, P. and {Shaw}, G. and {Bianchi}, S. and {Chatzikos}, M. and {Tsujimoto}, M. and {Ferland}, G.~J.},
        title = "{The 2025 release of Cloudy}",
      journal = {\rmxaa},
         year = 2025,
        month = nov,
       volume = {61},
        pages = {120-133},
          doi = {10.22201/ia.01851101p.2025.61.03.01},
archivePrefix = {arXiv},
       eprint = {2508.01102},
 primaryClass = {astro-ph.GA},
       adsurl = {https://ui.adsabs.harvard.edu/abs/2025RMxAA..61c.120G}
}

@ARTICLE{sdss_dr8,
       author = {{Aihara}, Hiroaki and {Allende Prieto}, Carlos and {An}, Deokkeun and {Anderson}, Scott F. and {Aubourg}, {\'E}ric and {Balbinot}, Eduardo and {Beers}, Timothy C. and {Berlind}, Andreas A. and {Bickerton}, Steven J. and {Bizyaev}, Dmitry and {Blanton}, Michael R. and {Bochanski}, John J. and {Bolton}, Adam S. and {Bovy}, Jo and {Brandt}, W.~N. and {Brinkmann}, J. and {Brown}, Peter J. and {Brownstein}, Joel R. and {Busca}, Nicolas G. and {Campbell}, Heather and {Carr}, Michael A. and {Chen}, Yanmei and {Chiappini}, Cristina and {Comparat}, Johan and {Connolly}, Natalia and {Cortes}, Marina and {Croft}, Rupert A.~C. and {Cuesta}, Antonio J. and {da Costa}, Luiz N. and {Davenport}, James R.~A. and {Dawson}, Kyle and {Dhital}, Saurav and {Ealet}, Anne and {Ebelke}, Garrett L. and {Edmondson}, Edward M. and {Eisenstein}, Daniel J. and {Escoffier}, Stephanie and {Esposito}, Massimiliano and {Evans}, Michael L. and {Fan}, Xiaohui and {Femen{\'\i}a Castell{\'a}}, Bruno and {Font-Ribera}, Andreu and {Frinchaboy}, Peter M. and {Ge}, Jian and {Gillespie}, Bruce A. and {Gilmore}, G. and {Gonz{\'a}lez Hern{\'a}ndez}, Jonay I. and {Gott}, J. Richard and {Gould}, Andrew and {Grebel}, Eva K. and {Gunn}, James E. and {Hamilton}, Jean-Christophe and {Harding}, Paul and {Harris}, David W. and {Hawley}, Suzanne L. and {Hearty}, Frederick R. and {Ho}, Shirley and {Hogg}, David W. and {Holtzman}, Jon A. and {Honscheid}, Klaus and {Inada}, Naohisa and {Ivans}, Inese I. and {Jiang}, Linhua and {Johnson}, Jennifer A. and {Jordan}, Cathy and {Jordan}, Wendell P. and {Kazin}, Eyal A. and {Kirkby}, David and {Klaene}, Mark A. and {Knapp}, G.~R. and {Kneib}, Jean-Paul and {Kochanek}, C.~S. and {Koesterke}, Lars and {Kollmeier}, Juna A. and {Kron}, Richard G. and {Lampeitl}, Hubert and {Lang}, Dustin and {Le Goff}, Jean-Marc and {Lee}, Young Sun and {Lin}, Yen-Ting and {Long}, Daniel C. and {Loomis}, Craig P. and {Lucatello}, Sara and {Lundgren}, Britt and {Lupton}, Robert H. and {Ma}, Zhibo and {MacDonald}, Nicholas and {Mahadevan}, Suvrath and {Maia}, Marcio A.~G. and {Makler}, Martin and {Malanushenko}, Elena and {Malanushenko}, Viktor and {Mandelbaum}, Rachel and {Maraston}, Claudia and {Margala}, Daniel and {Masters}, Karen L. and {McBride}, Cameron K. and {McGehee}, Peregrine M. and {McGreer}, Ian D. and {M{\'e}nard}, Brice and {Miralda-Escud{\'e}}, Jordi and {Morrison}, Heather L. and {Mullally}, F. and {Muna}, Demitri and {Munn}, Jeffrey A. and {Murayama}, Hitoshi and {Myers}, Adam D. and {Naugle}, Tracy and {Neto}, Angelo Fausti and {Nguyen}, Duy Cuong and {Nichol}, Robert C. and {O'Connell}, Robert W. and {Ogando}, Ricardo L.~C. and {Olmstead}, Matthew D. and {Oravetz}, Daniel J. and {Padmanabhan}, Nikhil and {Palanque-Delabrouille}, Nathalie and {Pan}, Kaike and {Pandey}, Parul and {P{\^a}ris}, Isabelle and {Percival}, Will J. and {Petitjean}, Patrick and {Pfaffenberger}, Robert and {Pforr}, Janine and {Phleps}, Stefanie and {Pichon}, Christophe and {Pieri}, Matthew M. and {Prada}, Francisco and {Price-Whelan}, Adrian M. and {Raddick}, M. Jordan and {Ramos}, Beatriz H.~F. and {Reyl{\'e}}, C{\'e}line and {Rich}, James and {Richards}, Gordon T. and {Rix}, Hans-Walter and {Robin}, Annie C. and {Rocha-Pinto}, Helio J. and {Rockosi}, Constance M. and {Roe}, Natalie A. and {Rollinde}, Emmanuel and {Ross}, Ashley J. and {Ross}, Nicholas P. and {Rossetto}, Bruno M. and {S{\'a}nchez}, Ariel G. and {Sayres}, Conor and {Schlegel}, David J. and {Schlesinger}, Katharine J. and {Schmidt}, Sarah J. and {Schneider}, Donald P. and {Sheldon}, Erin and {Shu}, Yiping and {Simmerer}, Jennifer and {Simmons}, Audrey E. and {Sivarani}, Thirupathi and {Snedden}, Stephanie A. and {Sobeck}, Jennifer S. and {Steinmetz}, Matthias and {Strauss}, Michael A. and {Szalay}, Alexander S. and {Tanaka}, Masayuki and {Thakar}, Aniruddha R. and {Thomas}, Daniel and {Tinker}, Jeremy L. and {Tofflemire}, Benjamin M. and {Tojeiro}, Rita and {Tremonti}, Christy A. and {Vandenberg}, Jan and {Vargas Maga{\~n}a}, M. and {Verde}, Licia and {Vogt}, Nicole P. and {Wake}, David A. and {Wang}, Ji and {Weaver}, Benjamin A. and {Weinberg}, David H. and {White}, Martin and {White}, Simon D.~M. and {Yanny}, Brian and {Yasuda}, Naoki and {Yeche}, Christophe and {Zehavi}, Idit},
        title = "{The Eighth Data Release of the Sloan Digital Sky Survey: First Data from SDSS-III}",
      journal = {\apjs},
         year = 2011,
        month = apr,
       volume = {193},
       number = {2},
          eid = {29},
        pages = {29},
          doi = {10.1088/0067-0049/193/2/29},
archivePrefix = {arXiv},
       eprint = {1101.1559},
 primaryClass = {astro-ph.IM},
       adsurl = {https://ui.adsabs.harvard.edu/abs/2011ApJS..193...29A}
}

@ARTICLE{sanders_resonance_2006,
       author = {{Sanders}, J.~S. and {Fabian}, A.~C.},
        title = "{Resonance scattering, absorption and off-centre abundance peaks in clusters of galaxies}",
      journal = {\mnras},
         year = 2006,
        month = jul,
       volume = {370},
       number = {1},
        pages = {63-73},
          doi = {10.1111/j.1365-2966.2006.10497.x},
archivePrefix = {arXiv},
       eprint = {astro-ph/0604575},
 primaryClass = {astro-ph},
       adsurl = {https://ui.adsabs.harvard.edu/abs/2006MNRAS.370...63S}
}

@article{polles_excitation_2021,
	title = {Excitation mechanisms in the intracluster filaments surrounding brightest cluster galaxies},
	volume = {651},
	copyright = {https://creativecommons.org/licenses/by/4.0},
	issn = {0004-6361, 1432-0746},
	url = {https://www.aanda.org/10.1051/0004-6361/202039730},
	doi = {10.1051/0004-6361/202039730},
	urldate = {2026-01-20},
	journal = {Astronomy \& Astrophysics},
	author = {Polles, F. L. and Salom{\'e}, P. and Guillard, P. and Godard, B. and Pineau Des Forêts, G. and Olivares, V. and Beckmann, R. S. and Canning, R. E. A. and Combes, F. and Dubois, Y. and Edge, A. C. and Fabian, A. C. and Ferland, G. J. and Hamer, S. L. and Lehnert, M. D.},
	month = jul,
	year = {2021},
	pages = {A13},
}

@article{mittal_herschel_2011,
	title = {Herschel★ observations of the {Centaurus} cluster - the dynamics of cold gas in a cool core: {Herschel} observations of the {Centaurus} cluster},
	volume = {418},
	issn = {00358711},
	shorttitle = {Herschel★ observations of the {Centaurus} cluster - the dynamics of cold gas in a cool core},
	url = {https://academic.oup.com/mnras/article-lookup/doi/10.1111/j.1365-2966.2011.19634.x},
	doi = {10.1111/j.1365-2966.2011.19634.x},
	language = {en},
	number = {4},
	urldate = {2026-07-07},
	journal = {Monthly Notices of the Royal Astronomical Society},
	author = {Mittal, R. and O’Dea, C. P. and Ferland, G. and Oonk, J. B. R. and Edge, A. C. and Canning, R. E. A. and Russell, H. and Baum, S. A. and Böhringer, H. and Combes, F. and Donahue, M. and Fabian, A. C. and Hatch, N. A. and Hoffer, A. and Johnstone, R. and McNamara, B. R. and Salom{\'e}, P. and Tremblay, G.},
	month = dec,
	year = {2011},
	pages = {2386--2402},
}

@PHDTHESIS{mcdonald_origins_2011,
       author = {{McDonald}, Michael Adam},
        title = "{On the origins and ionization mechanisms of H{\ensuremath{\alpha}} Filaments in the cool cores of galaxy groups and clusters}",
       school = {University of Maryland, College Park},
         year = 2011,
        month = jan,
       adsurl = {https://ui.adsabs.harvard.edu/abs/2011PhDT.......390M}
}

@software{davies_2025_15791615,
  author       = {Davies, James},
  title        = {snowblind},
  month        = jul,
  year         = 2025,
  publisher    = {Zenodo},
  version      = {0.2.1},
  doi          = {10.5281/zenodo.15791615},
  url          = {https://doi.org/10.5281/zenodo.15791615},
  swhid        = {swh:1:dir:addc34bf393efbafa15c043ed6c2014c8981dc02
                   ;origin=https://doi.org/10.5281/zenodo.15791614;vi
                   sit=swh:1:snp:0cb510ce88238e2a1d8b994e51073cdec91d
                   6062;anchor=swh:1:rel:bba76cd74a849d55d63653355389
                   e6364b9c02e0;path=mpi-astronomy-snowblind-2a5d076
                  },
}

@misc{bushouse_2025,
  title = {{{JWST}} Calibration Pipeline},
  author = {Bushouse, Howard and Eisenhamer, Jonathan and Dencheva, Nadia and Davies, James and Greenfield, Perry and Morrison, Jane and Hodge, Phil and Simon, Bernie and Grumm, David and Droettboom, Michael and Slavich, Edward and Sosey, Megan and Pauly, Tyler and Miller, Todd and Jedrzejewski, Robert and Hack, Warren and Davis, David and Crawford, Steven and Law, David and Gordon, Karl and Regan, Michael and Cara, Mihai and MacDonald, Ken and Bradley, Larry and Shanahan, Clare and Jamieson, William and Teodoro, Mairan and Williams, Thomas and {Pena-Guerrero}, Maria and Graham, Brett and Molter, Edward and Brandt, Timothy and Hayes, Christian and Cooper, Rachel and Clarke, Melanie and Filippazzo, Joseph},
  year = 2025,
  month = nov,
  doi = {10.5281/zenodo.17515973},
  howpublished = {Zenodo},
  swhid = {swh:1:dir:fc8e0b17375bd6292e6f2fe6b758a3e44b81aa01 ;origin=https://doi.org/10.5281/zenodo.6984365;vis it=swh:1:snp:ee0c72d562544a3226903308be9ab3969e139 858;anchor=swh:1:rel:4164a7a6a89b54a00ff5ed5b74c67 5e41f26b3fc;path=spacetelescope-jwst-ee52a96}
}

@article{desouza2017,
  title = {A Probabilistic Approach to Emission-Line Galaxy Classification},
  author = {De Souza, R. S. and Dantas, M. L. L. and {Costa-Duarte}, M. V. and Feigelson, E. D. and Killedar, M. and Lablanche, P.-Y. and Vilalta, R. and {Krone-Martins}, A. and Beck, R. and Gieseke, F.},
  year = 2017,
  month = dec,
  journal = {Monthly Notices of the Royal Astronomical Society},
  volume = {472},
  number = {3},
  pages = {2808--2822},
  issn = {0035-8711, 1365-2966},
  doi = {10.1093/mnras/stx2156},
  urldate = {2026-11-03},
  langid = {english}
}

@article{mcdonald_optical_2012,
	title = {{OPTICAL} {SPECTROSCOPY} {OF} {Hα} {FILAMENTS} {IN} {COOL} {CORE} {CLUSTERS}: {KINEMATICS}, {REDDENING}, {AND} {SOURCES} {OF} {IONIZATION}},
	volume = {746},
	issn = {0004-637X, 1538-4357},
	shorttitle = {{OPTICAL} {SPECTROSCOPY} {OF} {Hα} {FILAMENTS} {IN} {COOL} {CORE} {CLUSTERS}},
	url = {https://iopscience.iop.org/article/10.1088/0004-637X/746/2/153},
	doi = {10.1088/0004-637X/746/2/153},
	number = {2},
	urldate = {2026-07-07},
	journal = {The Astrophysical Journal},
	author = {McDonald, Michael and Veilleux, Sylvain and Rupke, David S. N.},
	month = feb,
	year = {2012},
	pages = {153},
}

@article{ferland_collisional_2009,
	title = {Collisional heating as the origin of filament emission in galaxy clusters},
	volume = {392},
	issn = {00358711, 13652966},
	url = {https://academic.oup.com/mnras/article-lookup/doi/10.1111/j.1365-2966.2008.14153.x},
	doi = {10.1111/j.1365-2966.2008.14153.x},
	language = {en},
	number = {4},
	urldate = {2026-07-07},
	journal = {Monthly Notices of the Royal Astronomical Society},
	author = {Ferland, G. J. and Fabian, A. C. and Hatch, N. A. and Johnstone, R. M. and Porter, R. L. and Van Hoof, P. A. M. and Williams, R. J. R.},
	month = feb,
	year = {2009},
	pages = {1475--1502},
}

@article{tremblay_multiphase_2012,
	title = {Multiphase signatures of active galactic nucleus feedback in {Abell} 2597: {AGN} feedback in {Abell} 2597},
	volume = {424},
	issn = {00358711},
	shorttitle = {Multiphase signatures of active galactic nucleus feedback in {Abell} 2597},
	url = {https://academic.oup.com/mnras/article-lookup/doi/10.1111/j.1365-2966.2012.21281.x},
	doi = {10.1111/j.1365-2966.2012.21281.x},
	language = {en},
	number = {2},
	urldate = {2026-07-07},
	journal = {Monthly Notices of the Royal Astronomical Society},
	author = {Tremblay, G. R. and O’Dea, C. P. and Baum, S. A. and Clarke, T. E. and Sarazin, C. L. and Bregman, J. N. and Combes, F. and Donahue, M. and Edge, A. C. and Fabian, A. C. and Ferland, G. J. and McNamara, B. R. and Mittal, R. and Oonk, J. B. R. and Quillen, A. C. and Russell, H. R. and Sanders, J. S. and Salom{\'e}, P. and Voit, G. M. and Wilman, R. J. and Wise, M. W.},
	month = aug,
	year = {2012},
	pages = {1026--1041},
}

@article{mcdonald_hst_2014,
	title = {\textit{{HST}} -{COS} {SPECTROSCOPY} {OF} {THE} {COOLING} {FLOW} {IN} {A1795}—{EVIDENCE} {FOR} {INEFFICIENT} {STAR} {FORMATION} {IN} {CONDENSING} {INTRACLUSTER} {GAS}},
	volume = {791},
	copyright = {http://iopscience.iop.org/info/page/text-and-data-mining},
	issn = {2041-8213},
	url = {https://iopscience.iop.org/article/10.1088/2041-8205/791/2/L30},
	doi = {10.1088/2041-8205/791/2/L30},
	number = {2},
	urldate = {2026-07-07},
	journal = {The Astrophysical Journal},
	author = {McDonald, Michael and Roediger, Joel and Veilleux, Sylvain and Ehlert, Steven},
	month = aug,
	year = {2014},
	pages = {L30},
}

@article{Crawford_1999, title={The ROSAT Brightest Cluster Sample -- III. Optical spectra of the central cluster galaxies}, volume={306}, ISSN={0035-8711, 1365-2966}, url={https://academic.oup.com/mnras/article/306/4/857/963324}, DOI={10.1046/j.1365-8711.1999.02583.x}, number={4}, journal={Monthly Notices of the Royal Astronomical Society}, author={Crawford, C. S. and Allen, S. W. and Ebeling, H. and Edge, A. C. and Fabian, A. C.}, year={1999}, pages={857–896}, language={en} }

@article{Olivares_2019, title={Ubiquitous cold and massive filaments in cool core clusters}, volume={631}, rights={http://creativecommons.org/licenses/by/4.0}, ISSN={0004-6361, 1432-0746}, url={https://www.aanda.org/10.1051/0004-6361/201935350}, DOI={10.1051/0004-6361/201935350}, abstractNote={Multi-phase filamentary structures around brightest cluster galaxies (BCG) are likely a key step of AGN-feedback. We observed molecular gas in three cool cluster cores, namely Centaurus, Abell S1101, and RXJ1539.5, and gathered ALMA (Atacama Large Millimeter/submillimeter Array) and MUSE (Multi Unit Spectroscopic Explorer) data for 12 other clusters. Those observations show clumpy, massive, and long (3−25 kpc) molecular filaments, preferentially located around the radio bubbles inflated by the AGN. Two objects show nuclear molecular disks. The optical nebula is certainly tracing the warm envelopes of cold molecular filaments. Surprisingly, the radial profile of the Hα/CO flux ratio is roughly constant for most of the objects, suggesting that (i) between 1.2 and 6 times more cold gas could be present and (ii) local processes must be responsible for the excitation. Projected velocities are between 100 and 400 km s−1, with disturbed kinematics and sometimes coherent gradients. This is likely due to the mixing in projection of several thin (and as yet) unresolved filaments. The velocity fields may be stirred by turbulence induced by bubbles, jets, or merger-induced sloshing. Velocity and dispersions are low, below the escape velocity. Cold clouds should eventually fall back and fuel the AGN. We compare the radial extent of the filaments,rfil, with the region where the X-ray gas can become thermally unstable. The filaments are always inside the low-entropy and short-cooling-time region, wheretcool/tff  <  20 (9 of 13 sources). The range oftcool/tffof 8−23 atrfil, is likely due to (i) a more complex gravitational potential affecting the free-fall timetff(sloshing, mergers, etc.) and (ii) the presence of inhomogeneities or uplifted gas in the ICM, affecting the cooling timetcool. For some of the sources,rfillies where the ratio of the cooling time to the eddy-turnover time,tcool/teddy, is approximately unity.}, journal={Astronomy \& Astrophysics}, author={Olivares, V. and Salome, P. and Combes, F. and Hamer, S. and Guillard, P. and Lehnert, M. D. and Polles, F. L. and Beckmann, R. S. and Dubois, Y. and Donahue, M. and Edge, A. and Fabian, A. C. and McNamara, B. and Rose, T. and Russell, H. R. and Tremblay, G. and Vantyghem, A. and Canning, R. E. A. and Ferland, G. and Godard, B. and Peirani, S. and Pineau Des Forets, G.}, year={2019}, month=nov, pages={A22} }

@article{Donahue_2011, title={POLYCYCLIC AROMATIC HYDROCARBONS, IONIZED GAS, AND MOLECULAR HYDROGEN IN BRIGHTEST CLUSTER GALAXIES OF COOL-CORE CLUSTERS OF GALAXIES}, volume={732}, ISSN={0004-637X, 1538-4357}, url={https://iopscience.iop.org/article/10.1088/0004-637X/732/1/40}, DOI={10.1088/0004-637X/732/1/40}, number={1}, journal={The Astrophysical Journal}, author={Donahue, Megan and De Messières, Geneviève E. and O’Connell, Robert W. and Mark Voit, G. and Hoffer, Aaron and McNamara, Brian R. and Nulsen, Paul E. J.}, year={2011}, month=may, pages={40} }

@article{Edwards_2009, title={The diverse nature of optical emission lines in brightest cluster galaxies: IFU observations of the central kiloparsec}, volume={396}, ISSN={00358711, 13652966}, url={https://academic.oup.com/mnras/article-lookup/doi/10.1111/j.1365-2966.2009.14897.x}, DOI={10.1111/j.1365-2966.2009.14897.x}, number={4}, journal={Monthly Notices of the Royal Astronomical Society}, author={Edwards, Louise O. V. and Robert, Carmelle and Mollá, Mercedes and McGee, Sean L.}, year={2009}, month={july}, pages={1953–1971}, language={en} }

@article{Cavagnolo_Donahue_Voit_Sun_2009, title={INTRACLUSTER MEDIUM ENTROPY PROFILES FOR A <i>CHANDRA</i> ARCHIVAL SAMPLE OF GALAXY CLUSTERS}, volume={182}, ISSN={0067-0049, 1538-4365}, url={https://iopscience.iop.org/article/10.1088/0067-0049/182/1/12}, DOI={10.1088/0067-0049/182/1/12}, number={1}, journal={The Astrophysical Journal Supplement Series}, author={Cavagnolo, Kenneth W. and Donahue, Megan and Voit, G. Mark and Sun, Ming}, year={2009}, month=may, pages={12–32} }

@article{Fabian_Johnstone_Sanders_Conselice_Crawford_Iii_Zweibel_2008, title={Magnetic support of the optical emission line filaments in NGC 1275}, volume={454}, rights={http://www.springer.com/tdm}, ISSN={0028-0836, 1476-4687}, url={https://www.nature.com/articles/nature07169}, DOI={10.1038/nature07169}, number={7207}, journal={Nature}, author={Fabian, A. C. and Johnstone, R. M. and Sanders, J. S. and Conselice, C. J. and Crawford, C. S. and Iii, J. S. Gallagher and Zweibel, E.}, year={2008}, month=aug, pages={968–970}, language={en} }

@article{Heckman_1989, title={Dynamical, physical, and chemical properties of emission-line nebulae in cooling flows}, volume={338}, ISSN={0004-637X, 1538-4357}, url={http://adsabs.harvard.edu/doi/10.1086/167181}, DOI={10.1086/167181}, journal={The Astrophysical Journal}, author={Heckman, Timothy M. and Baum, Stefi A. and Van Breugel, W. J. M. and McCarthy, Patrick}, year={1989}, month=mar, pages={48}, language={en} }

@article{Kirkpatrick_2009, title={A <i>CHANDRA</i> X-RAY ANALYSIS OF ABELL 1664: COOLING, FEEDBACK, AND STAR FORMATION IN THE CENTRAL CLUSTER GALAXY}, volume={697}, ISSN={0004-637X, 1538-4357}, url={https://iopscience.iop.org/article/10.1088/0004-637X/697/1/867}, DOI={10.1088/0004-637X/697/1/867}, number={1}, journal={The Astrophysical Journal}, author={Kirkpatrick, C. C. and McNamara, B. R. and Rafferty, D. A. and Nulsen, P. E. J. and Bîrzan, L. and Kazemzadeh, F. and Wise, M. W. and Gitti, M. and Cavagnolo, K. W.}, year={2009}, month=may, pages={867–879} }

@article{McDonald_2015, title={EXTENDED, DUSTY STAR FORMATION FUELED BY A RESIDUAL COOLING FLOW IN THE CLUSTER OF GALAXIES S{\'e}RSIC 159-03}, volume={804}, rights={http://iopscience.iop.org/info/page/text-and-data-mining}, ISSN={1538-4357}, url={https://iopscience.iop.org/article/10.1088/0004-637X/804/1/16}, DOI={10.1088/0004-637X/804/1/16}, number={1}, journal={The Astrophysical Journal}, author={McDonald, M. and Werner, N. and Oonk, J. B. R. and Veilleux, S.}, year={2015}, month=apr, pages={16} }

@article{Egami_2006, title={Spitzer Observations of the Brightest Galaxies in X‐Ray‐Luminous Clusters}, volume={647}, ISSN={0004-637X, 1538-4357}, url={https://iopscience.iop.org/article/10.1086/504519}, DOI={10.1086/504519}, number={2}, journal={The Astrophysical Journal}, author={Egami, E. and Misselt, K. A. and Rieke, G. H. and Wise, M. W. and Neugebauer, G. and Kneib, J.‐P. and Le Floc’h, E. and Smith, G. P. and Blaylock, M. and Dole, H. and Frayer, D. T. and Huang, J.‐S. and Krause, O. and Papovich, C. and Perez‐Gonzalez, P. G. and Rigby, J. R.}, year={2006}, month=aug, pages={922–933}, language={en} }

@article{cid_fernandes_alternative_2010,
	title = {Alternative diagnostic diagrams and the ‘forgotten’ population of weak line galaxies in the {SDSS}: {The} forgotten population of {WLGs} in the {SDSS}},
	volume = {403},
	issn = {00358711},
	shorttitle = {Alternative diagnostic diagrams and the ‘forgotten’ population of weak line galaxies in the {SDSS}},
	url = {https://academic.oup.com/mnras/article-lookup/doi/10.1111/j.1365-2966.2009.16185.x},
	doi = {10.1111/j.1365-2966.2009.16185.x},
	language = {en},
	number = {2},
	urldate = {2026-06-18},
	journal = {Monthly Notices of the Royal Astronomical Society},
	author = {Cid Fernandes, R. and Stasińska, G. and Schlickmann, M. S. and Mateus, A. and Vale Asari, N. and Schoenell, W. and Sodr{\'e}, L. and {(the SEAGal collaboration)}},
	month = feb,
	year = {2010},
	pages = {1036--1053},
}

@ARTICLE{Allard2012,
       author = {{Allard}, F. and {Homeier}, D. and {Freytag}, B.},
        title = "{Models of very-low-mass stars, brown dwarfs and exoplanets}",
      journal = {Philosophical Transactions of the Royal Society of London Series A},
         year = 2012,
        month = jun,
       volume = {370},
       number = {1968},
        pages = {2765-2777},
          doi = {10.1098/rsta.2011.0269},
archivePrefix = {arXiv},
       eprint = {1112.3591},
 primaryClass = {astro-ph.SR},
       adsurl = {https://ui.adsabs.harvard.edu/abs/2012RSPTA.370.2765A}
}

@INPROCEEDINGS{Markwardt2009,
       author = {{Markwardt}, C.~B.},
        title = "{Non-linear Least-squares Fitting in IDL with MPFIT}",
    booktitle = {Astronomical Data Analysis Software and Systems XVIII},
         year = 2009,
       editor = {{Bohlender}, D.~A. and {Durand}, D. and {Dowler}, P.},
       series = {Astronomical Society of the Pacific Conference Series},
       volume = {411},
        month = sep,
        pages = {251},
          doi = {10.48550/arXiv.0902.2850},
archivePrefix = {arXiv},
       eprint = {0902.2850},
 primaryClass = {astro-ph.IM},
       adsurl = {https://ui.adsabs.harvard.edu/abs/2009ASPC..411..251M}
}

@article{reefe_apj_2025,
	title = {Cold {Gas} and {Star} {Formation} in the {Phoenix} {Cluster} with {JWST}},
	volume = {989},
	issn = {0004-637X},
	url = {https://dx.doi.org/10.3847/1538-4357/aded90},
	doi = {10.3847/1538-4357/aded90},
	language = {en},
	number = {2},
	urldate = {2025-09-04},
	journal = {The Astrophysical Journal},
	author = {Reefe, Michael and McDonald, Michael and Chatzikos, Marios and Seebeck, Jerome and Mushotzky, Richard and Veilleux, Sylvain and Allen, Steven W. and Bayliss, Matthew and Calzadilla, Michael and Canning, Rebecca and Donahue, Megan and Floyd, Benjamin and Gaspari, Massimo and Hlavacek-Larrondo, Julie and McNamara, Brian and Russell, Helen and Sarkar, Arnab and Sharon, Keren and Somboonpanyakul, Taweewat},
	month = aug,
	year = {2025},
	note = {Publisher: The American Astronomical Society},
	pages = {156},
}

@article{turner_quadrupole_1977,
	title = {The {Quadrupole} {Vibration}-{Rotation} {Transition} {Probabilities} of {Molecular} {Hydrogen}},
	volume = {35},
	issn = {0067-0049, 1538-4365},
	url = {http://adsabs.harvard.edu/doi/10.1086/190481},
	doi = {10.1086/190481},
	language = {en},
	urldate = {2026-03-16},
	journal = {The Astrophysical Journal Supplement Series},
	author = {Turner, J. and Kirby-Docken, K. and Dalgarno, A.},
	month = jan,
	year = {1977},
	pages = {281},
}

@article{dabrowski_lyman_1984,
	title = {The {Lyman} and {Werner} bands of {H2}},
	volume = {62},
	issn = {0008-4204},
	url = {https://cdnsciencepub.com/doi/10.1139/p84-210},
	doi = {10.1139/p84-210},
	number = {12},
	urldate = {2026-03-16},
	journal = {Canadian Journal of Physics},
	publisher = {NRC Research Press},
	author = {Dabrowski, I.},
	month = dec,
	year = {1984},
	pages = {1639--1664},
}

@article{bailey_principal_2012,
	title = {Principal {Component} {Analysis} with {Noisy} and/or {Missing} {Data}},
	volume = {124},
	issn = {1538-3873},
	url = {https://iopscience.iop.org/article/10.1086/668105},
	doi = {10.1086/668105},
	language = {en},
	number = {919},
	urldate = {2026-03-15},
	journal = {Publications of the Astronomical Society of the Pacific},
	publisher = {IOP Publishing},
	author = {Bailey, Stephen},
	month = sep,
	year = {2012},
	pages = {1015},
}

@article{kewley_host_2006,
	title = {The host galaxies and classification of active galactic nuclei},
	volume = {372},
	issn = {0035-8711},
	url = {https://doi.org/10.1111/j.1365-2966.2006.10859.x},
	doi = {10.1111/j.1365-2966.2006.10859.x},
	number = {3},
	urldate = {2026-03-15},
	journal = {Monthly Notices of the Royal Astronomical Society},
	author = {Kewley, Lisa J. and Groves, Brent and Kauffmann, Guinevere and Heckman, Tim},
	month = nov,
	year = {2006},
	pages = {961--976},
}

@article{kewley_theoretical_2001,
	title = {Theoretical {Modeling} of {Starburst} {Galaxies}},
	volume = {556},
	issn = {0004-637X},
	url = {https://iopscience.iop.org/article/10.1086/321545},
	doi = {10.1086/321545},
	language = {en},
	number = {1},
	urldate = {2026-03-15},
	journal = {The Astrophysical Journal},
	publisher = {IOP Publishing},
	author = {Kewley, L. J. and Dopita, M. A. and Sutherland, R. S. and Heisler, C. A. and Trevena, J.},
	month = jul,
	year = {2001},
	pages = {121},
}

@article{kauffmann_host_2003,
	title = {The host galaxies of active galactic nuclei},
	volume = {346},
	issn = {0035-8711},
	url = {https://doi.org/10.1111/j.1365-2966.2003.07154.x},
	doi = {10.1111/j.1365-2966.2003.07154.x},
	number = {4},
	urldate = {2026-03-15},
	journal = {Monthly Notices of the Royal Astronomical Society},
	author = {Kauffmann, Guinevere and Heckman, Timothy M. and Tremonti, Christy and Brinchmann, Jarle and Charlot, St{\'e}phane and White, Simon D. M. and Ridgway, Susan E. and Brinkmann, Jon and Fukugita, Masataka and Hall, Patrick B. and Ivezić, Željko and Richards, Gordon T. and Schneider, Donald P.},
	month = dec,
	year = {2003},
	pages = {1055--1077},
}

@article{edge_detection_2001,
	title = {The detection of molecular gas in the central galaxies of cooling flow clusters},
	volume = {328},
	issn = {0035-8711},
	url = {https://ui.adsabs.harvard.edu/abs/2001MNRAS.328..762E/abstract},
	doi = {10.1046/j.1365-8711.2001.04802.x},
	language = {en},
	number = {3},
	urldate = {2026-04-13},
	journal = {Monthly Notices of the Royal Astronomical Society},
	author = {Edge, A. C.},
	month = dec,
	year = {2001},
	pages = {762--782},
}

@article{johnstone_discovery_2007,
	title = {Discovery of atomic and molecular mid-infrared emission lines in off-nuclear regions of {NGC} 1275 and {NGC} 4696 with the {Spitzer} {Space} {Telescope}},
	volume = {382},
	issn = {0035-8711},
	url = {https://doi.org/10.1111/j.1365-2966.2007.12460.x},
	doi = {10.1111/j.1365-2966.2007.12460.x},
	number = {3},
	urldate = {2026-03-30},
	journal = {Monthly Notices of the Royal Astronomical Society},
	author = {Johnstone, R. M. and Hatch, N. A. and Ferland, G. J. and Fabian, A. C. and Crawford, C. S. and Wilman, R. J.},
	month = nov,
	year = {2007},
	pages = {1246--1260},
}

@article{reefe_nature_2025,
	title = {Directly imaging the cooling flow in the {Phoenix} cluster},
	volume = {638},
	copyright = {2025 The Author(s), under exclusive licence to Springer Nature Limited},
	issn = {1476-4687},
	url = {https://www.nature.com/articles/s41586-024-08369-x},
	doi = {10.1038/s41586-024-08369-x},
	language = {en},
	number = {8050},
	urldate = {2025-09-04},
	journal = {Nature},
	author = {Reefe, Michael and McDonald, Michael and Chatzikos, Marios and Seebeck, Jerome and Mushotzky, Richard and Veilleux, Sylvain and Allen, Steven W. and Bayliss, Matthew and Calzadilla, Michael and Canning, Rebecca and Floyd, Benjamin and Gaspari, Massimo and Hlavacek-Larrondo, Julie and McNamara, Brian and Russell, Helen and Sharon, Keren and Somboonpanyakul, Taweewat},
	month = feb,
	year = {2025},
	note = {Publisher: Nature Publishing Group},
	pages = {360--364},
}

@article{togi_lighting_2016,
	title = {{LIGHTING} {THE} {DARK} {MOLECULAR} {GAS}: {H2} {AS} {A} {DIRECT} {TRACER}},
	volume = {830},
	issn = {0004-637X},
	shorttitle = {{LIGHTING} {THE} {DARK} {MOLECULAR} {GAS}},
	url = {https://doi.org/10.3847/0004-637X/830/1/18},
	doi = {10.3847/0004-637X/830/1/18},
	language = {en},
	number = {1},
	urldate = {2025-11-25},
	journal = {The Astrophysical Journal},
	publisher = {The American Astronomical Society},
	author = {Togi, Aditya and Smith, J. D. T.},
	month = oct,
	year = {2016},
	pages = {18},
}

@article{black_fluorescent_1987,
	title = {Fluorescent {Excitation} of {Interstellar} {H} 2},
	volume = {322},
	issn = {0004-637X},
	url = {https://ui.adsabs.harvard.edu/abs/1987ApJ...322..412B/abstract},
	doi = {10.1086/165740},
	language = {en},
	urldate = {2026-05-05},
	journal = {The Astrophysical Journal},
	author = {Black, John H. and van Dishoeck, Ewine F.},
	month = nov,
	year = {1987},
	pages = {412},
}

@article{guillard_h2_2009,
	title = {H$_{\textrm{2}}$ formation and excitation in the {Stephan}'s {Quintet} galaxy-wide collision},
	volume = {502},
	issn = {0004-6361, 1432-0746},
	url = {http://www.aanda.org/10.1051/0004-6361/200811263},
	doi = {10.1051/0004-6361/200811263},
	number = {2},
	urldate = {2026-05-20},
	journal = {Astronomy \& Astrophysics},
	author = {Guillard, P. and Boulanger, F. and Pineau Des Forêts, G. and Appleton, P. N.},
	month = aug,
	year = {2009},
	pages = {515--528},
}

@article{perna_ga-nifs_2023,
	title = {{GA}-{NIFS}: {The} ultra-dense, interacting environment of a dual {AGN} at \textit{z} ∼ 3.3 revealed by {JWST}/{NIRSpec} {IFS}},
	volume = {679},
	copyright = {https://creativecommons.org/licenses/by/4.0},
	issn = {0004-6361, 1432-0746},
	shorttitle = {{GA}-{NIFS}},
	url = {https://www.aanda.org/10.1051/0004-6361/202346649},
	doi = {10.1051/0004-6361/202346649},
	urldate = {2026-05-20},
	journal = {Astronomy \& Astrophysics},
	author = {Perna, M. and Arribas, S. and Marshall, M. and D’Eugenio, F. and Übler, H. and Bunker, A. and Charlot, S. and Carniani, S. and Jakobsen, P. and Maiolino, R. and Rodríguez Del Pino, B. and Willott, C. J. and Böker, T. and Circosta, C. and Cresci, G. and Curti, M. and Husemann, B. and Kumari, N. and Lamperti, I. and P{\'e}rez-González, P. G. and Scholtz, J.},
	month = nov,
	year = {2023},
	pages = {A89},
}

@article{russell_driving_2019,
	title = {Driving massive molecular gas flows in central cluster galaxies with {AGN} feedback},
	volume = {490},
	copyright = {https://academic.oup.com/journals/pages/open\_access/funder\_policies/chorus/standard\_publication\_model},
	issn = {0035-8711, 1365-2966},
	url = {https://academic.oup.com/mnras/article/490/3/3025/5579032},
	doi = {10.1093/mnras/stz2719},
	language = {en},
	number = {3},
	urldate = {2026-05-20},
	journal = {Monthly Notices of the Royal Astronomical Society},
	author = {Russell, H R and McNamara, B R and Fabian, A C and Nulsen, P E J and Combes, F and Edge, A C and Madar, M and Olivares, V and Salom{\'e}, P and Vantyghem, A N},
	month = dec,
	year = {2019},
	pages = {3025--3045},
}

@article{sanders_very_2016,
	title = {A very deep \textit{{Chandra}} view of metals, sloshing and feedback in the {Centaurus} cluster of galaxies},
	volume = {457},
	issn = {0035-8711, 1365-2966},
	url = {https://academic.oup.com/mnras/article-lookup/doi/10.1093/mnras/stv2972},
	doi = {10.1093/mnras/stv2972},
	language = {en},
	number = {1},
	urldate = {2026-07-08},
	journal = {Monthly Notices of the Royal Astronomical Society},
	author = {Sanders, J. S. and Fabian, A. C. and Taylor, G. B. and Russell, H. R. and Blundell, K. M. and Canning, R. E. A. and Hlavacek-Larrondo, J. and Walker, S. A. and Grimes, C. K.},
	month = mar,
	year = {2016},
	pages = {82--109},
}

@article{sparks_imaging_1989,
	title = {Imaging observations of gas and dust in {NGC} 4696 and implications for cooling flow models},
	volume = {345},
	issn = {0004-637X, 1538-4357},
	url = {http://adsabs.harvard.edu/doi/10.1086/167890},
	doi = {10.1086/167890},
	language = {en},
	urldate = {2026-07-08},
	journal = {The Astrophysical Journal},
	author = {Sparks, W. B. and Macchetto, F. and Golombek, D.},
	month = oct,
	year = {1989},
	pages = {153},
}

@article{fabian_optical_1982,
	title = {Optical filaments around {NGC} 4696 in the {Centaurus} cluster},
	volume = {201},
	issn = {0035-8711, 1365-2966},
	url = {https://academic.oup.com/mnras/article-lookup/doi/10.1093/mnras/201.1.17P},
	doi = {10.1093/mnras/201.1.17P},
	language = {en},
	number = {1},
	urldate = {2026-07-08},
	journal = {Monthly Notices of the Royal Astronomical Society},
	author = {Fabian, A. C. and Atherton, P. D. and Taylor, K. and Nulsen, P. E. J.},
	month = nov,
	year = {1982},
	pages = {17P--19P},
}

@article{collaboration_bulk_2025,
	title = {The bulk motion of gas in the core of the {Centaurus} galaxy cluster},
	volume = {638},
	issn = {0028-0836, 1476-4687},
	url = {http://arxiv.org/abs/2502.08722},
	doi = {10.1038/s41586-024-08561-z},
	number = {8050},
	urldate = {2026-01-13},
	journal = {Nature},
	author = {XRISM Collaboration},
	month = feb,
	year = {2025},
	note = {arXiv:2502.08722 [astro-ph]},
	pages = {365--369},
}

@article{matilsky_x-ray_1985,
	title = {An {X}-ray study of the {Centaurus} {Cluster} of galaxies using {Einstein}},
	volume = {291},
	issn = {0004-637X, 1538-4357},
	url = {http://adsabs.harvard.edu/doi/10.1086/163102},
	doi = {10.1086/163102},
	language = {en},
	urldate = {2026-07-08},
	journal = {The Astrophysical Journal},
	author = {Matilsky, T. and Jones, C. and Forman, W.},
	month = apr,
	year = {1985},
	pages = {621},
}

@article{hamer_discovery_2019,
	title = {Discovery of a diffuse optical line emitting halo in the core of the {Centaurus} cluster of galaxies: line emission outside the protection of the filaments},
	volume = {483},
	copyright = {https://academic.oup.com/journals/pages/open\_access/funder\_policies/chorus/standard\_publication\_model},
	issn = {0035-8711, 1365-2966},
	shorttitle = {Discovery of a diffuse optical line emitting halo in the core of the {Centaurus} cluster of galaxies},
	url = {https://academic.oup.com/mnras/article/483/4/4984/5257854},
	doi = {10.1093/mnras/sty3459},
	language = {en},
	number = {4},
	urldate = {2026-05-29},
	journal = {Monthly Notices of the Royal Astronomical Society},
	author = {Hamer, S L and Fabian, A C and Russell, H R and Salom{\'e}, P and Combes, F and Olivares, V and Polles, F L and Edge, A C and Beckmann, R S},
	month = mar,
	year = {2019},
	pages = {4984--4998},
}

@article{mcdonald_massive_2012,
	title = {A massive, cooling-flow-induced starburst in the core of a luminous cluster of galaxies},
	volume = {488},
	copyright = {http://www.springer.com/tdm},
	issn = {0028-0836, 1476-4687},
	url = {https://www.nature.com/articles/nature11379},
	doi = {10.1038/nature11379},
	language = {en},
	number = {7411},
	urldate = {2026-05-20},
	journal = {Nature},
	author = {McDonald, M. and Bayliss, M. and Benson, B. A. and Foley, R. J. and Ruel, J. and Sullivan, P. and Veilleux, S. and Aird, K. A. and Ashby, M. L. N. and Bautz, M. and Bazin, G. and Bleem, L. E. and Brodwin, M. and Carlstrom, J. E. and Chang, C. L. and Cho, H. M. and Clocchiatti, A. and Crawford, T. M. and Crites, A. T. and De Haan, T. and Desai, S. and Dobbs, M. A. and Dudley, J. P. and Egami, E. and Forman, W. R. and Garmire, G. P. and George, E. M. and Gladders, M. D. and Gonzalez, A. H. and Halverson, N. W. and Harrington, N. L. and High, F. W. and Holder, G. P. and Holzapfel, W. L. and Hoover, S. and Hrubes, J. D. and Jones, C. and Joy, M. and Keisler, R. and Knox, L. and Lee, A. T. and Leitch, E. M. and Liu, J. and Lueker, M. and Luong-Van, D. and Mantz, A. and Marrone, D. P. and McMahon, J. J. and Mehl, J. and Meyer, S. S. and Miller, E. D. and Mocanu, L. and Mohr, J. J. and Montroy, T. E. and Murray, S. S. and Natoli, T. and Padin, S. and Plagge, T. and Pryke, C. and Rawle, T. D. and Reichardt, C. L. and Rest, A. and Rex, M. and Ruhl, J. E. and Saliwanchik, B. R. and Saro, A. and Sayre, J. T. and Schaffer, K. K. and Shaw, L. and Shirokoff, E. and Simcoe, R. and Song, J. and Spieler, H. G. and Stalder, B. and Staniszewski, Z. and Stark, A. A. and Story, K. and Stubbs, C. W. and Šuhada, R. and Van Engelen, A. and Vanderlinde, K. and Vieira, J. D. and Vikhlinin, A. and Williamson, R. and Zahn, O. and Zenteno, A.},
	month = aug,
	year = {2012},
	pages = {349--352},
}

@article{ubler_ga-nifs_2024,
	title = {{GA}-{NIFS}: {NIRSpec} reveals evidence for non-circular motions and {AGN} feedback in {GN20}},
	volume = {533},
	copyright = {https://creativecommons.org/licenses/by/4.0/},
	issn = {0035-8711, 1365-2966},
	shorttitle = {{GA}-{NIFS}},
	url = {https://academic.oup.com/mnras/article/533/4/4287/7737681},
	doi = {10.1093/mnras/stae1993},
	language = {en},
	number = {4},
	urldate = {2026-05-20},
	journal = {Monthly Notices of the Royal Astronomical Society},
	author = {Übler, Hannah and D’Eugenio, Francesco and Perna, Michele and Arribas, Santiago and Jones, Gareth C and Bunker, Andrew J and Carniani, Stefano and Charlot, St{\'e}phane and Maiolino, Roberto and Rodríguez del Pino, Bruno and Willott, Chris J and Böker, Torsten and Cresci, Giovanni and Kumari, Nimisha and Lamperti, Isabella and Parlanti, Eleonora and Scholtz, Jan and Venturi, Giacomo},
	month = sep,
	year = {2024},
	pages = {4287--4299},
}

@article{appleton_powerful_2017,
	title = {Powerful {H}$_{\textrm{2}}$ {Line} {Cooling} in {Stephan}’s {Quintet}. {II}. {Group}-wide {Gas} and {Shock} {Modeling} of the {Warm} {H}$_{\textrm{2}}$ and a {Comparison} with [{C} ii] 157.7 μm {Emission} and {Kinematics}},
	volume = {836},
	issn = {0004-637X, 1538-4357},
	url = {https://iopscience.iop.org/article/10.3847/1538-4357/836/1/76},
	doi = {10.3847/1538-4357/836/1/76},
	number = {1},
	urldate = {2026-05-20},
	journal = {The Astrophysical Journal},
	author = {Appleton, P. N. and Guillard, P. and Togi, A. and Alatalo, K. and Boulanger, F. and Cluver, M. and Pineau Des Forêts, G. and Lisenfeld, U. and Ogle, P. and Xu, C. K.},
	month = feb,
	year = {2017},
	pages = {76},
}

@article{sternberg_infrared_1989,
	title = {The infrared response of molecular hydrogen gas to ultraviolet radiation - {High}-density regions},
	volume = {338},
	issn = {0004-637X, 1538-4357},
	url = {http://adsabs.harvard.edu/doi/10.1086/167193},
	doi = {10.1086/167193},
	language = {en},
	urldate = {2026-05-20},
	journal = {The Astrophysical Journal},
	author = {Sternberg, A. and Dalgarno, A.},
	month = mar,
	year = {1989},
	pages = {197},
}

@article{hollenbach_molecule_1989,
	title = {Molecule formation and infrared emission in fast interstellar shocks. {III} - {Results} for {J} shocks in molecular clouds},
	volume = {342},
	issn = {0004-637X, 1538-4357},
	url = {http://adsabs.harvard.edu/doi/10.1086/167595},
	doi = {10.1086/167595},
	language = {en},
	urldate = {2026-05-20},
	journal = {The Astrophysical Journal},
	author = {Hollenbach, David and McKee, Christopher F.},
	month = jul,
	year = {1989},
	pages = {306},
}

@article{harris_array_2020,
  title = {Array Programming with {{NumPy}}},
  author = {Harris, Charles R. and Millman, K. Jarrod and Van Der Walt, St{\'e}fan J. and Gommers, Ralf and Virtanen, Pauli and Cournapeau, David and Wieser, Eric and Taylor, Julian and Berg, Sebastian and Smith, Nathaniel J. and Kern, Robert and Picus, Matti and Hoyer, Stephan and Van Kerkwijk, Marten H. and Brett, Matthew and Haldane, Allan and Del R{\'i}o, Jaime Fern{\'a}ndez and Wiebe, Mark and Peterson, Pearu and {G{\'e}rard-Marchant}, Pierre and Sheppard, Kevin and Reddy, Tyler and Weckesser, Warren and Abbasi, Hameer and Gohlke, Christoph and Oliphant, Travis E.},
  year = 2020,
  month = sep,
  journal = {Nature},
  volume = {585},
  number = {7825},
  pages = {357--362},
  issn = {0028-0836, 1476-4687},
  doi = {10.1038/s41586-020-2649-2},
  urldate = {2026-04-27},
  langid = {english}
}

@ARTICLE{begelman_1990,
       author = {{Begelman}, Mitchell C. and {Fabian}, A.~C.},
        title = "{Turbulent mixing layers in the interstellar and intracluster medium.}",
      journal = {\mnras},
         year = 1990,
        month = may,
       volume = {244},
        pages = {26P-29},
       adsurl = {https://ui.adsabs.harvard.edu/abs/1990MNRAS.244P..26B}
}

@ARTICLE{slavin_1993,
       author = {{Slavin}, Jonathan D. and {Shull}, J.~M. and {Begelman}, Mitchell C.},
        title = "{Turbulent Mixing Layers in the Interstellar Medium of Galaxies}",
      journal = {\apj},
         year = 1993,
        month = apr,
       volume = {407},
        pages = {83},
          doi = {10.1086/172494},
       adsurl = {https://ui.adsabs.harvard.edu/abs/1993ApJ...407...83S}
}

@ARTICLE{voit_donahue_1997,
       author = {{Voit}, G. Mark and {Donahue}, Megan},
        title = "{A Deep Look at the Emission-Line Nebula in Abell 2597}",
      journal = {\apj},
         year = 1997,
        month = sep,
       volume = {486},
       number = {1},
        pages = {242-252},
          doi = {10.1086/304509},
archivePrefix = {arXiv},
       eprint = {astro-ph/9706107},
 primaryClass = {astro-ph},
       adsurl = {https://ui.adsabs.harvard.edu/abs/1997ApJ...486..242V}
}

@article{hunter_matplotlib_2007,
  title = {Matplotlib: {{A 2D}} Graphics Environment},
  author = {Hunter, J. D.},
  year = 2007,
  journal = {Computing in Science \& Engineering},
  volume = {9},
  number = {3},
  pages = {90--95},
  publisher = {IEEE COMPUTER SOC},
  doi = {10.1109/MCSE.2007.55}
}

@article{theastropycollaboration_astropy_2022,
  title = {The {{Astropy Project}}: {{Sustaining}} and {{Growing}} a {{Community-oriented Open-source Project}} and the {{Latest Major Release}} (v5.0) of the {{Core Package}}*},
  shorttitle = {The {{Astropy Project}}},
  author = {{The Astropy Collaboration} and {Price-Whelan}, Adrian M. and Lim, Pey Lian and Earl, Nicholas and Starkman, Nathaniel and Bradley, Larry and Shupe, David L. and Patil, Aarya A. and Corrales, Lia and Brasseur, C. E. and N{\"o}the, Maximilian and Donath, Axel and Tollerud, Erik and Morris, Brett M. and Ginsburg, Adam and Vaher, Eero and Weaver, Benjamin A. and Tocknell, James and Jamieson, William and Van Kerkwijk, Marten H. and Robitaille, Thomas P. and Merry, Bruce and Bachetti, Matteo and G{\"u}nther, H. Moritz and {Paper Authors} and Aldcroft, Thomas L. and {Alvarado-Montes}, Jaime A. and Archibald, Anne M. and B{\'o}di, Attila and Bapat, Shreyas and Barentsen, Geert and Baz{\'a}n, Juanjo and Biswas, Manish and Boquien, M{\'e}d{\'e}ric and Burke, D. J. and Cara, Daria and Cara, Mihai and Conroy, Kyle E and Conseil, Simon and Craig, Matthew W. and Cross, Robert M. and Cruz, Kelle L. and D'Eugenio, Francesco and Dencheva, Nadia and Devillepoix, Hadrien A. R. and Dietrich, J{\"o}rg P. and Eigenbrot, Arthur Davis and Erben, Thomas and Ferreira, Leonardo and {Foreman-Mackey}, Daniel and Fox, Ryan and Freij, Nabil and Garg, Suyog and Geda, Robel and Glattly, Lauren and Gondhalekar, Yash and Gordon, Karl D. and Grant, David and Greenfield, Perry and Groener, Austen M. and Guest, Steve and Gurovich, Sebastian and Handberg, Rasmus and Hart, Akeem and {Hatfield-Dodds}, Zac and Homeier, Derek and Hosseinzadeh, Griffin and Jenness, Tim and Jones, Craig K. and Joseph, Prajwel and Kalmbach, J. Bryce and Karamehmetoglu, Emir and Ka{\l}uszy{\'n}ski, Miko{\l}aj and Kelley, Michael S. P. and Kern, Nicholas and Kerzendorf, Wolfgang E. and Koch, Eric W. and Kulumani, Shankar and Lee, Antony and Ly, Chun and Ma, Zhiyuan and MacBride, Conor and Maljaars, Jakob M. and Muna, Demitri and Murphy, N. A. and Norman, Henrik and O'Steen, Richard and Oman, Kyle A. and Pacifici, Camilla and Pascual, Sergio and {Pascual-Granado}, J. and Patil, Rohit R. and Perren, Gabriel I and Pickering, Timothy E. and Rastogi, Tanuj and Roulston, Benjamin R. and Ryan, Daniel F and Rykoff, Eli S. and Sabater, Jose and Sakurikar, Parikshit and Salgado, Jes{\'u}s and Sanghi, Aniket and Saunders, Nicholas and Savchenko, Volodymyr and Schwardt, Ludwig and {Seifert-Eckert}, Michael and Shih, Albert Y. and Jain, Anany Shrey and Shukla, Gyanendra and Sick, Jonathan and Simpson, Chris and Singanamalla, Sudheesh and Singer, Leo P. and Singhal, Jaladh and Sinha, Manodeep and Sip{\H o}cz, Brigitta M. and Spitler, Lee R. and Stansby, David and Streicher, Ole and {\v S}umak, Jani and Swinbank, John D. and Taranu, Dan S. and Tewary, Nikita and Tremblay, Grant R. and {Val-Borro}, Miguel De and Van Kooten, Samuel J. and Vasovi{\'c}, Zlatan and Verma, Shresth and De Miranda Cardoso, Jos{\'e} Vin{\'i}cius and Williams, Peter K. G. and Wilson, Tom J. and Winkel, Benjamin and {Wood-Vasey}, W. M. and Xue, Rui and Yoachim, Peter and Zhang, Chen and Zonca, Andrea and {Astropy Project Contributors}},
  year = 2022,
  month = aug,
  journal = {\apj},
  volume = {935},
  number = {2},
  pages = {167},
  issn = {0004-637X, 1538-4357},
  doi = {10.3847/1538-4357/ac7c74},
  urldate = {2026-04-27},
  langid = {english}
}

@article{olivares_hx-ray_2025,
	title = {An {Hα}–{X}-ray surface-brightness correlation for filaments in cooling-flow clusters},
	volume = {9},
	issn = {2397-3366},
	url = {https://www.nature.com/articles/s41550-024-02473-8},
	doi = {10.1038/s41550-024-02473-8},
	language = {en},
	number = {3},
	urldate = {2026-07-07},
	journal = {Nature Astronomy},
	author = {Olivares, Valeria and Picquenot, Adrien and Su, Yuanyuan and Gaspari, Massimo and Gendron-Marsolais, Marie-Lou and Polles, Fiorella L. and Nulsen, Paul},
	month = jan,
	year = {2025},
	pages = {449--457},
}

@article{taylor_radio_2006,
	title = {The low-power nucleus of {PKS} 1246-410 in the {Centaurus} {Cluster}},
	volume = {365},
	issn = {0035-8711, 1365-2966},
	url = {http://arxiv.org/abs/astro-ph/0510247},
	doi = {10.1111/j.1365-2966.2005.09705.x},
	number = {3},
	urldate = {2025-10-28},
	journal = {Monthly Notices of the Royal Astronomical Society},
	author = {Taylor, G. B. and Sanders, J. S. and Fabian, A. C. and Allen, S. W.},
	month = jan,
	year = {2006},
	note = {arXiv:astro-ph/0510247},
	pages = {705--711},
}

@article{ogle_jet-powered_2010,
	title = {Jet-powered {Molecular} {Hydrogen} {Emission} from {Radio} {Galaxies}},
	volume = {724},
	issn = {0004-637X},
	url = {https://ui.adsabs.harvard.edu/abs/2010ApJ...724.1193O/abstract},
	doi = {10.1088/0004-637X/724/2/1193},
	language = {en},
	number = {2},
	urldate = {2026-05-05},
	journal = {The Astrophysical Journal, Volume 724, Issue 2, pp. 1193-1217 (2010).},
	author = {Ogle, Patrick and Boulanger, Francois and Guillard, Pierre and Evans, Daniel A. and Antonucci, Robert and Appleton, P. N. and Nesvadba, Nicole and Leipski, Christian},
	month = dec,
	year = {2010},
	pages = {1193},
}

@article{davies_fueling_2014,
	title = {{FUELING} {ACTIVE} {GALACTIC} {NUCLEI}. {II}. {SPATIALLY} {RESOLVED} {MOLECULAR} {INFLOWS} {AND} {OUTFLOWS}},
	volume = {792},
	issn = {0004-637X},
	url = {https://doi.org/10.1088/0004-637X/792/2/101},
	doi = {10.1088/0004-637X/792/2/101},
	language = {en},
	number = {2},
	urldate = {2026-05-05},
	journal = {The Astrophysical Journal},
	publisher = {The American Astronomical Society},
	author = {Davies, R. I. and Maciejewski, W. and Hicks, E. K. S. and Emsellem, E. and Erwin, P. and Burtscher, L. and Dumas, G. and Lin, M. and Malkan, M. A. and Müller-Sánchez, F. and Orban de Xivry, G. and Rosario, D. J. and Schnorr-Müller, A. and Tran, A.},
	month = aug,
	year = {2014},
	pages = {101},
}

@article{ghodsi_molecular_2024,
	title = {Molecular gas excitation in the circumgalactic medium of {MACS1931}–26},
	volume = {689},
	copyright = {https://creativecommons.org/licenses/by/4.0},
	issn = {0004-6361, 1432-0746},
	url = {https://www.aanda.org/10.1051/0004-6361/202449642},
	doi = {10.1051/0004-6361/202449642},
	urldate = {2025-09-04},
	journal = {Astronomy \& Astrophysics},
	author = {Ghodsi, L. and Zhou, J. and Andreani, P. and De Breuck, C. and Man, A. W. S. and Miyamoto, Y. and Bisbas, T. G. and Lundgren, A. and Zhang, Z.-Y.},
	month = sep,
	year = {2024},
	pages = {A67},
}

@article{tremblay_galaxy-scale_2018,
	title = {A {Galaxy}-{Scale} {Fountain} of {Cold} {Molecular} {Gas} {Pumped} by a {Black} {Hole}},
	volume = {865},
	issn = {0004-637X, 1538-4357},
	url = {http://arxiv.org/abs/1808.00473},
	doi = {10.3847/1538-4357/aad6dd},
	number = {1},
	urldate = {2025-09-10},
	journal = {The Astrophysical Journal},
	author = {Tremblay, Grant R. and Combes, Françoise and Oonk, J. B. Raymond and Russell, Helen R. and McDonald, Michael A. and Gaspari, Massimo and Husemann, Bernd and Nulsen, Paul E. J. and McNamara, Brian R. and Hamer, Stephen L. and O'Dea, Christopher P. and Baum, Stefi A. and Davis, Timothy A. and Donahue, Megan and Voit, G. Mark and Edge, Alastair C. and Blanton, Elizabeth L. and Bremer, Malcolm N. and Bulbul, Esra and Clarke, Tracy E. and David, Laurence P. and Edwards, Louise O. V. and Eggerman, Dominic A. and Fabian, Andrew C. and Forman, William R. and Jones, Christine and Kerman, Nathaniel and Kraft, Ralph P. and Li, Yuan and Powell, Meredith C. and Randall, Scott W. and Salom{\'e}, Philippe and Simionescu, Aurora and Su, Yuanyuan and Sun, Ming and Urry, C. Megan and Vantyghem, Adrian N. and Wilkes, Belinda J. and ZuHone, John A.},
	month = sep,
	year = {2018},
	note = {arXiv:1808.00473 [astro-ph]},
	pages = {13},
}

@article{fabian_hst_2016,
	title = {\textit{{HST}} imaging of the dusty filaments and nucleus swirl in {NGC4696} at the centre of the {Centaurus} {Cluster}},
	volume = {461},
	issn = {0035-8711, 1365-2966},
	url = {https://academic.oup.com/mnras/article-lookup/doi/10.1093/mnras/stw1350},
	doi = {10.1093/mnras/stw1350},
	language = {en},
	number = {1},
	urldate = {2025-09-04},
	journal = {Monthly Notices of the Royal Astronomical Society},
	author = {Fabian, A. C. and Walker, S. A. and Russell, H. R. and Pinto, C. and Canning, R. E. A. and Salome, P. and Sanders, J. S. and Taylor, G. B. and Zweibel, E. G. and Conselice, C. J. and Combes, F. and Crawford, C. S. and Ferland, G. J. and Gallagher, J. S. and Hatch, N. A. and Johnstone, R. M. and Reynolds, C. S.},
	month = sep,
	year = {2016},
	pages = {922--928},
}

@article{danforth_far-uv_2016,
	title = {{FAR}-{UV} {EMISSION} {PROPERTIES} {OF} {FR1} {RADIO} {GALAXIES}},
	volume = {832},
	issn = {0004-637X, 1538-4357},
	url = {https://iopscience.iop.org/article/10.3847/0004-637X/832/1/76},
	doi = {10.3847/0004-637X/832/1/76},
	number = {1},
	urldate = {2025-09-04},
	journal = {The Astrophysical Journal},
	author = {Danforth, Charles W. and Stocke, John T. and France, Kevin and Begelman, Mitchell C. and Perlman, Eric},
	month = nov,
	year = {2016},
	pages = {76},
}

@article{panagoulia_searching_2013,
	title = {Searching for the missing iron mass in the core of the {Centaurus} cluster},
	volume = {433},
	issn = {1365-2966, 0035-8711},
	url = {http://academic.oup.com/mnras/article/433/4/3290/1750333/Searching-for-the-missing-iron-mass-in-the-core-of},
	doi = {10.1093/mnras/stt969},
	language = {en},
	number = {4},
	urldate = {2025-09-04},
	journal = {Monthly Notices of the Royal Astronomical Society},
	author = {Panagoulia, E. K. and Fabian, A. C. and Sanders, J. S.},
	month = aug,
	year = {2013},
	pages = {3290--3296},
}

@article{canning_deep_2011,
	title = {A deep spectroscopic study of the filamentary nebulosity in {NGC} 4696, the brightest cluster galaxy in the {Centaurus} cluster: {The} filaments of {NGC} 4696},
	volume = {417},
	issn = {00358711},
	shorttitle = {A deep spectroscopic study of the filamentary nebulosity in {NGC} 4696, the brightest cluster galaxy in the {Centaurus} cluster},
	url = {https://academic.oup.com/mnras/article-lookup/doi/10.1111/j.1365-2966.2011.19470.x},
	doi = {10.1111/j.1365-2966.2011.19470.x},
	language = {en},
	number = {4},
	urldate = {2025-09-04},
	journal = {Monthly Notices of the Royal Astronomical Society},
	author = {Canning, R. E. A. and Fabian, A. C. and Johnstone, R. M. and Sanders, J. S. and Crawford, C. S. and Ferland, G. J. and Hatch, N. A.},
	month = nov,
	year = {2011},
	pages = {3080--3099},
}

@article{leftley_jwstnirspec_2024,
	title = {{JWST}/{NIRSpec} and {MIRI} observations of an expanding, jet-driven bubble of warm {H}{\textless}{SUB}{\textgreater}2{\textless}/{SUB}{\textgreater} in the radio galaxy {3C} 326 {N}},
	volume = {689},
	issn = {0004-6361},
	url = {https://ui.adsabs.harvard.edu/abs/2024A&A...689A.314L/abstract},
	doi = {10.1051/0004-6361/202449848},
	language = {en},
	urldate = {2025-07-03},
	journal = {Astronomy and Astrophysics},
	author = {Leftley, J. H. and Nesvadba, N. P. H. and Bicknell, G. V. and Janssen, R. M. J. and Mukherjee, D. and Petrov, R. and Shende, M. B. and Zovaro, H. R. M.},
	month = sep,
	year = {2024},
	pages = {A314},
}

@article{federle_turbulent_2024,
	title = {The turbulent life of {NGC} 4696 as told by its globular cluster system},
	volume = {689},
	copyright = {https://creativecommons.org/licenses/by/4.0},
	issn = {0004-6361, 1432-0746},
	url = {https://www.aanda.org/10.1051/0004-6361/202348235},
	doi = {10.1051/0004-6361/202348235},
	urldate = {2025-09-04},
	journal = {Astronomy \& Astrophysics},
	author = {Federle, S. and Gómez, M. and Mieske, S. and Harris, W. E. and Hilker, M. and Yegorova, I. A. and Harris, G. L. H.},
	month = sep,
	year = {2024},
	pages = {A342},
}

@article{churazov_evidence_1999,
	title = {Evidence for {Merging} in the {Centaurus} {Cluster}},
	volume = {520},
	issn = {0004-637X, 1538-4357},
	url = {https://iopscience.iop.org/article/10.1086/307421},
	doi = {10.1086/307421},
	language = {en},
	number = {1},
	urldate = {2025-09-04},
	journal = {The Astrophysical Journal},
	author = {Churazov, E. and Gilfanov, M. and Forman, W. and Jones, C.},
	month = jul,
	year = {1999},
	pages = {105--110},
}

@article{anderson_searching_2016,
	title = {Searching for {FUV} line emission from 10$^{\textrm{7}}$ {K} gas in massive elliptical galaxies and galaxy clusters as a tracer of turbulent velocities},
	volume = {459},
	issn = {0035-8711, 1365-2966},
	url = {https://academic.oup.com/mnras/article-lookup/doi/10.1093/mnras/stw822},
	doi = {10.1093/mnras/stw822},
	language = {en},
	number = {3},
	urldate = {2025-09-04},
	journal = {Monthly Notices of the Royal Astronomical Society},
	author = {Anderson, Michael E. and Sunyaev, Rashid},
	month = jul,
	year = {2016},
	pages = {2806--2821},
}

@article{mur_near-infrared_2006,
	title = {The {Near}-{Infrared} {Surface} {Brightness} {Distribution} of {NGC} 4696},
	volume = {23},
	copyright = {https://www.cambridge.org/core/terms},
	issn = {1323-3580, 1448-6083},
	url = {https://www.cambridge.org/core/product/identifier/S1323358000002800/type/journal_article},
	doi = {10.1071/AS05037},
	language = {en},
	number = {1},
	urldate = {2025-09-04},
	journal = {Publications of the Astronomical Society of Australia},
	author = {Mur, Pablo Arnalte and Ellis, Simon C. and Colless, Matthew},
	year = {2006},
	pages = {33--37},
}

@article{crawford_extended_2005,
	title = {The extended {H} -emitting filaments surrounding {NGC} 4696, the central galaxy of the {Centaurus} cluster},
	volume = {363},
	issn = {0035-8711, 1365-2966},
	url = {https://academic.oup.com/mnras/article-lookup/doi/10.1111/j.1365-2966.2005.09463.x},
	doi = {10.1111/j.1365-2966.2005.09463.x},
	language = {en},
	number = {1},
	urldate = {2025-09-04},
	journal = {Monthly Notices of the Royal Astronomical Society},
	author = {Crawford, C. S. and Hatch, N. A. and Fabian, A. C. and Sanders, J. S.},
	month = oct,
	year = {2005},
	pages = {216--222},
}

@article{fabian_observational_2012,
	title = {Observational {Evidence} of {Active} {Galactic} {Nuclei} {Feedback}},
	volume = {50},
	issn = {0066-4146, 1545-4282},
	url = {https://www.annualreviews.org/doi/10.1146/annurev-astro-081811-125521},
	doi = {10.1146/annurev-astro-081811-125521},
	language = {en},
	number = {1},
	urldate = {2025-11-06},
	journal = {Annual Review of Astronomy and Astrophysics},
	author = {Fabian, A.C.},
	month = sep,
	year = {2012},
	pages = {455--489},
}

@article{mcnamara_mechanical_2012,
	title = {Mechanical feedback from active galactic nuclei in galaxies, groups and clusters},
	volume = {14},
	issn = {1367-2630},
	url = {https://iopscience.iop.org/article/10.1088/1367-2630/14/5/055023},
	doi = {10.1088/1367-2630/14/5/055023},
	number = {5},
	urldate = {2025-11-06},
	journal = {New Journal of Physics},
	author = {McNamara, B R and Nulsen, P E J},
	month = may,
	year = {2012},
	pages = {055023},
}

@article{gaspari_constraining_2013,
	title = {Constraining turbulence and conduction in the hot {ICM} through density perturbations},
	volume = {559},
	issn = {0004-6361, 1432-0746},
	url = {http://www.aanda.org/10.1051/0004-6361/201322295},
	doi = {10.1051/0004-6361/201322295},
	urldate = {2025-11-06},
	journal = {Astronomy \& Astrophysics},
	author = {Gaspari, M. and Churazov, E.},
	month = nov,
	year = {2013},
	pages = {A78},
}

@article{voit_precipitation-regulated_2015,
	title = {{PRECIPITATION}-{REGULATED} {STAR} {FORMATION} {IN} {GALAXIES}},
	volume = {808},
	copyright = {http://iopscience.iop.org/info/page/text-and-data-mining},
	issn = {2041-8213},
	url = {https://iopscience.iop.org/article/10.1088/2041-8205/808/1/L30},
	doi = {10.1088/2041-8205/808/1/L30},
	number = {1},
	urldate = {2025-11-06},
	journal = {The Astrophysical Journal},
	author = {Voit, G. Mark and Bryan, Greg L. and O’Shea, Brian W. and Donahue, Megan},
	month = jul,
	year = {2015},
	pages = {L30},
}

@article{allen_rosat_1994,
	title = {A {ROSAT} {PSPC} investigation of the {Centaurus} cluster of galaxies},
	volume = {269},
	issn = {0035-8711, 1365-2966},
	url = {https://academic.oup.com/mnras/article-lookup/doi/10.1093/mnras/269.2.409},
	doi = {10.1093/mnras/269.2.409},
	language = {en},
	number = {2},
	urldate = {2026-07-08},
	journal = {Monthly Notices of the Royal Astronomical Society},
	author = {Allen, S. W. and Fabian, A. C.},
	month = jul,
	year = {1994},
	pages = {409--426},
}

@article{sanders_constraints_2010,
	title = {Constraints on turbulent velocity broadening for a sample of clusters, groups and elliptical galaxies using {XMM}-{Newton}: {Turbulence} in galaxy clusters},
	issn = {00358711},
	shorttitle = {Constraints on turbulent velocity broadening for a sample of clusters, groups and elliptical galaxies using {XMM}-{Newton}},
	url = {https://academic.oup.com/mnras/article-lookup/doi/10.1111/j.1365-2966.2010.17561.x},
	doi = {10.1111/j.1365-2966.2010.17561.x},
	language = {en},
	urldate = {2025-11-11},
	journal = {Monthly Notices of the Royal Astronomical Society},
	author = {Sanders, J. S. and Fabian, A. C. and Smith, R. K.},
	month = oct,
	year = {2010},
	pages = {no--no},
}

@ARTICLE{Rauscher2024,
       author = {{Rauscher}, Bernard J.},
        title = "{NSClean: An Algorithm for Removing Correlated Noise from JWST NIRSpec Images}",
      journal = {\pasp},
         year = 2024,
        month = jan,
       volume = {136},
       number = {1},
          eid = {015001},
        pages = {015001},
          doi = {10.1088/1538-3873/ad1b36},
archivePrefix = {arXiv},
       eprint = {2306.03250},
 primaryClass = {astro-ph.IM},
       adsurl = {https://ui.adsabs.harvard.edu/abs/2024PASP..136a5001R}
}

@article{cappellari_parametric_2004,
    title = {Parametric {Recovery} of {Line}‐of‐{Sight} {Velocity} {Distributions} from {Absorption}‐{Line} {Spectra} of {Galaxies} via {Penalized} {Likelihood}},
    volume = {116},
    issn = {0004-6280, 1538-3873},
    url = {http://iopscience.iop.org/article/10.1086/381875},
    doi = {10.1086/381875},
    language = {en},
    number = {816},
    urldate = {2026-07-27},
    journal = {Publications of the Astronomical Society of the Pacific},
    author = {Cappellari, Michele and Emsellem, Eric},
    month = feb,
    year = {2004},
    pages = {138--147},
}

@article{vazdekis_uv-extended_2016,
    title = {{UV}-extended {E}-{MILES} stellar population models: young components in massive early-type galaxies},
    volume = {463},
    issn = {0035-8711, 1365-2966},
    shorttitle = {{UV}-extended {E}-{MILES} stellar population models},
    url = {https://academic.oup.com/mnras/article-lookup/doi/10.1093/mnras/stw2231},
    doi = {10.1093/mnras/stw2231},
    language = {en},
    number = {4},
    urldate = {2026-07-27},
    journal = {Monthly Notices of the Royal Astronomical Society},
    author = {Vazdekis, A. and Koleva, M. and Ricciardelli, E. and Röck, B. and Falcón-Barroso, J.},
    month = dec,
    year = {2016},
    pages = {3409--3436},
}

@article{hitomicollaborationQuiescentIntraclusterMedium2016,
    title = {The quiescent intracluster medium in the core of the {Perseus} cluster},
    volume = {535},
    issn = {0028-0836, 1476-4687},
    url = {https://www.nature.com/articles/nature18627},
    doi = {10.1038/nature18627},
    language = {en},
    number = {7610},
    urldate = {2024-11-17},
    journal = {Nature},
    author = {{Hitomi Collaboration} and Aharonian, Felix and Akamatsu, Hiroki and Akimoto, Fumie and Allen, Steven W. and Anabuki, Naohisa and Angelini, Lorella and Arnaud, Keith and Audard, Marc and Awaki, Hisamitsu and Axelsson, Magnus and Bamba, Aya and Bautz, Marshall and Blandford, Roger and Brenneman, Laura and Brown, Gregory V. and Bulbul, Esra and Cackett, Edward and Chernyakova, Maria and Chiao, Meng and Coppi, Paolo and Costantini, Elisa and De Plaa, Jelle and Den Herder, Jan-Willem and Done, Chris and Dotani, Tadayasu and Ebisawa, Ken and Eckart, Megan and Enoto, Teruaki and Ezoe, Yuichiro and Fabian, Andrew C. and Ferrigno, Carlo and Foster, Adam and Fujimoto, Ryuichi and Fukazawa, Yasushi and Furuzawa, Akihiro and Galeazzi, Massimiliano and Gallo, Luigi and Gandhi, Poshak and Giustini, Margherita and Goldwurm, Andrea and Gu, Liyi and Guainazzi, Matteo and Haba, Yoshito and Hagino, Kouichi and Hamaguchi, Kenji and Harrus, Ilana and Hatsukade, Isamu and Hayashi, Katsuhiro and Hayashi, Takayuki and Hayashida, Kiyoshi and Hiraga, Junko and Hornschemeier, Ann and Hoshino, Akio and Hughes, John and Iizuka, Ryo and Inoue, Hajime and Inoue, Yoshiyuki and Ishibashi, Kazunori and Ishida, Manabu and Ishikawa, Kumi and Ishisaki, Yoshitaka and Itoh, Masayuki and Iyomoto, Naoko and Kaastra, Jelle and Kallman, Timothy and Kamae, Tuneyoshi and Kara, Erin and Kataoka, Jun and Katsuda, Satoru and Katsuta, Junichiro and Kawaharada, Madoka and Kawai, Nobuyuki and Kelley, Richard and Khangulyan, Dmitry and Kilbourne, Caroline and King, Ashley and Kitaguchi, Takao and Kitamoto, Shunji and Kitayama, Tetsu and Kohmura, Takayoshi and Kokubun, Motohide and Koyama, Shu and Koyama, Katsuji and Kretschmar, Peter and Krimm, Hans and Kubota, Aya and Kunieda, Hideyo and Laurent, Philippe and Lebrun, François and Lee, Shiu-Hang and Leutenegger, Maurice and Limousin, Olivier and Loewenstein, Michael and Long, Knox S. and Lumb, David and Madejski, Grzegorz and Maeda, Yoshitomo and Maier, Daniel and Makishima, Kazuo and Markevitch, Maxim and Matsumoto, Hironori and Matsushita, Kyoko and McCammon, Dan and McNamara, Brian and Mehdipour, Missagh and Miller, Eric and Miller, Jon and Mineshige, Shin and Mitsuda, Kazuhisa and Mitsuishi, Ikuyuki and Miyazawa, Takuya and Mizuno, Tsunefumi and Mori, Hideyuki and Mori, Koji and Moseley, Harvey and Mukai, Koji and Murakami, Hiroshi and Murakami, Toshio and Mushotzky, Richard and Nagino, Ryo and Nakagawa, Takao and Nakajima, Hiroshi and Nakamori, Takeshi and Nakano, Toshio and Nakashima, Shinya and Nakazawa, Kazuhiro and Nobukawa, Masayoshi and Noda, Hirofumi and Nomachi, Masaharu and O’Dell, Steve and Odaka, Hirokazu and Ohashi, Takaya and Ohno, Masanori and Okajima, Takashi and Ota, Naomi and Ozaki, Masanobu and Paerels, Frits and Paltani, Stephane and Parmar, Arvind and Petre, Robert and Pinto, Ciro and Pohl, Martin and Porter, F. Scott and Pottschmidt, Katja and Ramsey, Brian and Reynolds, Christopher and Russell, Helen and Safi-Harb, Samar and Saito, Shinya and Sakai, Kazuhiro and Sameshima, Hiroaki and Sato, Goro and Sato, Kosuke and Sato, Rie and Sawada, Makoto and Schartel, Norbert and Serlemitsos, Peter and Seta, Hiromi and Shidatsu, Megumi and Simionescu, Aurora and Smith, Randall and Soong, Yang and Stawarz, Lukasz and Sugawara, Yasuharu and Sugita, Satoshi and Szymkowiak, Andrew and Tajima, Hiroyasu and Takahashi, Hiromitsu and Takahashi, Tadayuki and Takeda, Shin’ichiro and Takei, Yoh and Tamagawa, Toru and Tamura, Keisuke and Tamura, Takayuki and Tanaka, Takaaki and Tanaka, Yasuo and Tanaka, Yasuyuki and Tashiro, Makoto and Tawara, Yuzuru and Terada, Yukikatsu and Terashima, Yuichi and Tombesi, Francesco and Tomida, Hiroshi and Tsuboi, Yohko and Tsujimoto, Masahiro and Tsunemi, Hiroshi and Tsuru, Takeshi and Uchida, Hiroyuki and Uchiyama, Hideki and Uchiyama, Yasunobu and Ueda, Shutaro and Ueda, Yoshihiro and Ueno, Shiro and Uno, Shin’ichiro and Urry, Meg and Ursino, Eugenio and De Vries, Cor and Watanabe, Shin and Werner, Norbert and Wik, Daniel and Wilkins, Dan and Williams, Brian and Yamada, Shinya and Yamaguchi, Hiroya and Yamaoka, Kazutaka and Yamasaki, Noriko Y. and Yamauchi, Makoto and Yamauchi, Shigeo and Yaqoob, Tahir and Yatsu, Yoichi and Yonetoku, Daisuke and Yoshida, Atsumasa and Yuasa, Takayuki and Zhuravleva, Irina and Zoghbi, Abderahmen},
    month = jul,
    year = {2016},
    pages = {117--121},
}
\bibliographystyle{aasjournalv7}

\end{CJK}
\end{document}